\documentclass[aps,preprint,english]{revtex4}
\usepackage[T1]{fontenc}
\usepackage[latin9]{inputenc}
\usepackage{amsmath}
\usepackage{amssymb}
\usepackage{graphicx}

\makeatletter

\providecommand{\tabularnewline}{\\}

\@ifundefined{textcolor}{}
{%
 \definecolor{BLACK}{gray}{0}
 \definecolor{WHITE}{gray}{1}
 \definecolor{RED}{rgb}{1,0,0}
 \definecolor{GREEN}{rgb}{0,1,0}
 \definecolor{BLUE}{rgb}{0,0,1}
 \definecolor{CYAN}{cmyk}{1,0,0,0}
 \definecolor{MAGENTA}{cmyk}{0,1,0,0}
 \definecolor{YELLOW}{cmyk}{0,0,1,0}
}

\usepackage{adjustbox}

\usepackage[position=top,caption=false]{subfig}

\@ifundefined{showcaptionsetup}{}{%
 \PassOptionsToPackage{caption=false}{subfig}}
\usepackage{subfig}
\makeatother

\begin{document}

\preprint{This line only printed with preprint option}

\title{Analytical Properties of Linear Electrostatic
Waves in Two-Component Quantum and Classical Plasmas}

\author{Shane Rightley}
\email{shane.rightley@colorado.edu}

\affiliation{Center for Integrated Plasma Studies, University of Colorado, Boulder}

\author{Dmitri Uzdensky}
\email{uzdensky@colorado.edu}

\affiliation{Center for Integrated Plasma Studies, University of Colorado, Boulder}
\begin{abstract}
We examine the properties of linear electrostatic waves in unmagnetized
quantum and classical plasmas consisting of one or two populations
of electrons with analytically tractable distribution functions in
the presence of a stationary neutralising ion background. Beginning
with the kinetic quantum plasma longitudinal susceptibility, we assess
the effects due to increasing complexity of the background distribution
function. Firstly, we analyse dispersion and Landau damping in one-component
plasmas and consider distribution functions with a variety of analytical
properties: the Dirac delta function, the Cauchy profile with two
complex first-order poles, the squared Cauchy profile with two second
order poles, and the inverse-quartic profile with four first-order
poles; we also briefly discuss the non-meromorphic totally- and arbitrarily-degenerate
Fermi-Dirac distribution. In order to study electrostatic instabilities,
we then turn to plasmas with two populations of electrons streaming
relative to each other in two cases: a symmetric case of two counter-streaming
identical populations and a bump-on-tail case with a primary population
and a delta-function beam. We obtain the corresponding linear kinetic
dispersion relations and evaluate the properties of instabilities
when the electron distribution functions are of the delta function,
Cauchy, squared-Cauchy, or inverse-quartic types. In agreement with
other studies, we find that in general quantum effects reduce the
range of wavelengths for unstable modes at long wavelengths. We also
find a second window of instability at shorter wavelengths and elucidate
its nature as being due to quantum recoil. We note the possible implications
for studies of laboratory and astrophysical quantum plasmas.
\end{abstract}
\maketitle

\section{Introduction}

Quantum effects in plasmas have been addressed since at least the
mid-1900s \citep{Bohm1950,Silin1958,Pines1961}, and have also been
a subject of renewed interest recently \citep{Marklund2008,Shukla2011,Tyshetskiy2011,Vladimirov2011}.
The reason for this renewed interest is twofold: it is due to the
increasing importance of plasma effects in certain quantum solid-state
systems \citep{Marklund2008} on the one hand, and the increasing
importance of quantum effects in dense plasmas in the laboratory \citep{Glenzer2007}
and in astrophysics \citep{Chabrier2002} on the other hand. As quantum
effects generally occur at small scales, it is reasonable to first
apply quantum mechanics to problems in kinetic plasma theory. Important
topics which require kinetic physics include transport theory and
the linear modes and stability properties of a plasma. This paper
is concerned with the latter. Linear physics is also important because
linear problems can often be solved analytically and are amenable
to simple interpretation. For this reason, much of our understanding
of complex processes in classical plasmas is founded upon knowledge
of the linear properties, and it can be expected that this will hold
true for quantum plasmas as well. Additionally, the comparative simplicity
of non-relativistic quantum plasma physics serves as a useful baseline
for more difficult problems in quantum plasma physics, such as relativistic
quantum-electrodynamic (QED) plasmas with significance to fundamental
theoretical physics and important astrophysical applications. It is
therefore advantageous to understand the linear kinetic physics of
non-relativistic quantum plasma waves.

The theory of linear waves in quantum plasmas has been studied alongside
classical theory during the early development of plasma physics \citep{Bohm1950,Bohm1953,Pines1962,Pines1961},
as well as in more recent works \citep{Bonitz1993b,Haas2001,Haas2009a,Melrose2010,Bonitz1994,Eliasson2010}.
Studies of one-dimensional electrostatic waves in quantum plasmas
have demonstrated unique effects due to quantum mechanics, including
dependence of Landau damping rates on quantum effects \citep{Krivitskii1991,Rightley2016,Melrose2010}
and the existence of entirely new modes that do not appear in classical
plasmas \citep{Krivitskii1991}. The introduction of a second, drifting
population of particles broadens the parameter space by introducing
a new density and temperature, and the separation velocity between
the two populations. In classical plasmas, this introduces the possibility
for growing unstable modes such as the Buneman, bump-on-tail, and
ion-acoustic instabilities \citep{Stix1992}. The quantum counterparts
of these streaming instabilities have been subject to a few isolated
studies in the past decades, although so far no comprehensive systematic
approach has yet been applied to kinetic quantum plasma instabilities
so far. For instance, instabilities in one-dimensional quantum wires
have been investigated by \citet{Bonitz1993,Bonitz1994}. Additionally,
the Nyquist method has been applied to two-population quantum plasmas,
and it has been demonstrated that a Penrose criterion does not exist
for a quantum plasma \citep{Haas2001}. Thus, no general rule has
been developed to show whether or not a two-population quantum plasma
will be unstable. Furthermore, the existence of a second region of
instability for sufficiently strong quantum effects has been demonstrated
for a two-stream plasma by \citet{Haas2009a}. Despite these interesting
results, so far there has been no comprehensive mapping of the different
instability types in quantum plasmas, in contrast to classical plasmas
as discussed, for example, by \citet{Lapuerta2002}. This is due in
part to the difficulty of working with the Fermi-Dirac distribution
function in the quantum kinetic theory, as noted by \citet{Krivitskii1991}
and by \citet{Vladimirov2011}. 

The purpose of this paper is to explore quantum dynamical effects
on linear electrostatic perturbations without the complexity of quantum
statistics by utilising a number of simplified, analytically convenient
distribution functions. We study a sequence of increasingly complex
distribution functions in one- and two-population classical and quantum
unmagnetized plasmas in order to look for regularities which may guide
further study of more realistic Fermi-Dirac plasmas. We analyse
the effects of quantum mechanics on the behaviour
of linear perturbations in single-population plasmas and in plasmas
consisting of two populations of electrons drifting relative to each
other with a stationary neutralising ion
background. After briefly describing the foundations of quantum kinetic
theory in~section \ref{sec:background}, we examine the dispersion
relation of waves in plasmas consisting of one electron population
using several toy equilibrium Wigner distribution functions which
allow the susceptibility to be integrated exactly~(section \ref{sec:onecomp}).
Then, we consider potentially unstable plasmas with two electron populations
in~section \ref{sec:twocomp}. For each toy Wigner function we examine
two cases: (1) the case of identical counter-drifting populations
and (2) the case of a main population with a zero-width drifting beam.
We analyse and plot the dispersion relations, map the wavelengths
and drift velocities that allow instabilities for different parameters
of the distribution functions, and additionally map the growth rates
and wavelengths of the fastest-growing modes. Each case is compared
to the similar situation in a classical plasma in order to elucidate
the differences due to quantum mechanics. Section \ref{sec:discussion}
concerns the limitations and possible applications and extensions
of this work. Finally, we draw conclusions as to the general effects
of the shape of the Wigner function and of quantum recoil on streaming
instabilities in~section~\ref{sec:conclusions}.

\section{Formalism of Non-Relativistic Quantum Plasma Physics\label{sec:background}}

In this section a quantum kinetic theory based on Wigner functions
is briefly reviewed in section \ref{sec:quantumkinetictheory}. We
then write down and discuss the longitudinal susceptibility for an
unmagnetized collisionless plasma of arbitrary composition including
the influence of quantum recoil and tunnelling in section \ref{sec:susceptibility}.
Subsequently, in section \ref{sec:parameters} we proceed to discuss
the characteristic scales and important parameters of the system,
and finally non-dimensionalize the susceptibility in preparation for
the analysis that follows in section~\ref{sec:onecomp}.

\subsection{Quantum Kinetic Theory}

\label{sec:quantumkinetictheory}

In quantum mechanics, the state of a many-body system of particles
may be described by its density operator~$\hat{\rho}$. The time
evolution of this operator under the influence of Hamiltonian $\hat{H}$
in the Heisenberg formalism is given by $\text{i}\hbar\partial\hat{\rho}/\partial t=\left[\hat{H},\hat{\rho}\right]$.

This operator theory may be cast in an equivalent theory of functions
in phase space $\left(\boldsymbol{x},\boldsymbol{p}\right)$ by introducing
the Wigner transformation 
\begin{equation}
W\left[\hat{A}\right]\equiv\int{\rm d}^{3}\boldsymbol{y}\thinspace\text{exp}\left(-2\text{i}\boldsymbol{p}\cdot\boldsymbol{y}/\hbar\right)\left\langle \boldsymbol{x}+\boldsymbol{y}\left|\hat{A}\right|\boldsymbol{x}-\boldsymbol{y}\right\rangle .
\end{equation}
 The Wigner transformation of the density operator is the Wigner quasi-probability
distribution function \citep{Wigner1932}, 
\begin{equation}
f\left(\boldsymbol{x},\boldsymbol{p}\right)\equiv W\left[\hat{\rho}\right]
\end{equation}
and the Wigner transformation of the equation of motion of the density
operator is the Moyal equation \citep{Moyal1949} 
\begin{equation}
\frac{\partial f}{\partial t}=-\frac{2}{\hbar}\left\{ f\left(\boldsymbol{x},\boldsymbol{p}\right)\text{sin}\left[\frac{\hbar}{2}\left(\overleftarrow{\partial_{\boldsymbol{x}}}\overrightarrow{\partial_{\boldsymbol{p}}}-\overleftarrow{\partial_{\boldsymbol{p}}}\overrightarrow{\partial_{\boldsymbol{x}}}\right)\right]H\left(\boldsymbol{x},\boldsymbol{p}\right)\right\} ,
\end{equation}
where the subscripts on the $\partial$ operators indicate the relevant
variable of differentiation, the arrows above the $\partial$ operators
indicate the function upon which they operate ($f$ on the left and
~$H$ on the right), and the $\text{sin}$ function with operator
argument is to be understood in terms of its Taylor expansion. The
Wigner function plays the role of the classical distribution function
in that its moments result in measurables such as the particle density,
velocity and current. The Moyal equation thus plays the role of the
Liouville equation and for a 1-particle Wigner function it plays the
role of the Vlasov equation. Details of this theory have been reviewed
by \citet{Liboff2003}.

\subsection{The Quantum Plasma Longitudinal Susceptibility}

\label{sec:susceptibility}

For easy comparison to wave phase-velocities, we work in the $\left(\boldsymbol{x},\boldsymbol{v}\right)$
phase space, where momentum is replaced with velocity through~$\boldsymbol{v}=\boldsymbol{p}/m$.
The longitudinal susceptibility for a population~$s$ drifting at
velocity~$\boldsymbol{U}_{s}$ with respect to a given reference
frame in an unmagnetized plasma \citep{Klimontovich1960,Lindhard1954}
is 
\begin{equation}
\chi_{s}\left(\omega,\boldsymbol{k}\right)=\frac{m_{s}\omega_{ps}^{2}}{2\hbar k^{2}}\int{\rm d}^{3}\boldsymbol{v}\frac{F_{0s}\left(\boldsymbol{v}+\hbar\boldsymbol{k}/m_{s}-\boldsymbol{U_{s}}\right)-F_{0s}\left(\boldsymbol{v}-\hbar\boldsymbol{k}/m_{s}-\boldsymbol{U_{s}}\right)}{\omega-\boldsymbol{k}\cdot\boldsymbol{v}},
\end{equation}
where $F_{0s}\left(\boldsymbol{v}\right)$ is the normalised background
Wigner quasi-probability distribution function for species~$s$,
$\omega_{ps}\equiv\left(4\pi n_{s}e^{2}/m_{s}\right)^{1/2}$ is the
population-specific plasma frequency.

For simplicity, we will assume modes parallel to $\boldsymbol{U}_{s}$:
$\boldsymbol{k}\parallel\boldsymbol{U}_{s}$. Modes propagating at
angle $\theta$ with respect to $\boldsymbol{U}$ can be accounted
for through the substitution $U\rightarrow U{\rm cos}\theta$ and
a variable substitution in the integration over velocities perpendicular
to $\boldsymbol{k}$. For $\boldsymbol{k}\parallel\boldsymbol{U}_{s}$,
performing the integral over directions perpendicular to $\boldsymbol{k}$
and re-labelling $v$ as the parallel velocity component, one obtains
\begin{equation}
\chi_{s}\left(\omega,k\right)=\frac{m_{s}\omega_{ps}^{2}}{2\hbar k^{3}}\int_{C}{\rm d}v\thinspace\frac{f_{0s}\left(v+\hbar k/m_{s}-U_{s}\right)-f_{0s}\left(v-\hbar k/m_{s}-U_{s}\right)}{\omega/k-v},
\end{equation}
where $f_{0s}\left(v\right)$ is now the reduced distribution function
\begin{equation}
f_{0s}\left(v\right)=\int{\rm d}^{2}v_{\perp}F_{0s}\left(\boldsymbol{v}\right)
\end{equation}
which is normalised to unity. The integral is performed along contour
along the real-$v$ axis so as to pass below the pole at $v=\omega/k$. 

For a multi-component plasma, we define the total plasma frequency
\begin{equation}
\omega_{p}^{2}\equiv\sum_{s}\omega_{ps}^{2}=4\pi e^{2}n\sum_{s}\frac{\tilde{n}_{s}}{m_{s}},
\end{equation}
where we have defined 
\begin{equation}
\tilde{n}_{s}\equiv\frac{n_{s}}{n},
\end{equation}
to be the fraction of total particles in population~$s$, where $n$
is the total number density and $\sum_{s}\tilde{n}_{s}=1$. At this
point we assume that only electrons are mobile and that the ions form
only a stationary neutralising background so that we can write 
\begin{equation}
m_{s}=m_{e}\equiv m.
\end{equation}
We then rewrite the susceptibility as $\chi_{s}=\tilde{n}_{s}\tilde{\chi}_{s}$
where 
\begin{equation}
\tilde{\chi}_{s}\left(\omega,k\right)=\frac{m\omega_{p}^{2}}{2\hbar k^{3}}\int_{-\infty}^{\infty}{\rm d}v\frac{f_{0s}\left(v+\hbar k/m_{s}-U_{s}\right)-f_{0s}\left(v-\hbar k/m_{s}-U_{s}\right)}{\omega/k-v}\label{eq:susceptibility_wunits}
\end{equation}
is the susceptibility normalised to the total density. We then obtain
the dispersion relation by solving the equation 
\begin{equation}
\epsilon\left(\omega,k\right)\equiv1+\sum_{s}\chi_{s}=0,\label{eq:dispeq}
\end{equation}
where $\epsilon$ is the dielectric function, and write solutions
of this equation as $\omega=\omega_{r}+{\rm i}\gamma$.

\subsection{Comments on Parameters and Scales}

\label{sec:parameters}

Before considering solutions of equation \ref{eq:dispeq}, we discuss
the important scales involved in this study, the relevant parameter
regimes, and the issue of non-dimensionalising the arguments and parameters
of the dielectric function. The fundamental parameters that describe
a one\textendash component unmagnetized plasma are the total density
$n$ and temperature~$T$. The density defines a length-scale: the
inter-particle spacing~$n^{-1/3}$, and the temperature determines
the characteristic particle velocity $\mathcal{V}=\left(2T/m\right)^{1/2}$.
In addition the density determines the plasma frequency $\omega_{p}=\left(4\pi ne^{2}/m\right)^{1/2}$,
which in turn determines the plasmon energy $\epsilon_{p}=\hbar\omega_{p}$.
Another important velocity scale is the velocity of a particle with
energy equal to the plasmon energy $\eta=\left(2\hbar\omega_{p}/m\right)^{1/2}$,
and an additional energy scale is the average electrostatic interaction
energy $u=e^{2}/n^{-1/3}$. Two final length scales that may be defined
are the thermal de Broglie wavelength of the particles $\lambda_{dB}=\hbar/\left(8\pi^{2}mT\right)^{1/2}$
and the Debye length $\lambda_{D}=\mathcal{V}/\omega_{p}=\left(T/4\pi ne^{2}\right)^{1/2}$.
For Fermi-Dirac electrons, there additionally exists the Fermi energy
$E_{F}=\hbar^{2}\left(3\pi n\right)^{2/3}/2m$ and, as the characteristic
velocity depends on the level of degeneracy, we define degeneracy-dependent
speed and screening length~$V_{*}$ and~$\lambda_{*}$ respectively,
as in \citet{Rightley2016}. These characteristic scales are summarised
in table \ref{tab:scales}. From these scales we can define a number
of dimensionless parameters, the values of which determine the relative
importance of different processes in the plasma. Length scales can
be compared to the inter-particle spacing by examining the number
of particles in a cubic volume bounded by the relevant length: $\Gamma\equiv n\lambda_{D}^{3}$,
which is the classical plasma parameter describing the relative importance
of electrostatic interactions and $\Theta\equiv n\lambda_{dB}^{3}\propto\left(T/E_{F}\right)^{-3/2}$,
which is a proxy for the level of degeneracy. We can also consider
the ratio $\lambda_{dB}/\lambda_{D}$ which determines the relative
importance of quantum effects for wave phenomena. We further create
three dimensionless numbers using the three energy scales: $H\equiv\lambda_{dB}/\lambda_{*}=\hbar\omega_{p}/4\pi T$,
$N_{B}\equiv a_{B}n^{1/3}=\left(\hbar\omega_{p}/u\right)^{2}/4\pi$
and $u/T\propto\Gamma^{-2/3}.$ These dimensionless parameters are
summarised in table \ref{tab:parameters}.

It is computationally convenient to remove dimensions from the susceptibility.
This task can be accomplished using two schemes. In both schemes a
logical time scale is the plasma frequency~$\omega_{p}$. There are
two ways, however, of introducing a velocity (and by proxy, length)
scale: (1) use the characteristic particle velocity of the background
distribution function~$\mathcal{V}$, and (2) use the velocity~$\eta$,
which is independent of the normalised distribution function. In scheme
$1$ we introduce variables $\tilde{v}\equiv v/\mathcal{V}_{s},$
$\Omega\equiv\omega/\omega_{p}$, $K\equiv k\mathcal{V}_{s}/\omega_{p}$,
$H\equiv\hbar\omega_{p}/m_{s}\mathcal{V}_{s}^{2}$, and $U_{s}\equiv U_{s}/\mathcal{V}_{s}$,
where $\mathcal{V}_{s}/\omega_{p}$ defines a length scale $\lambda_{s}$
which is the Debye length for the given distribution function, to
obtain the following: 
\begin{equation}
\tilde{\chi}_{s}^{(1)}\left(K,\Omega\right)=\frac{\mathcal{V}_{s}}{2HK^{3}}\tilde{n}_{s}\int_{C}{\rm d}\tilde{v}\thinspace\frac{f_{0s}\left(\tilde{v}+HK-U_{s}\right)-f_{0s}\left(\tilde{v}-HK-U_{s}\right)}{\Omega/K-\tilde{v}}.\label{eq:generalsusceptibility}
\end{equation}
Here, quantum effects are due to the parameter $H$, which is the
ratio of the plasmon energy to the average energy of a plasma particle.

In scheme 2 we use dimensionless parameters $\tilde{v}\equiv v/\eta$,
$\Omega\equiv\omega/\omega_{ps}$, $K\equiv\eta k/\omega_{p}$, $U_{s}\equiv U_{s}/\eta$,
$\mathcal{V}_{s}\equiv\mathcal{V}_{s}/\eta$. This results in 
\[
\tilde{\chi}_{s}^{(2)}\left(K,\Omega\right)=\frac{\eta}{2K^{3}}\tilde{n}_{s}\int_{C}{\rm d}\tilde{v}\thinspace\frac{f_{0s}\left(\tilde{v}+K-U_{s}\right)-f_{0s}\left(\tilde{v}-K-U_{s}\right)}{\Omega/K-\tilde{v}},
\]
where the quantum recoil effects have been subsumed into the dimensionless
wavenumber~$K$. In the rest of this work we will utilise whichever
scheme is most convenient for the given situation.

We can centre our frame of reference on the distribution's peak through
the transformation $\omega\rightarrow\omega+k\cdot U_{s}$ or, inversely,
for a population centred on our frame of reference, we can transform
with $\omega\rightarrow\omega-k\cdot U_{s}$ to obtain a drifting
population. We will take advantage of this when we turn to systems
of two populations drifting with respect to one another.

\begin{table}
\begin{center}%
\begin{tabular}{cccc}
Symbol & Name  & Expression & Numerical Value\tabularnewline
\hline 
\hline 
$\omega_{pe}$ & Electron Plasma Frequency & $\left(4\pi e^{2}n/m\right)^{1/2}$ & $5.64\times10^{4}\sqrt{n}\thinspace\text{rad/s}$\tabularnewline
\hline 
$v_{Te}$ & Electron Thermal Speed & $\left(2T/m\right)^{1/2}$ & $5.93\times10^{7}T\thinspace\text{cm/s}$\tabularnewline
$\eta$ & Particle plasmon Velocity  & $\left(2\hbar\omega_{p}/m\right)^{1/2}$ & $361n^{1/4}\thinspace\text{cm/s}$\tabularnewline
\hline 
$E_{p}$ & Plasmon Energy  & $\hbar\omega_{p}$ & $3.71\times10^{-4}n^{1/2}\thinspace\text{eV}$\tabularnewline
$u$ & Average Electrostatic Energy  & $e^{2}/n^{-1/3}$ & $1.44\times10^{-7}n^{1/3}\thinspace\text{eV}$\tabularnewline
\hline 
$a_{B}$ & Bohr Radius  & $\hbar^{2}/me^{2}$ & $5.29\times10^{-9}\thinspace\text{cm}$\tabularnewline
$\lambda_{D}$ & Debye length & $\left(T/4\pi ne^{2}\right)^{1/2}$ & $7.43\times10^{2}T^{1/2}n^{-1/2}\thinspace\text{cm}$\tabularnewline
$l_{int}$ & Inter-particle Spacing & $n^{-1/3}$ & $n^{-1/3}\thinspace\text{cm}$\tabularnewline
$\lambda_{dB}$ & de Broglie Wavelength & $\hbar/\left(mT\right)^{1/2}$ & $2.76\times10^{-8}T^{-1/2}\thinspace\text{cm}$\tabularnewline
\end{tabular}\end{center}

\caption{Table of relevant scales, with $T$ measured in $\text{eV}$ and $n$
measured in $\text{cm}^{-3}$.\label{tab:scales}}
\end{table}

\begin{table}
\begin{center}%
\begin{tabular}{cccc}
Symbol  & Name  & Definition  & Numerical Value\tabularnewline
\hline 
\hline 
$\Theta$  & Degeneracy Parameter  & $n\lambda_{dB}^{3}$  & $2.10\times10^{-23}T^{-3/2}n$\tabularnewline
$\Gamma$  & Plasma Parameter  & $n\lambda_{D}^{3}$  & $1.72\times10^{9}T^{3/2}n^{-1/2}$\tabularnewline
$H$  & Quantum Recoil Parameter  & $\hbar\omega_{ps}/m_{s}\mathcal{V}_{s}^{2}$ & $1.86\times10^{-4}T^{-1}n^{1/2}$\tabularnewline
$N_{B}$  & Number of Particles in Bohr Sphere  & $na_{B}^{3}$ & $1.48\times10^{-25}n$\tabularnewline
\end{tabular}\end{center}\caption{Table of dimensionless parameters, with $T$ measured in $\text{eV}$
and $n$ measured in $\text{cm}^{-3}$. \label{tab:parameters}}
\end{table}

\section{One-Component Plasmas\label{sec:onecomp}}

Before considering two-component plasmas and the possibility of instabilities
in them, we analyse the case of a one-component plasma; that is, a
plasma consisting of a single population of particles described by
an equilibrium distribution function $f_{0}\left(v\right)$, in the
presence of a stationary neutralising background. This discussion
will provide insight into the properties of the susceptibility equation
\ref{eq:susceptibility_wunits}, motivate certain distribution functions
which should be considered, and provide a baseline against which to
compare the results for more complex plasmas.

For a distribution function that vanishes at $\left|v\right|\rightarrow\infty$
and that has a finite number of simple poles, the integral in equation
\ref{eq:susceptibility_wunits} can be performed exactly using the
residue theorem. Each pole in the distribution function contributes
a term in the susceptibility with a pole at the same point in the
complex $\omega/k$ plane, and this results in a complex root of the
dielectric function $1+\sum_{s}\chi_{s}$. For this reason, the analytic
structure of the distribution function is the determining factor in
the number of modes present.

\subsection{Equilibrium Distribution Functions}

\label{sec:distributions}

In order to understand the influence of the quantum recoil on the
susceptibility, it is useful to consider cases for which the susceptibility
may be integrated analytically. There are several distribution functions
for which this is possible. The most simple is the \textit{Dirac delta-function
distribution}, 
\begin{equation}
f_{\delta}\left(v\right)=\delta\left(v\right),\label{eq:distdel}
\end{equation}
which effectively models particles with zero velocity spread. Due
to the absence of a finite width, classically the delta-function distribution
does not allow for Landau damping as there is no possibility of particles
moving in resonance with the wave phase speed, nor does it allow for
wave dispersion as the sound speed is zero.

A slightly more realistic function is the \textit{Cauchy distribution}
with width~$\mathcal{V}$ 
\begin{equation}
f_{C}\left(v\right)=\frac{\mathcal{V}}{\pi\left(v^{2}+\mathcal{V}^{2}\right)}=\frac{\mathcal{V}}{\pi(v-\text{i}\mathcal{V})(v+\text{i}\mathcal{V})}.\label{eq:distcau}
\end{equation}
For this and the related distributions we show the function in a simple
form and in a form demonstrating the complex roots. This distribution
has a finite width which allows for Landau damping. However, $f_{C}\left(v\right)$
does not have a finite second moment and hence still does not account
for wave dispersion, as there is not a well-defined pressure and thus
no well-defined sound speed.

In order to accommodate this effect, one can consider the \textit{squared
Cauchy distribution} 
\begin{equation}
f_{C2}\left(v\right)=\frac{2\mathcal{V}^{3}}{\pi\left(v^{2}+\mathcal{V}^{2}\right)^{2}}=\frac{2\mathcal{V}^{3}}{\pi(v+\text{i}\mathcal{V})^{2}(v-\text{i}\mathcal{V})^{2}},\label{eq:distcau2}
\end{equation}
which can be used to define a finite pressure and for which the susceptibility
may still be integrated analytically. The squared Cauchy distribution
also has poles at $v=\pm iV$, but unlike in the previous case the
poles are of second order. However, the poles exist only at two points.
In order to further elucidate the effect of the complex structure
of the distribution function, we define an ``\textit{inverse-quartic}''
or $f_{4}\left(v\right)$ function 
\begin{equation}
f_{4}\left(v\right)=\frac{\sqrt{2}\mathcal{V}^{3}}{\pi\left(v^{4}+\mathcal{V}^{4}\right)}=\frac{\sqrt{2}\mathcal{V}^{3}}{\pi\left(v-\frac{1-\text{i}}{\sqrt{2}}\mathcal{V}\right)\left(v+\frac{1-\text{i}}{\sqrt{2}}\mathcal{V}\right)\left(v-\frac{1+\text{i}}{\sqrt{2}}\mathcal{V}\right)\left(v+\frac{1+\text{i}}{\sqrt{2}}\mathcal{V}\right)},\label{eq:dist2}
\end{equation}
which has a denominator of the same order as~$f_{C2}$ but four simple
poles instead of two second order poles.

Continuing in this fashion, we can generalise to two different functions:
the Cauchy to the power~$J$-function 
\begin{equation}
f_{CJ}\left(v\right)\propto\frac{1}{\left(v^{2}+\mathcal{V}^{2}\right)^{J}}
\end{equation}
which has two poles of order $J$ at~$v=\pm iV$. This function can
be normalised for an arbitrary value of~$J$: 
\begin{equation}
f_{CJ}\left(v\right)=\frac{\mathcal{V}^{2J-1}\Gamma(J)}{\sqrt{\pi}\Gamma\left(J-\frac{1}{2}\right)}\frac{1}{(v-\text{i}\mathcal{V})^{J}(v+\text{i}\mathcal{V})^{J}}.\label{eq:distcauJ}
\end{equation}
This is a special case of the ``Kappa Distribution'' with integer
power of~$\kappa$. We can further generalise the Cauchy distribution
with the inverse $J'th$ function 
\begin{equation}
f_{J}\left(v\right)\propto\frac{1}{v^{J}+\mathcal{V}^{J}}
\end{equation}
which has~$J$ first-order poles at $v=\left(-1\right)^{j/J}$ with
$1\leq j\leq J$. This can also be normalised for an arbitrary~$J$
value: 
\begin{equation}
f_{J}\left(v\right)=\frac{J\mathcal{V}^{J-1}}{(2\pi)\csc\left(\frac{\pi}{J}\right)}\prod_{j=1}^{J}\frac{1}{v-\mathcal{V}(-1)^{\left(2j-1\right)/J}}.\label{eq:distJ}
\end{equation}

We will use the functions $f_{CJ}\left(v\right)$ and $f_{J}\left(v\right)$
in an attempt to describe both qualitative and quantitative effects
of poles in the distribution function. A completely degenerate population
of fermions follows the reduced \textit{totally degenerate Fermi-Dirac
distribution} function 
\begin{equation}
f_{D}\left(v\right)=\begin{cases}
\begin{array}{cc}
\mathcal{V}_{F}^{-2}\left(\mathcal{V}_{F}^{2}-v^{2}\right) & v\leq\mathcal{V}_{F}\\
0 & v>\mathcal{V}_{F}
\end{array}\end{cases},\label{eq:distDeg}
\end{equation}
where the parabolic shape is a result of $v$ being only the velocity
component parallel to the wave vector. In this case, the characteristic
velocity $\mathcal{V}_{F}\equiv\hbar\left(3\pi^{2}n\right)^{1/3}/m$
is the Fermi velocity (the speed of a particle with kinetic energy
equal to the Fermi energy). The function \eqref{eq:distDeg} is not
meromorphic due to the discontinuities at $v=\pm\mathcal{V}_{F}$,
and this strongly influences wave properties; the consequences of
this are investigated in Refs. \citep{Krivitskii1991,Eliasson2010,Vladimirov2011}.

The \textit{totally degenerate Fermi-Dirac distribution }is a limit
of a more general \textit{arbitrarily degenerate Fermi-Dirac distribution}
\begin{equation}
f_{FD}\left(v\right)=\frac{1}{\sqrt{\pi}\textrm{Li}_{3/2}\left(-\textrm{e}^{\mu}\right)}\frac{1}{\mathcal{V}}\textrm{Ln}\left(1+\textrm{e}^{-v^{2}/\mathcal{V}^{2}+\mu/T}\right),\label{eq:distFD}
\end{equation}
which is obtained by integrating the general Fermi-Dirac distribution
over perpendicular velocities. In equation \ref{eq:distFD}, $\text{Li}$
is the polylogarithm and here~$\mathcal{V}$ is the classical thermal
velocity. The normalisation can be obtained by expanding in the classical
limit $\mu/T\rightarrow-\infty$, integrating term-by-term, and then
re-summing to all orders in $\mu/T$ as performed by \citet{Melrose2010a}.

\begin{figure}
\begin{center}\subfloat[]{\includegraphics[width=0.40\columnwidth]{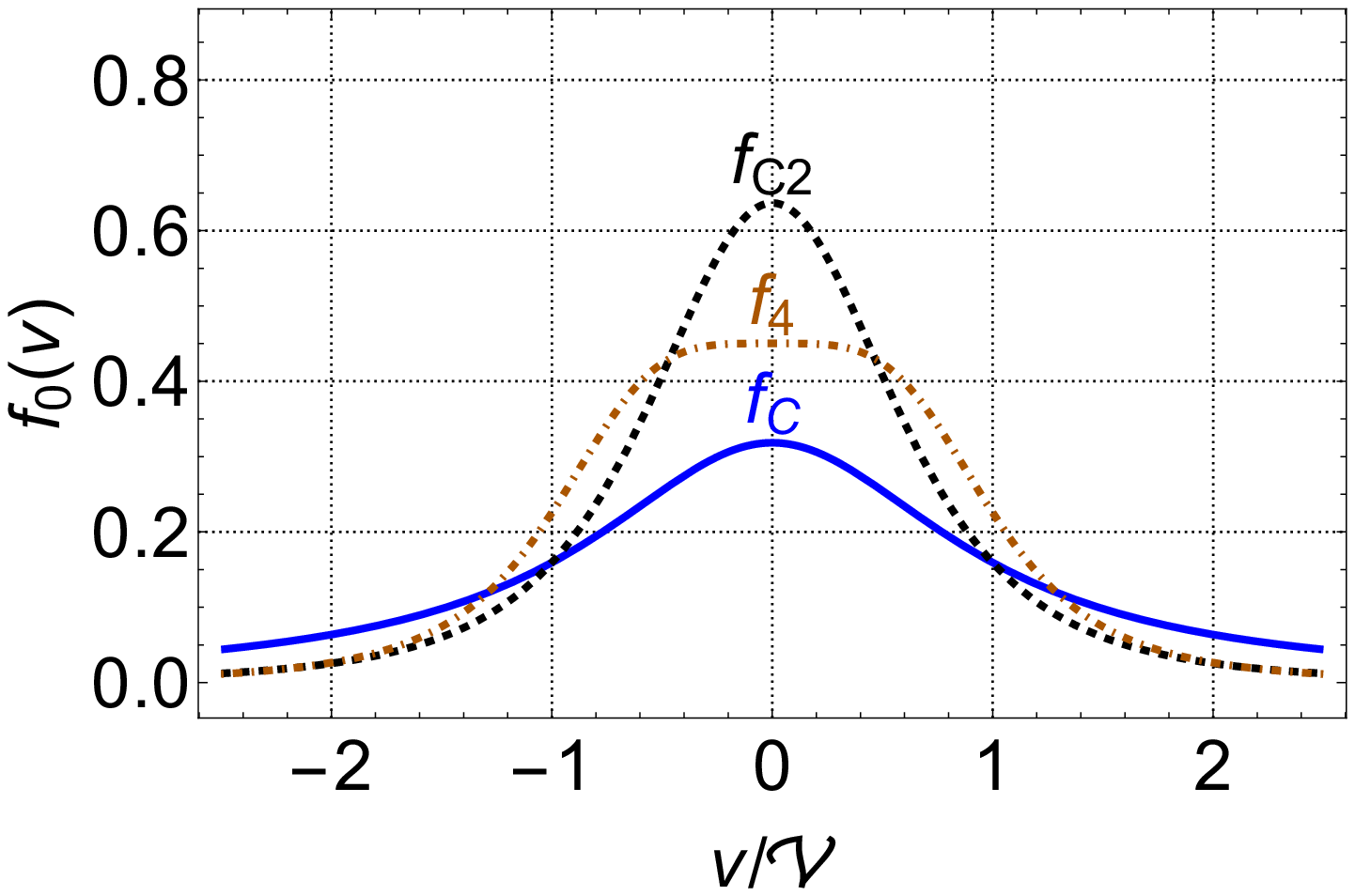}}\hspace{0.02\columnwidth}\subfloat[]{\includegraphics[width=0.40\columnwidth]{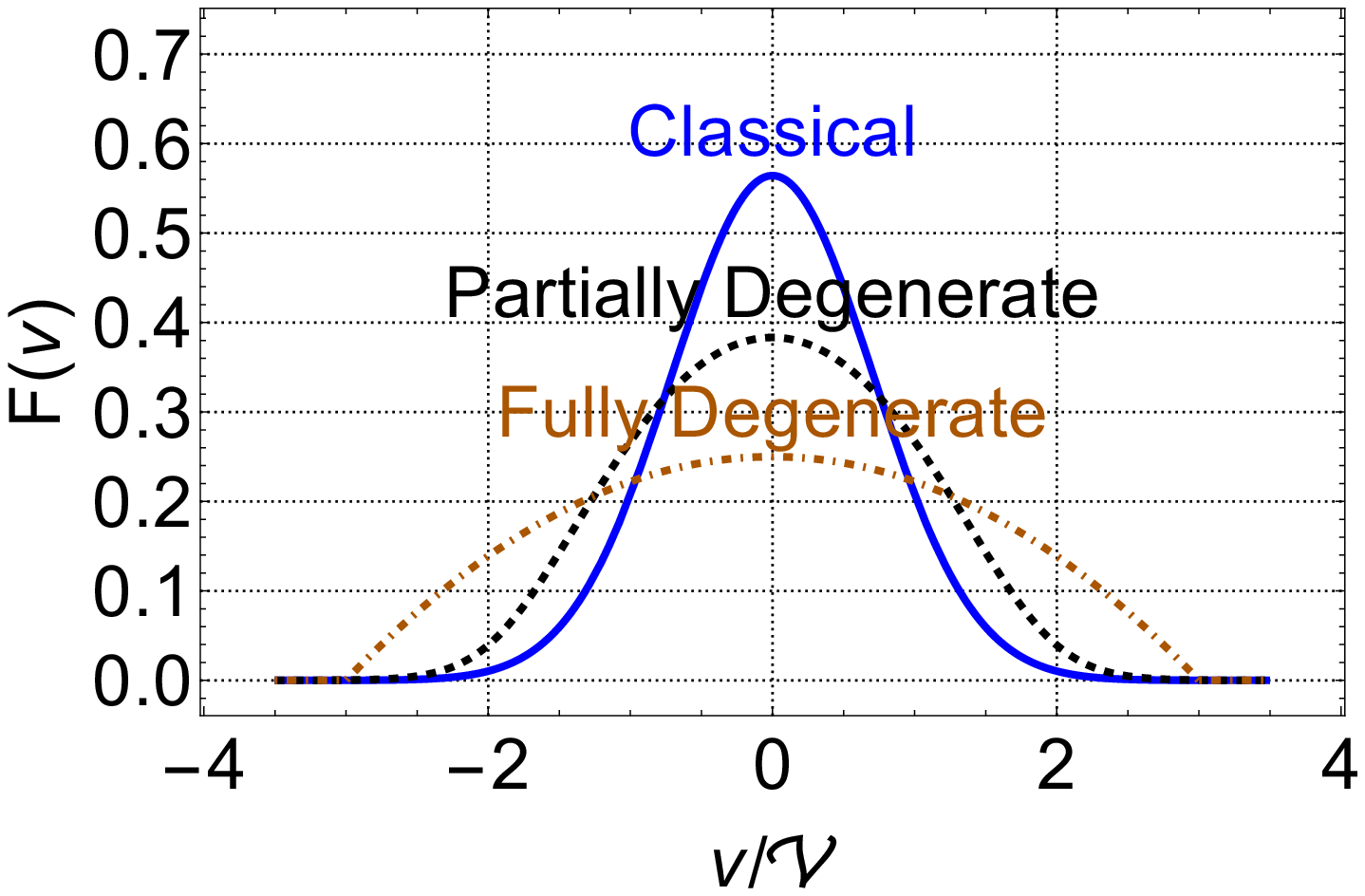}}

\end{center}\caption{Distribution functions that are considered in this paper. Cauchy-type
functions are shown in panel (a). The~$f_{C2}$ function is steeper
and narrower than the Cauchy function, and the~$f_{N}$ function
is broader and flatter on the top. Fermi-Dirac type distribution functions
are shown in panel (b), with degeneracy level ranging from classical
Maxwellian to fully degenerate truncated distribution. \label{fig:distributions-1}}
\end{figure}

\begin{figure}
\begin{center}\subfloat[]{\includegraphics[width=0.40\columnwidth]{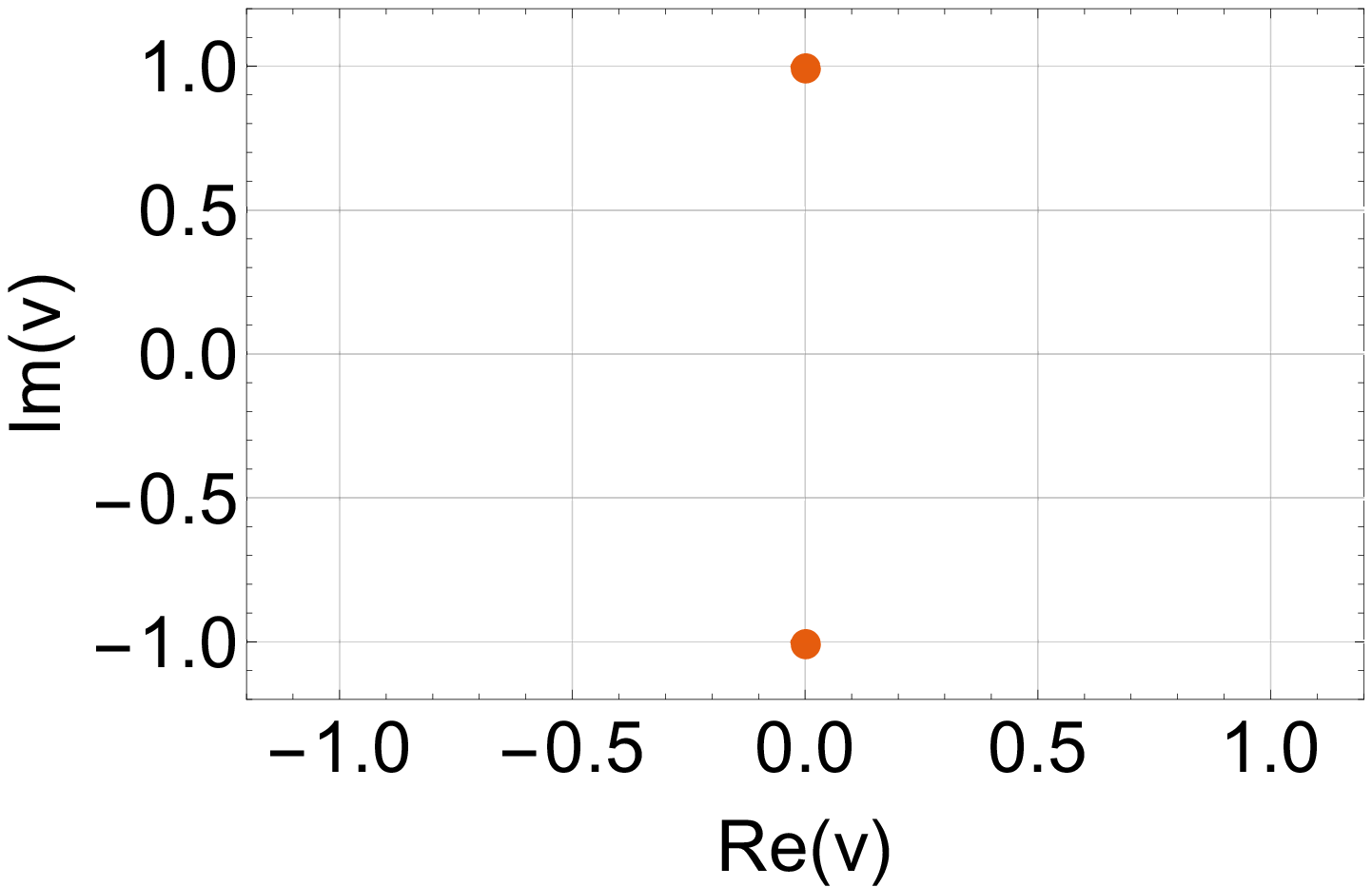}}\hspace{0.02\columnwidth}\subfloat[]{\includegraphics[width=0.40\columnwidth]{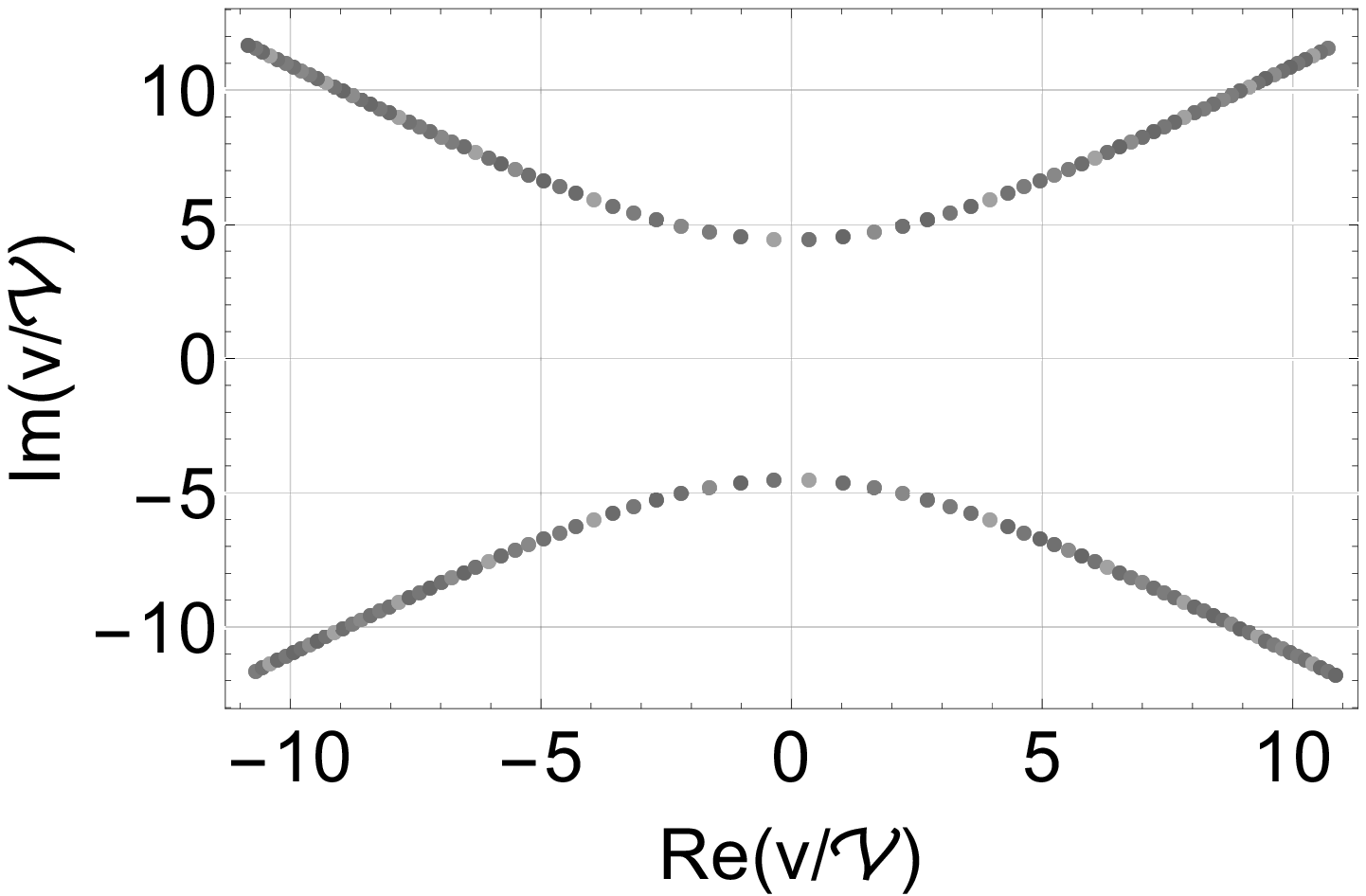}}

\vspace{1mm}
\subfloat[]{\includegraphics[width=0.40\columnwidth]{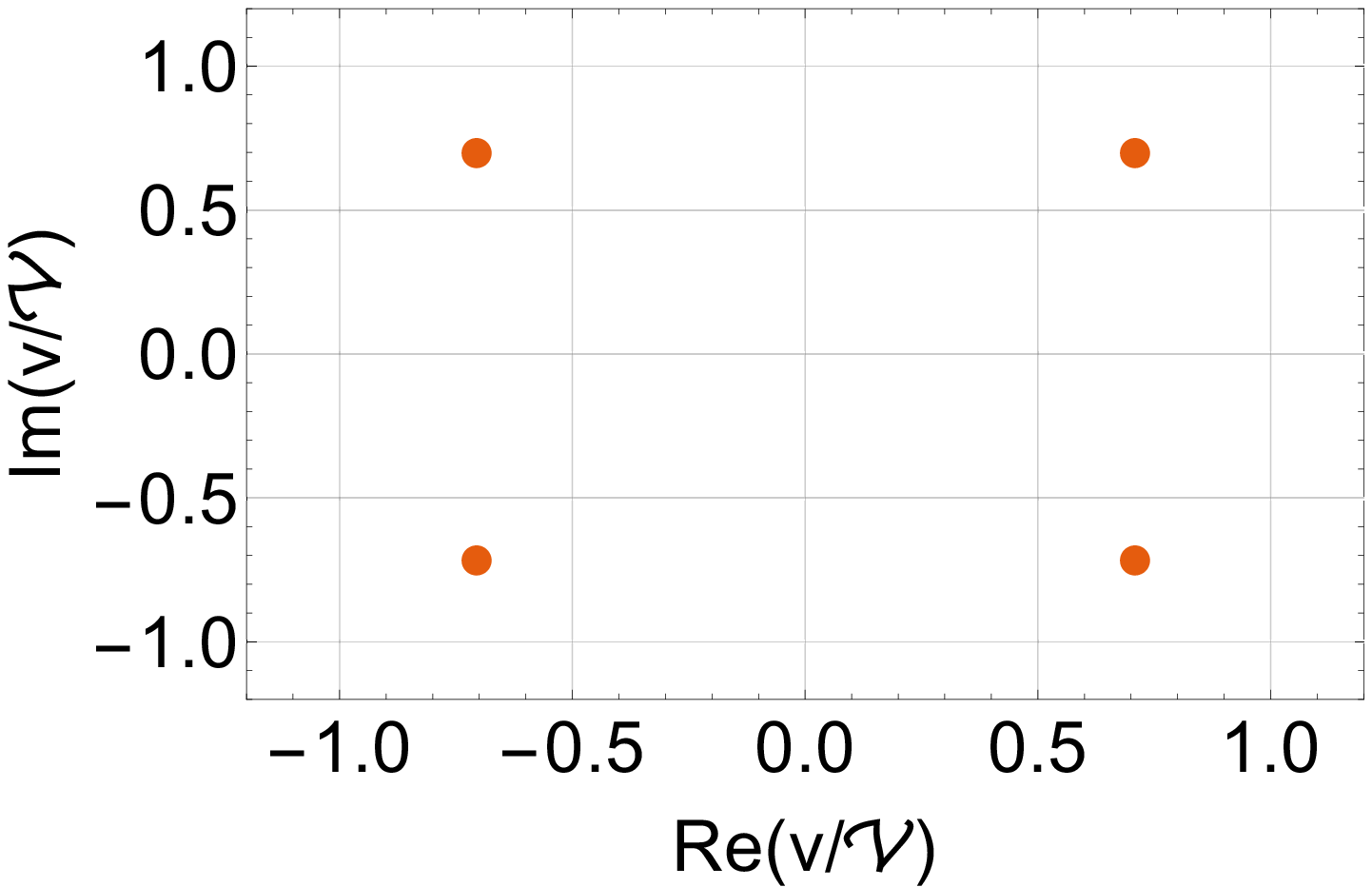}}\hspace{0.02\columnwidth}\subfloat[]{\includegraphics[width=0.40\columnwidth]{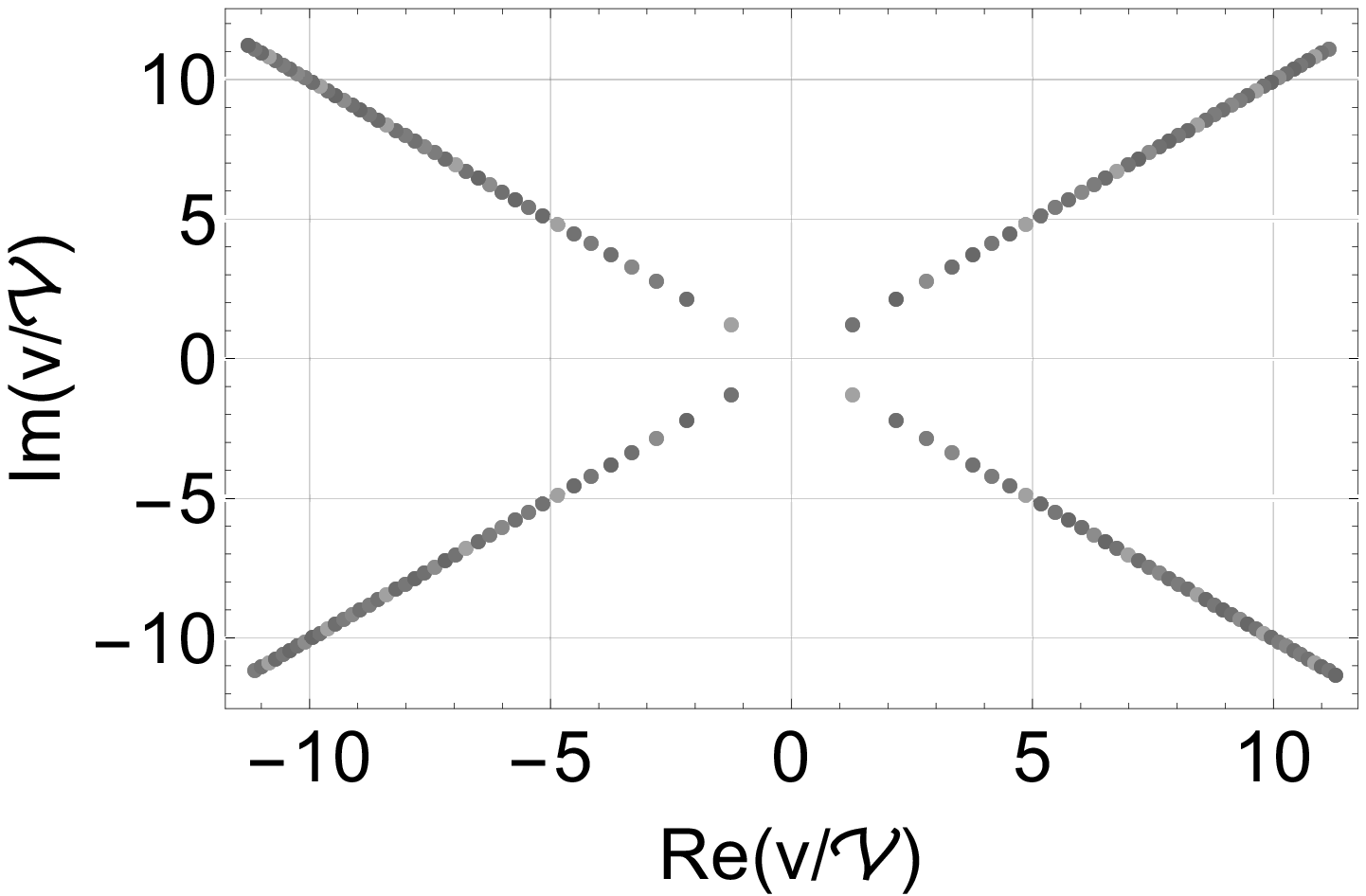}}

\vspace{1mm}
\subfloat[]{\includegraphics[width=0.40\columnwidth]{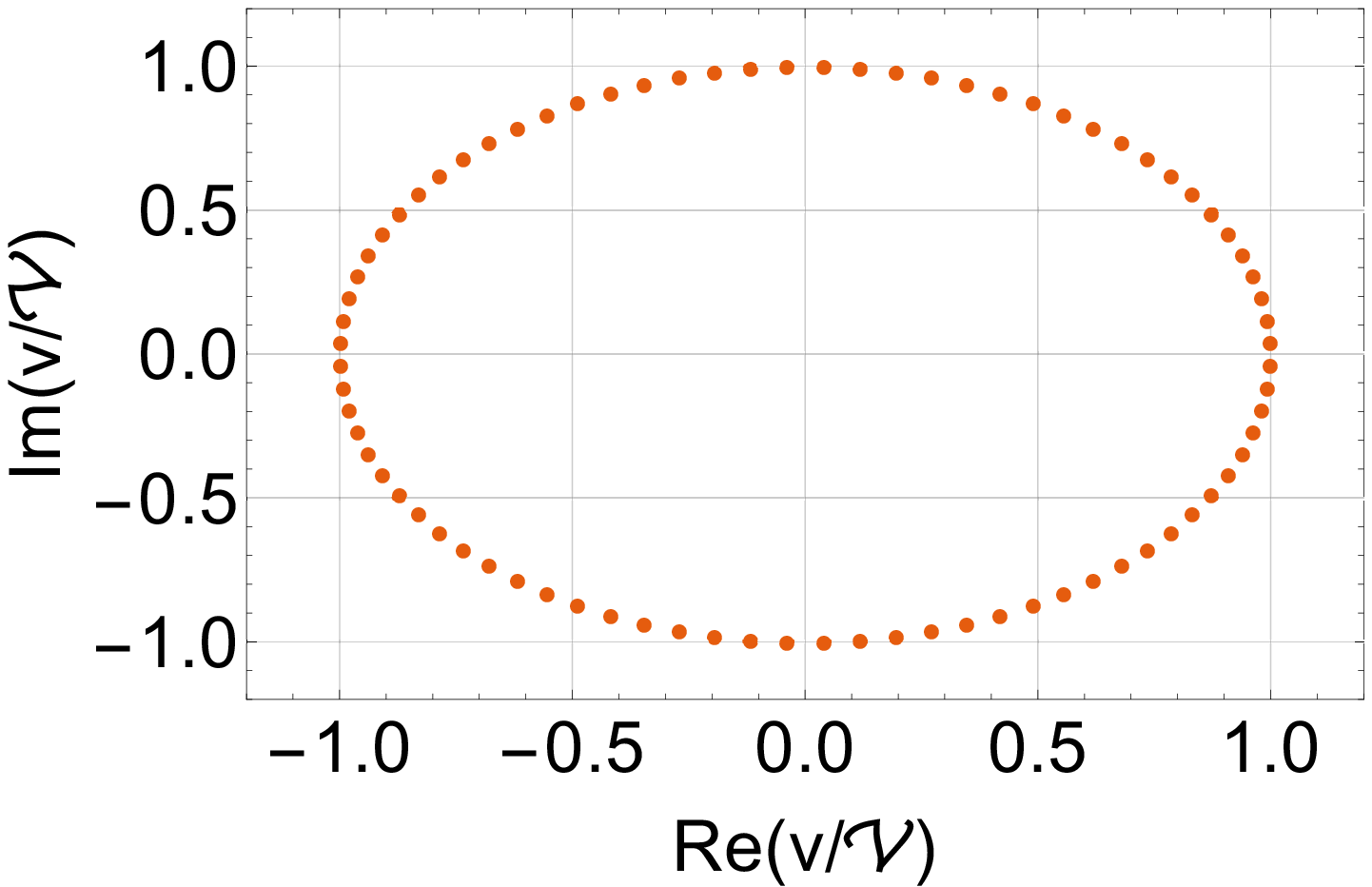}}\hspace{0.02\columnwidth}\subfloat[]{\includegraphics[width=0.40\columnwidth]{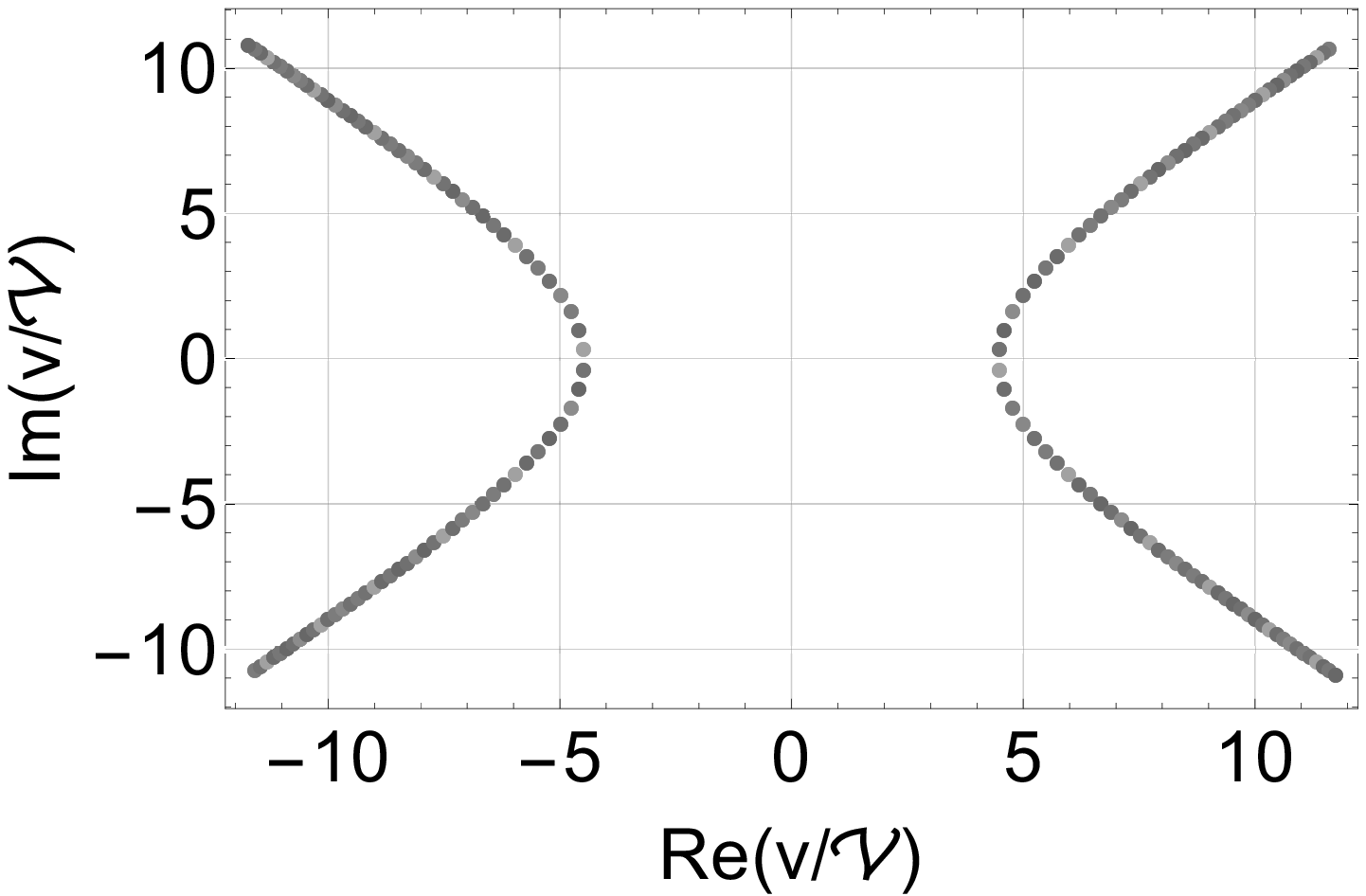}}

\end{center}

\caption{Poles of the distribution functions considered in this paper in the
complex-velocity space. Panels on the left (red points) show poles
of Cauchy-type distributions $\left(v^{J}+\mathcal{V}^{J}\right)^{-1}$
for $J=2,\thinspace4,\thinspace80$ (a, c and e) and panels on the
right (gray points) show poles of the Fermi-Dirac $f_{FD}\left(v\right)$
(equation \ref{eq:distFD}) for $\mu/T=-20,\thinspace0,\thinspace20$
(b, d and f). \label{fig:distributions-3}}
\end{figure}

In order to better understand these distribution functions, they are
plotted in figures \ref{fig:distributions-1}a (Cauchy-type functions)
and \ref{fig:distributions-1}b (Fermi-Dirac type functions). However,
as it is necessary to evaluate these distribution functions with complex
arguments, we also illustrate their structure in the complex $v$
plane in figure \ref{fig:distributions-3}.

We now proceed to present the susceptibilities obtained by inserting
the above distribution functions into equation \ref{eq:generalsusceptibility}.
With the exception of the arbitrarily degenerate Fermi-Dirac distribution,
the distributions have been chosen so that the integration of equation
\ref{eq:generalsusceptibility} can be carried out analytically by
making use of the residue theorem. In the case of $f_{\delta}\left(v\right)$
the integration is trivial, and for ~$f_{D}$ it is carried out by
standard methods with a well-defined result when the phase velocity
is greater than the Fermi velocity. Smaller phase velocities result
in complex logarithms for which the branch cuts must be considered
carefully. The following susceptibilities are the result of these
integrations:

\begin{equation}
\chi_{\delta}\left(\omega,k\right)=\frac{\omega_{p}^{2}}{k^{4}\hbar^{2}/m^{2}-\omega^{2}};\label{eq:chidel}
\end{equation}
\begin{equation}
\chi_{C}\left(\omega,k\right)=\frac{\omega_{p}^{2}}{(k\mathcal{V}-\text{i}\omega)^{2}+k^{4}\hbar^{2}/m^{2}};\label{eq:chicau}
\end{equation}
\begin{equation}
\chi_{C2}\left(\omega,k\right)=\frac{\omega_{p}^{2}\left(k^{4}\hbar^{2}/m^{2}+3k^{2}\mathcal{V}^{2}-4\text{i}k\mathcal{V}\omega-\omega^{2}\right)}{\left[(k\mathcal{V}-\text{i}\omega)^{2}+k^{4}\hbar^{2}/m^{2}\right]^{2}};\label{eq:chic2}
\end{equation}

\begin{gather}
\chi_{4}\left(\omega,k\right)=\frac{\text{i}m^{2}\omega_{p}^{2}}{\sqrt{2}}\times\nonumber \\
\biggl[\left((-1)^{3/4}k^{4}\hbar^{2}+\sqrt[4]{-1}k^{2}m^{2}\mathcal{V}^{2}+2km^{2}\mathcal{V}\omega-(-1)^{3/4}m^{2}\omega^{2}\right)^{-1}\nonumber \\
+\left(\sqrt[4]{-1}k^{4}\hbar^{2}+(-1)^{3/4}k^{2}m^{2}\mathcal{V}^{2}+2km^{2}\mathcal{V}\omega-\sqrt[4]{-1}m^{2}\omega^{2}\right)^{-1}\biggr];\label{eq:chi2}
\end{gather}

\begin{equation}
\chi_{CJ}\left(\omega,k\right)=\pi\text{i}\frac{\omega_{p}^{2}}{k^{3}}\frac{\mathcal{V}^{2J-1}\Gamma(J)}{\sqrt{\pi}\Gamma\left(J-\frac{1}{2}\right)}\text{Res}\left[\frac{1}{\left(v^{2}+\mathcal{V}^{2}\right)^{J}}\frac{1}{v-\omega/k},i\mathcal{V}\right];\label{eq:chicauJ}
\end{equation}
\begin{gather}
\begin{gathered}\begin{array}{c}
\chi_{J}\left(\omega,k\right)=\sum_{j=1}^{\frac{J}{2}}\left(-\text{i}\right)\sin\left(\frac{\pi}{J}\right)\times\end{array}\\
\left[(-1)^{\frac{1-2j}{J}}\left(\frac{k^{4}\hbar^{2}}{m^{2}\omega_{p}^{2}}-\frac{\omega^{2}}{\omega_{p}^{2}}\right)+\frac{k^{2}\mathcal{V}^{2}(-1)^{\frac{2j+J-1}{J}}}{\omega_{p}^{2}}-\frac{2k\mathcal{V}\omega}{\omega_{p}^{2}}\right]^{-1}
\end{gathered}
;\label{eq:chiJ}
\end{gather}
and, in agreement with \citep{Krivitskii1991,Eliasson2010,Vladimirov2011}
\begin{gather}
\begin{gathered}\chi_{D}\left(\omega,k\right)=\frac{3m}{8\hbar k\mathcal{V}}\left(\frac{\omega_{p}}{k\mathcal{V}}\right)^{2}\times\\
\biggl[4\hbar k\mathcal{V}/m+\left(\omega/k-\hbar k/m+\mathcal{V}\right)(\omega/k-\hbar k/m-\mathcal{V})\log\left(\frac{\hbar k/m-\mathcal{V}-\omega/k}{\hbar k/m+\mathcal{V}-\omega/k}\right)\\
-\left(\omega/k+\hbar k/m-\mathcal{V}\right)\left(\omega/k+\hbar k/m+\mathcal{V}\right)\log\left(\frac{\hbar k/m+\mathcal{V}+\omega/k}{\hbar k/m-\mathcal{V}+\omega/k}\right)\biggr]
\end{gathered}
.\label{eq:chiDeg}
\end{gather}

\subsection{Dispersion Relations in Single-Population Plasmas}

Before proceeding to the case of multiple populations, we present
the dispersion relation in single population plasmas obtained from
solving 
\begin{equation}
0=1+\chi_{s}.
\end{equation}
For the delta-function we obtain

\begin{equation}
\frac{\omega_{\delta}}{\omega_{p}}=\begin{cases}
\begin{array}{c}
\pm\sqrt{1+\frac{k^{4}\hbar^{2}}{m^{2}\omega_{p}^{2}}}\\
\pm\frac{k^{2}\hbar}{m\omega_{p}}
\end{array}\end{cases},\label{eq:dispdel}
\end{equation}
in agreement with \citet{Haas2009a}. Note that there is indeed no
Landau damping, but there is now wave dispersion which is due entirely
to quantum dynamical effects. Additionally, the second pair of modes
is called zero-sound (see e.g. \citet{Krivitskii1991}) and is a purely
quantum effect. Due to the lack of a characteristic particle velocity
scale~$\mathcal{V}$, for plotting we non-dimensionalize according
to the second scheme by introducing variables $\omega\equiv\omega/\omega_{p}$,
$k\equiv\sqrt{2\hbar/m\omega_{p}}k$ where we utilise the velocity
scale $\eta\equiv\sqrt{2\hbar\omega_{p}/m}$ which is the speed of
an electron with kinetic energy equal to the plasmon energy~$\hbar\omega_{p}$.

\begin{figure}
\begin{center}\subfloat[]{\includegraphics[width=0.40\columnwidth]{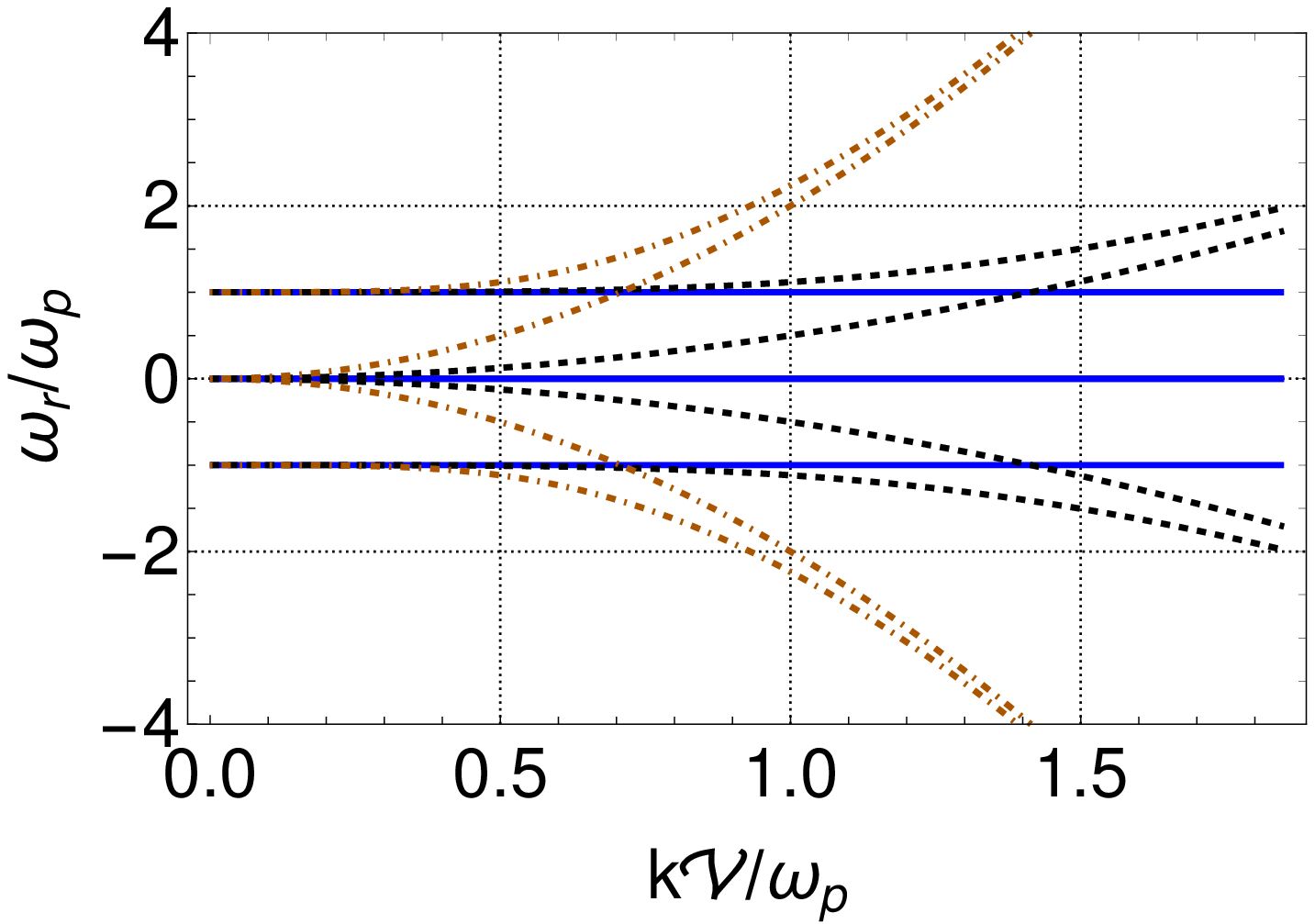}}\hspace{0.02\columnwidth}\subfloat[]{\includegraphics[width=0.40\columnwidth]{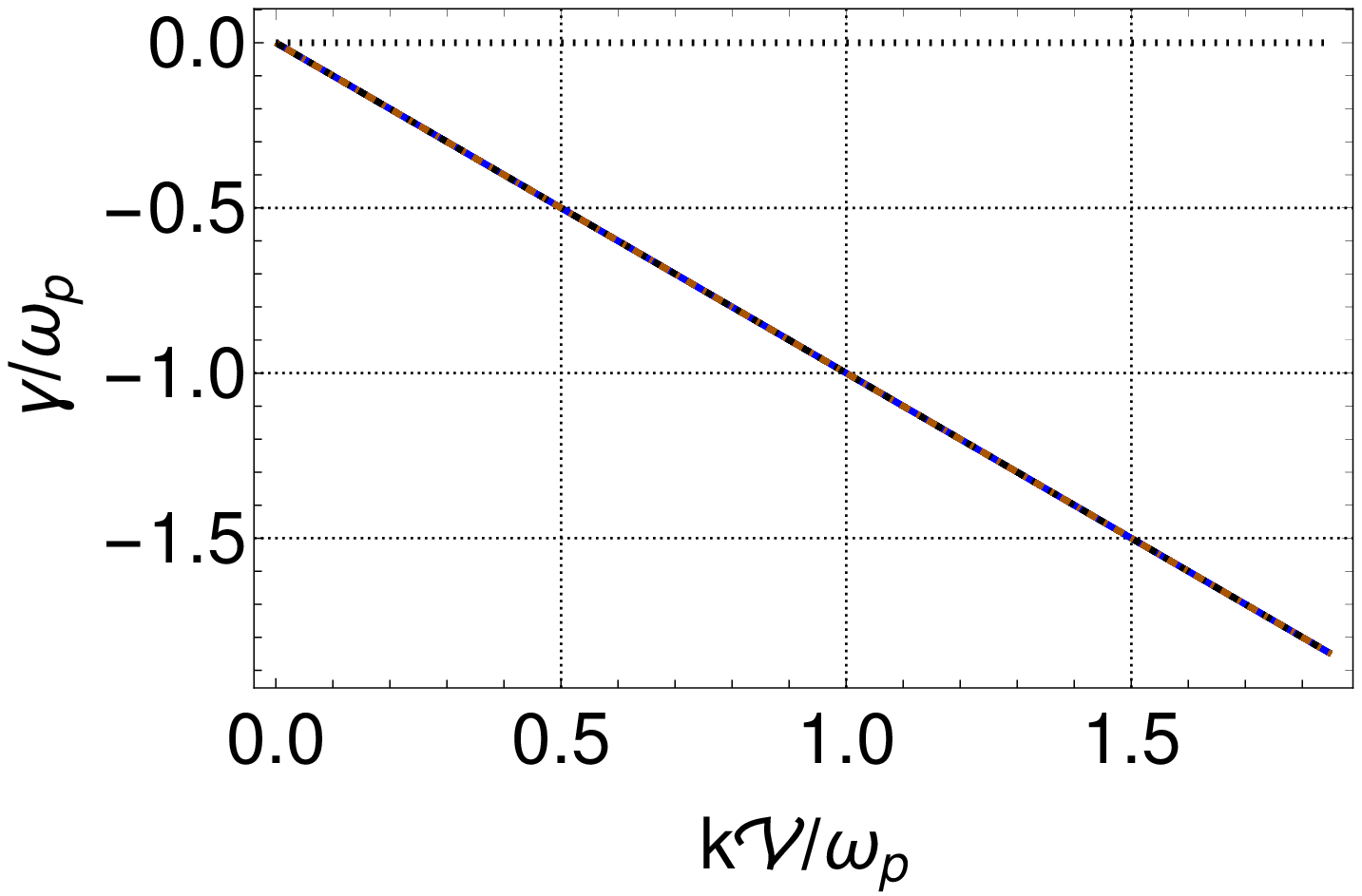}}\end{center}

\caption{Dispersion relation (equation \ref{eq:dispcau}) for single-population
plasma with Cauchy distribution function~$f_{C}$ (equation \ref{eq:distcau})
for three values of~$H$: $H=0$ (solid, blue), $H=0.5$ (dashed,
black), and $H=2$ (dot-dashed, orange). The quantum parameter~$H$
causes dispersion in the real part (left panel) of the frequency but
the damping rate~$-\gamma$ (right panel) is independent of~$H$
and equal to $k\mathcal{V}/\omega_{p}$. \label{fig:1popcau}}
\end{figure}

For the Cauchy distribution we obtain 
\begin{equation}
\frac{\omega_{C}}{\omega_{p}}=\begin{cases}
\begin{array}{c}
\pm\sqrt{1+\frac{k^{4}\hbar^{2}}{m^{2}\omega_{p}^{2}}}-\frac{ik\mathcal{V}}{\omega_{p}}=\pm\sqrt{1+H^{2}K^{4}}-\text{i}K\\
\pm\frac{k^{2}\hbar}{m\omega_{p}}-\frac{ik\mathcal{V}}{\omega_{p}}=HK^{2}-\text{i}K
\end{array}\end{cases}\label{eq:dispcau}
\end{equation}
in agreement with \citet{Haas2001}. There is now Landau damping,
with the damping rate simply equal to the dimensionless wavenumber~$K$,
but the real part of the frequency is the same as for the delta-function
case. This dispersion relation is plotted in figure \ref{fig:1popcau}.

\begin{figure}
\begin{center}\subfloat[]{\includegraphics[width=0.40\columnwidth]{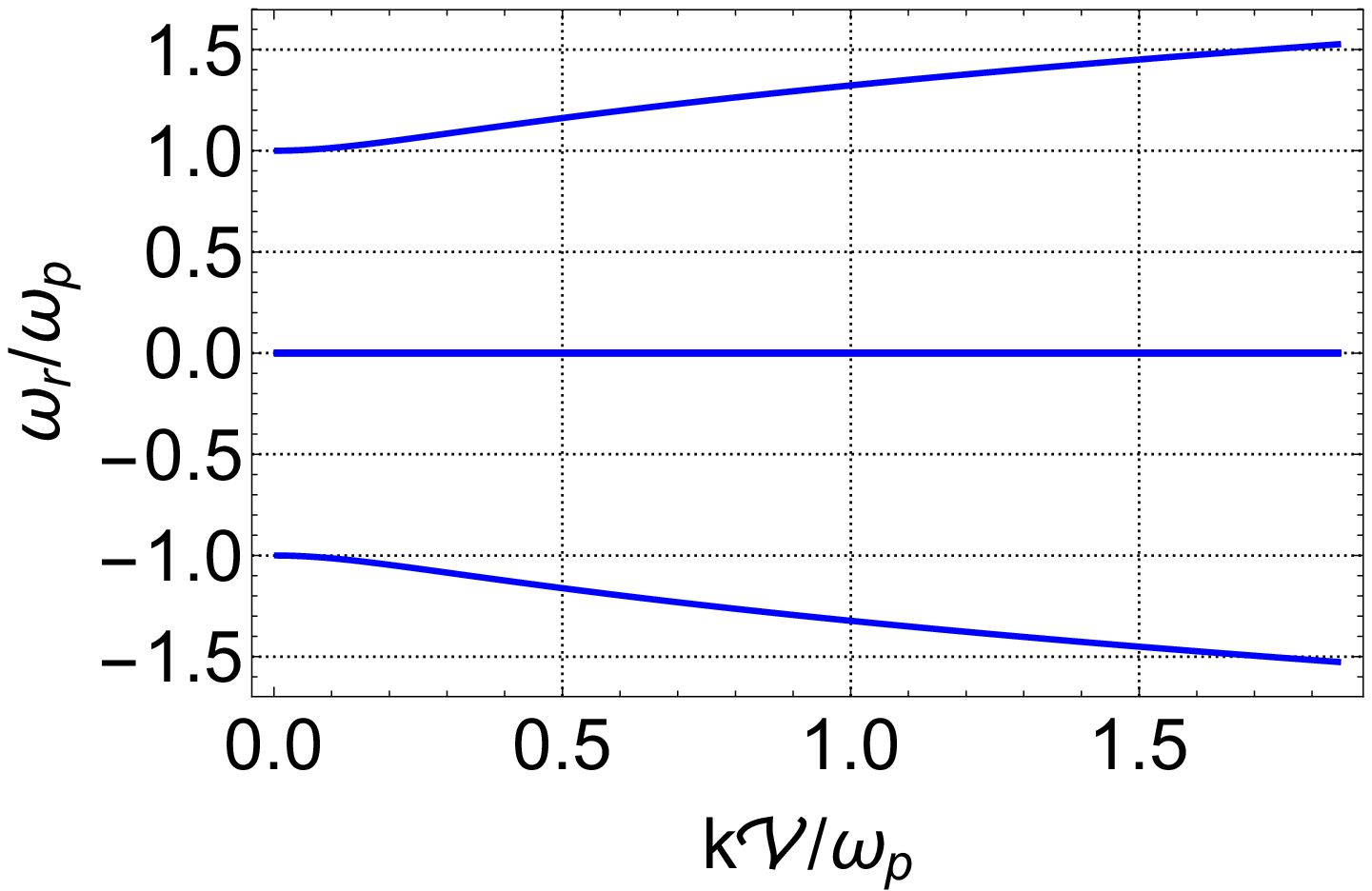}}\hspace{0.02\columnwidth}\subfloat[]{\includegraphics[width=0.40\columnwidth]{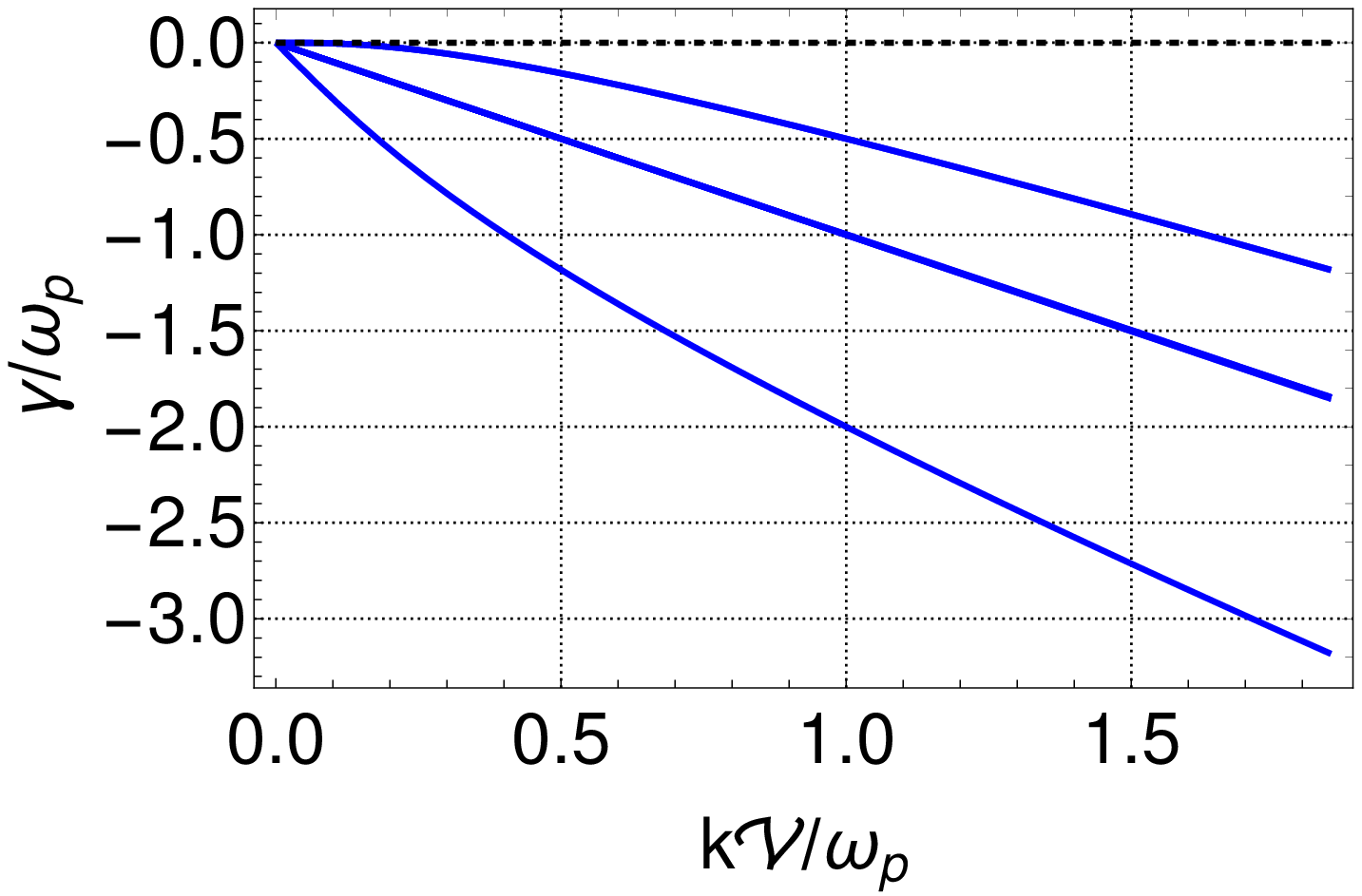}}

\vspace{1mm}
\subfloat[]{\includegraphics[width=0.40\columnwidth]{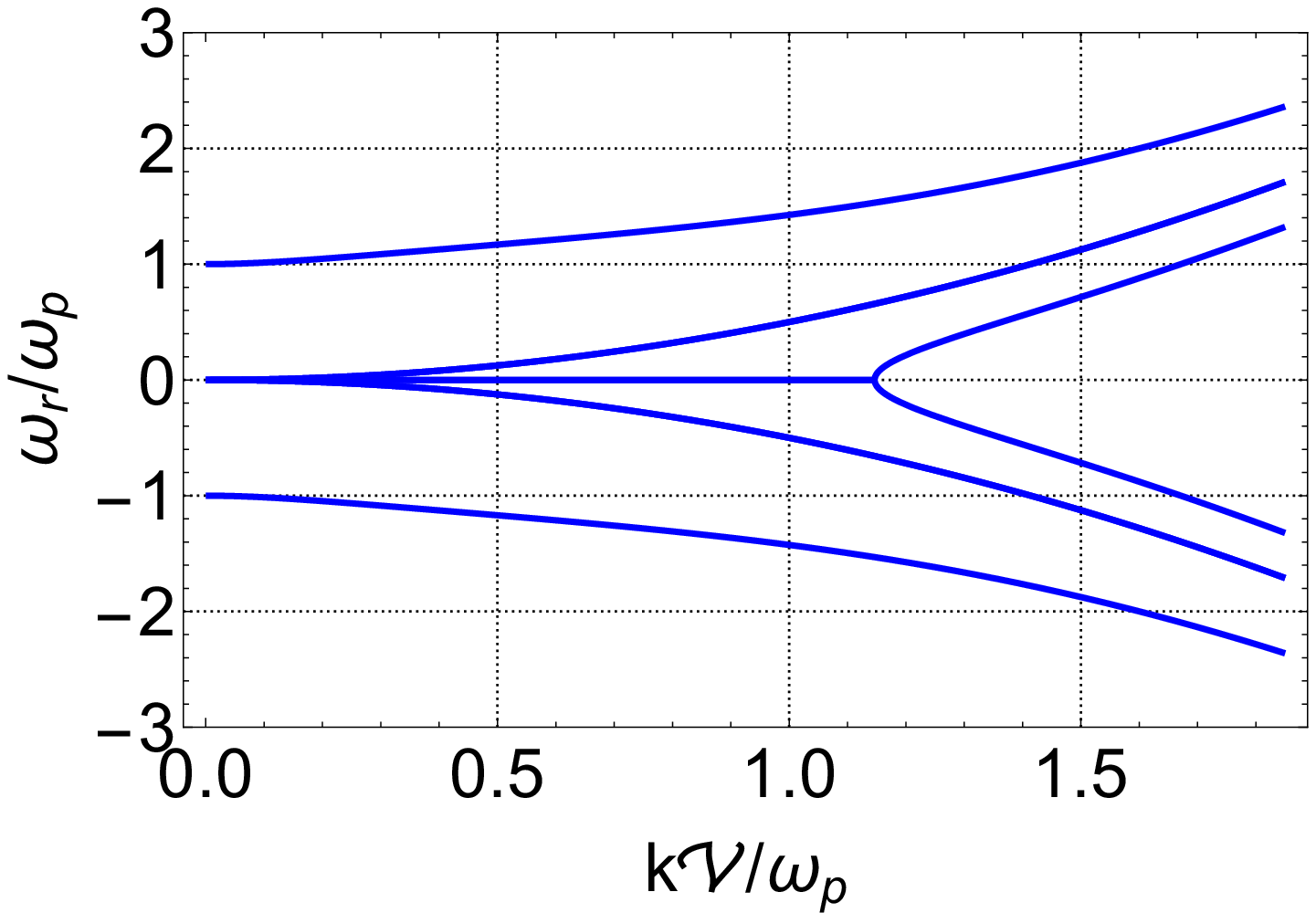}}\hspace{0.02\columnwidth}\subfloat[]{\includegraphics[width=0.40\columnwidth]{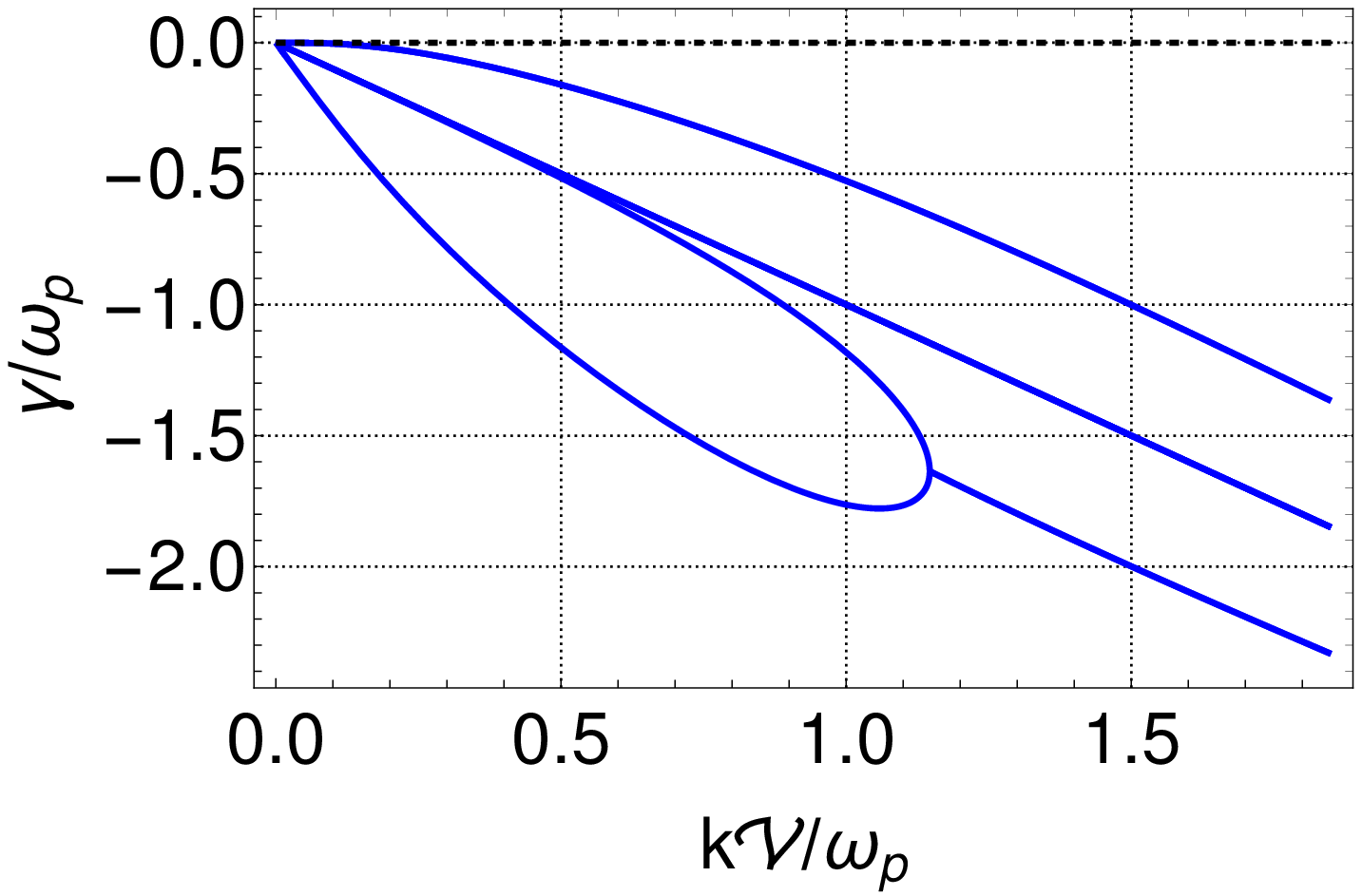}}

\vspace{1mm}
\subfloat[]{\includegraphics[width=0.40\columnwidth]{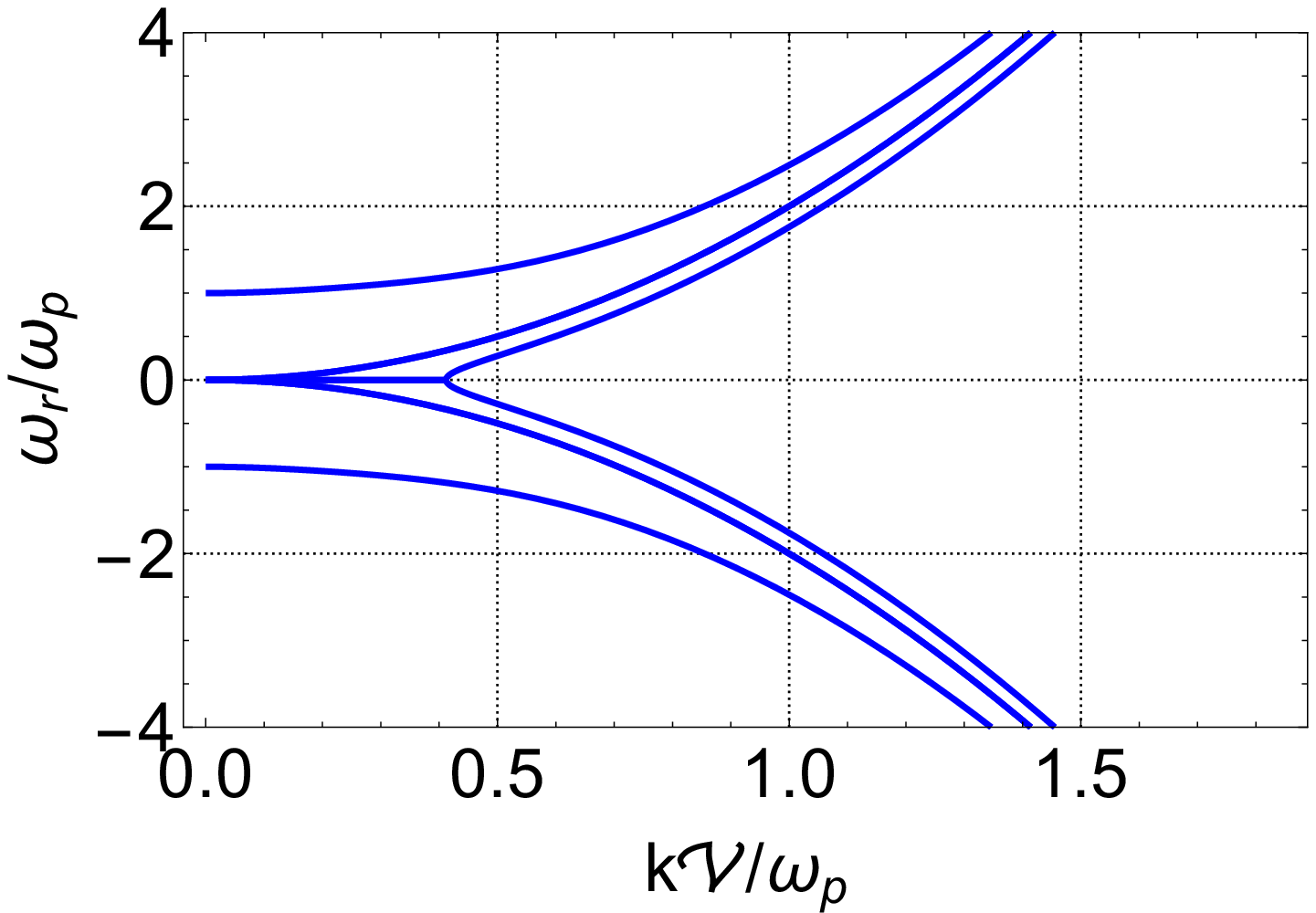}}\hspace{0.02\columnwidth}\subfloat[]{\includegraphics[width=0.40\columnwidth]{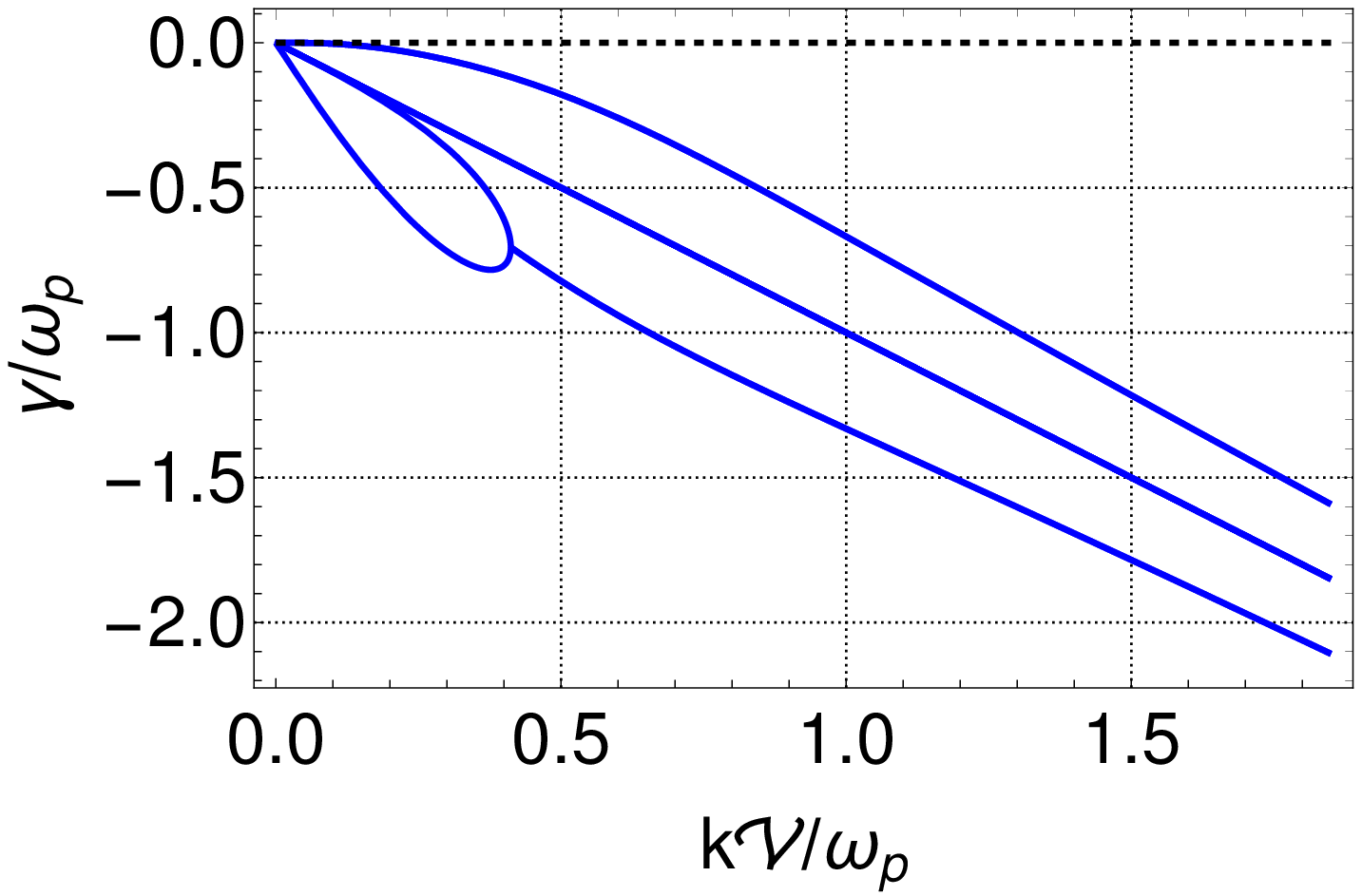}}\end{center}

\caption{Dispersion relation for single-population plasma with squared Cauchy
or~$f_{C2}$ distribution function (equation \ref{eq:distcau2})
for three values of~$H$:~$0$ (panels a and b),~$0.5$ (panels
c and d), and~$2$ (panels e and f). \label{fig:1popcau2} }
\end{figure}

\begin{figure}
\begin{center}\subfloat[]{\includegraphics[width=0.40\columnwidth]{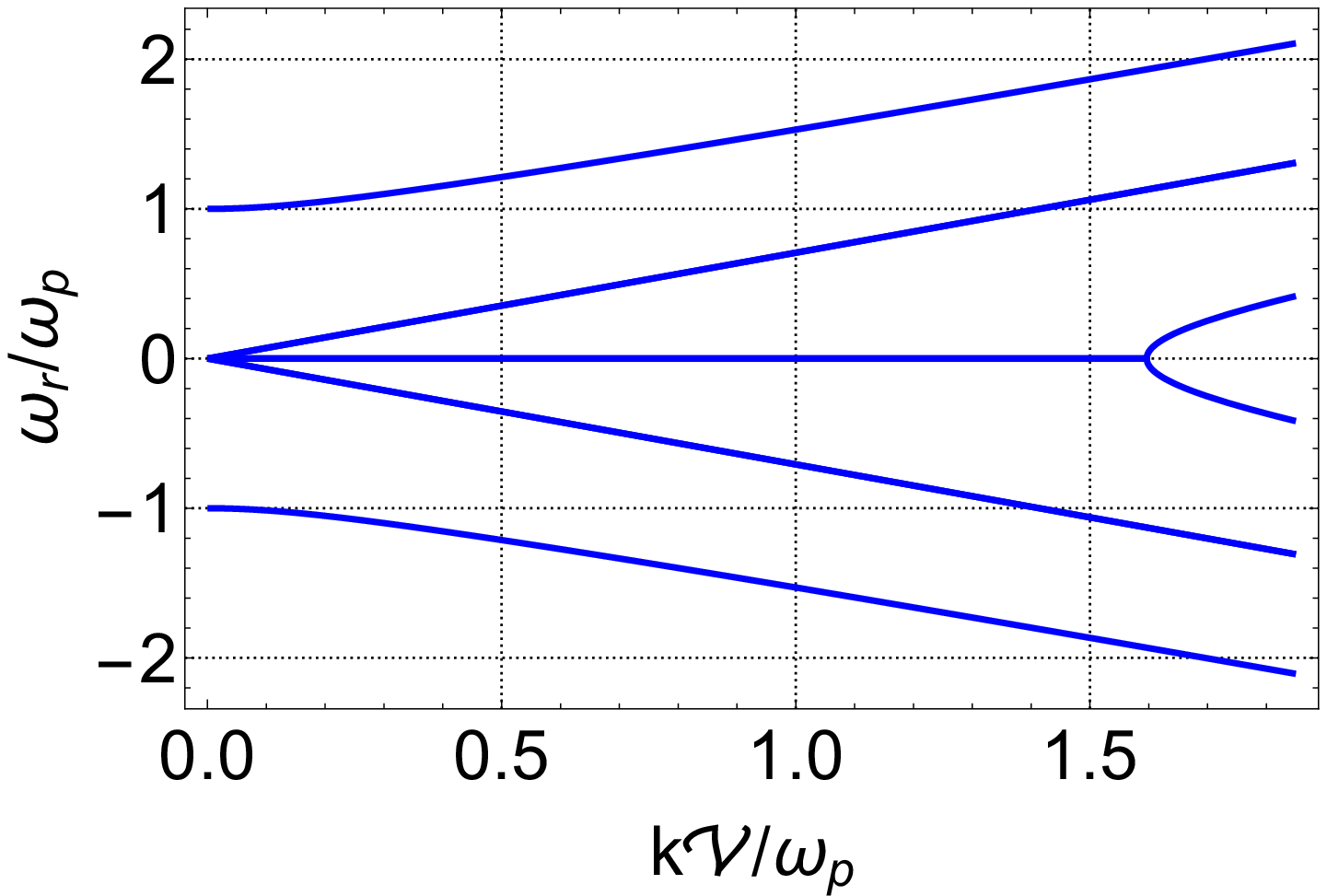}}\hspace{0.02\columnwidth}\subfloat[]{\includegraphics[width=0.40\columnwidth]{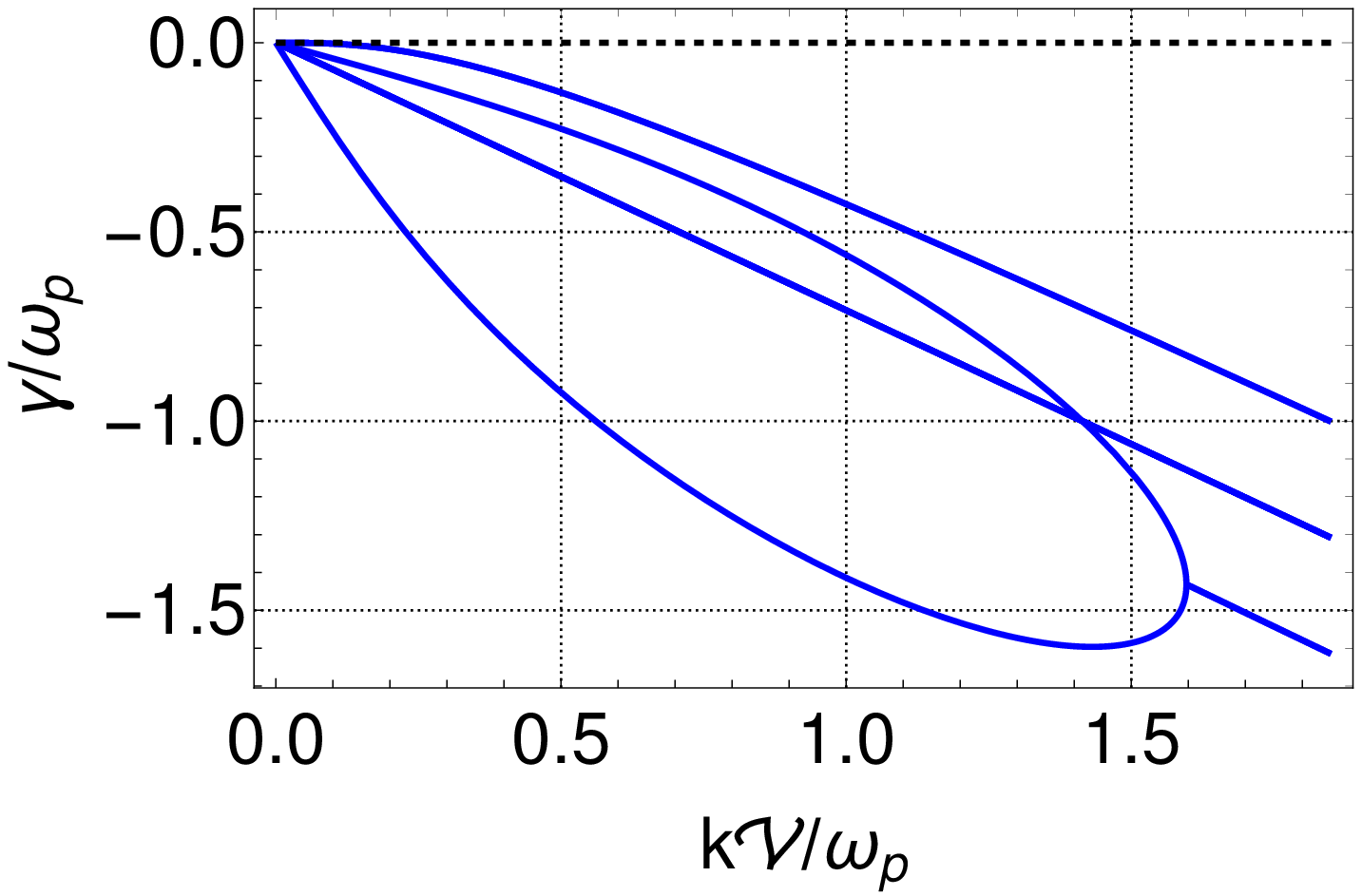}}

\vspace{1mm}
\subfloat[]{\includegraphics[width=0.40\columnwidth]{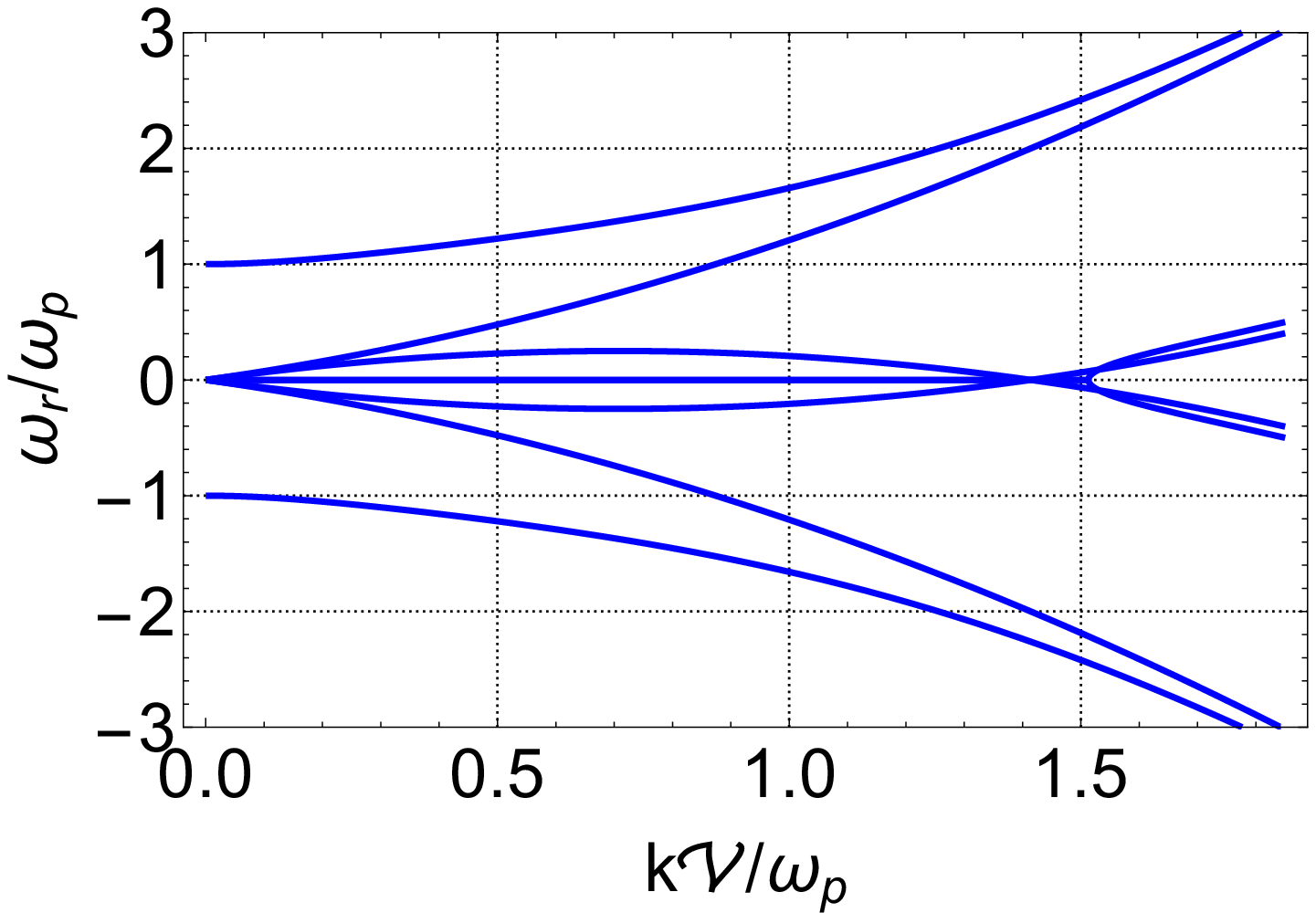}}\hspace{0.02\columnwidth}\subfloat[]{\includegraphics[width=0.40\columnwidth]{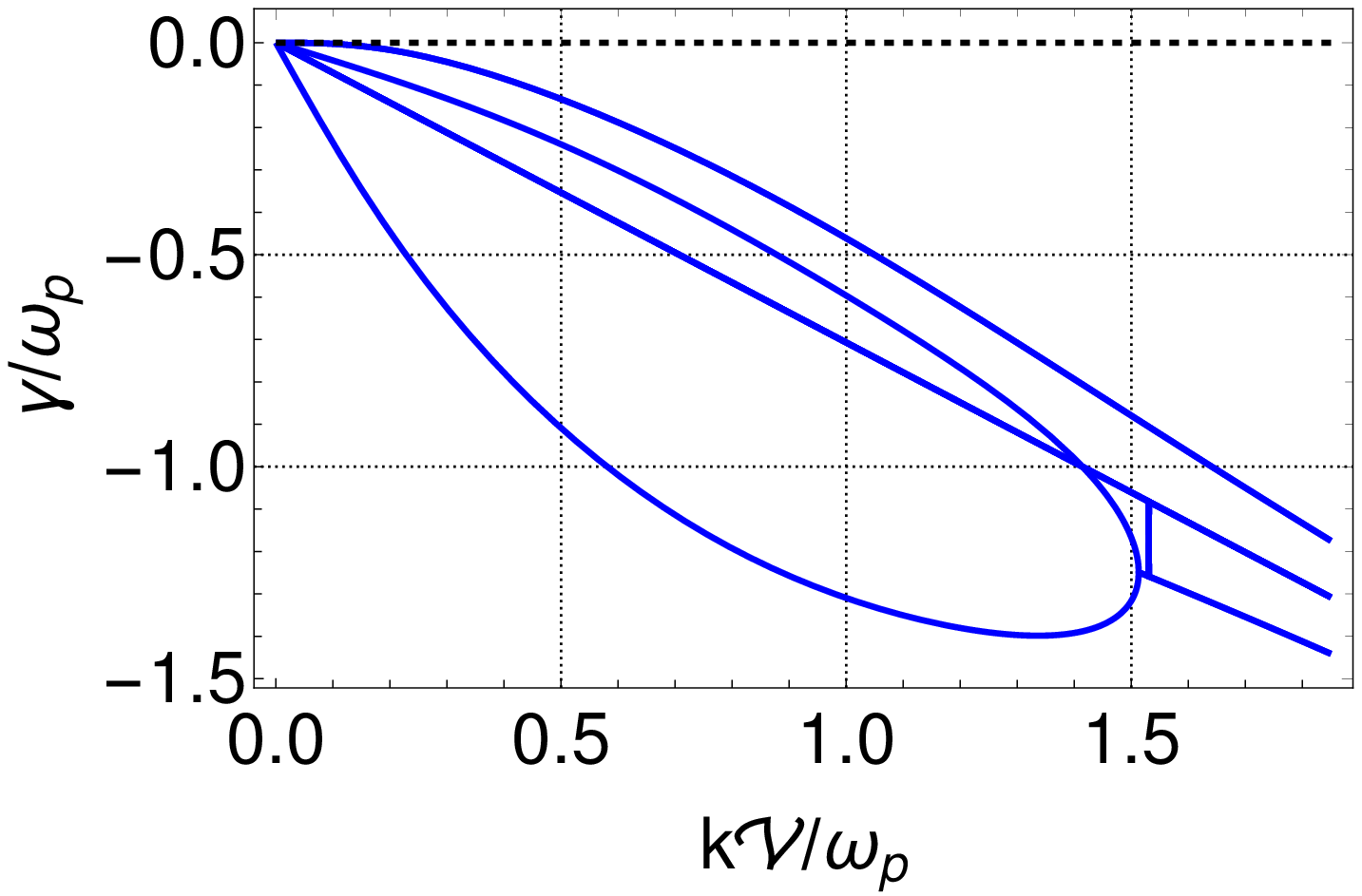}}

\vspace{1mm}
\subfloat[]{\includegraphics[width=0.40\columnwidth]{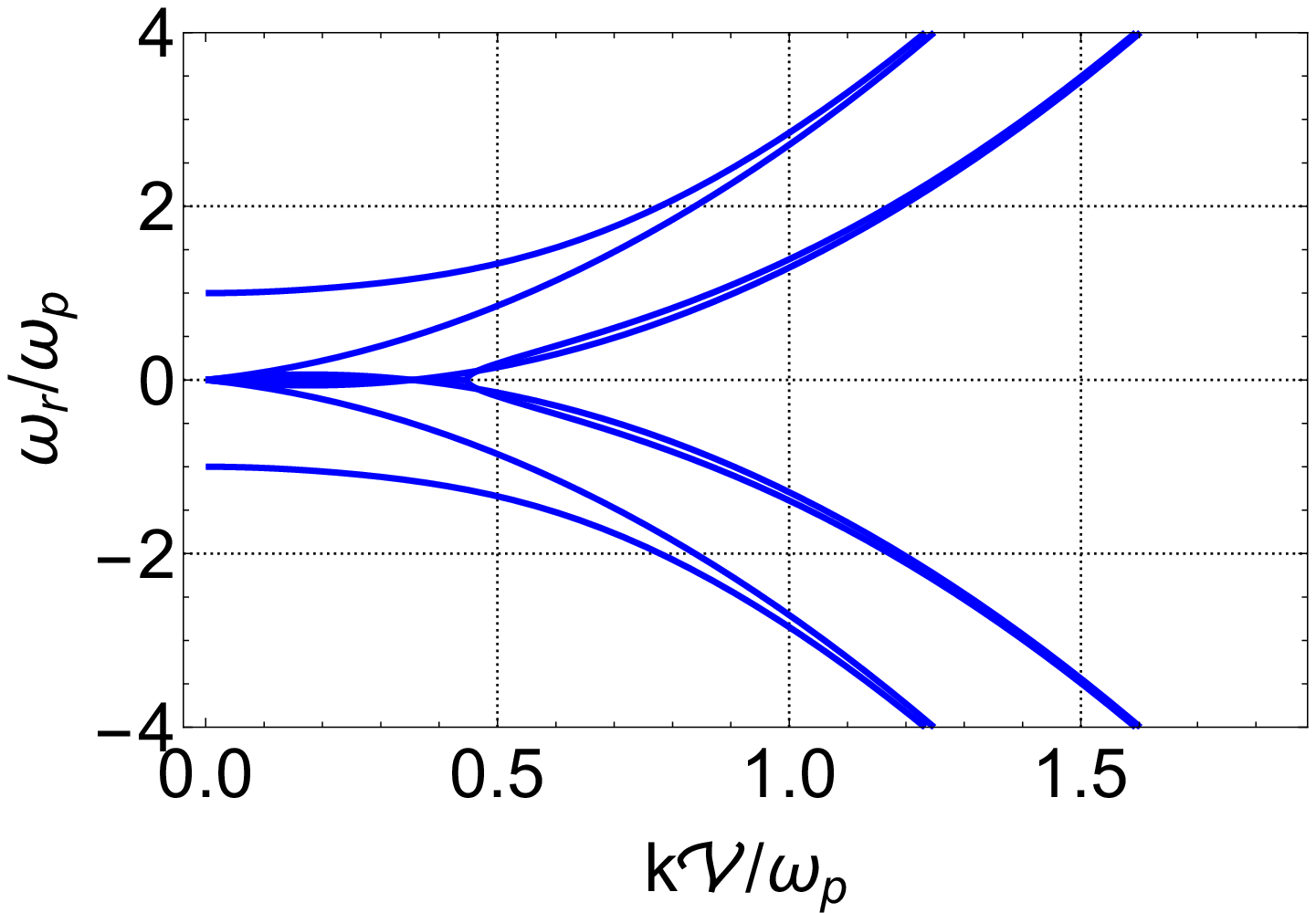}}\hspace{0.02\columnwidth}\subfloat[]{\includegraphics[width=0.40\columnwidth]{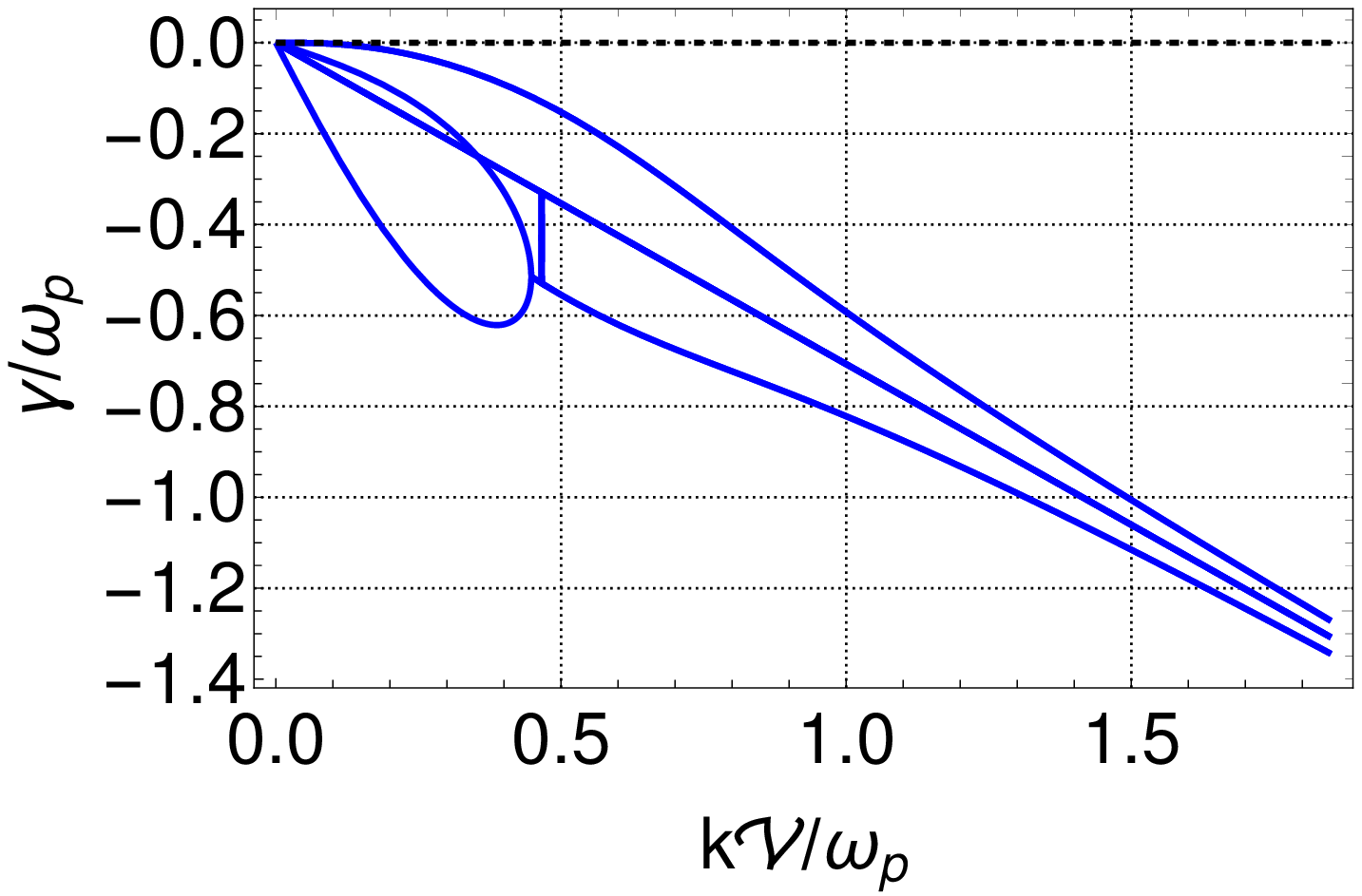}}\end{center}

\caption{Dispersion relation for single-population plasma with inverse-quartic
or~$f_{4}$ distribution function (equation \ref{eq:dist2}) for
three values of~$H$:~$0$ (panels a and b),~$0.5$ (panels c and
d), and~$2$ (panels e and f).\label{fig:1popcaudub}}
\end{figure}

For the more complicated distribution functions~$\chi_{CJ}$ and~$\chi_{J}$,
there is not an explicit algebraic solution for the frequency as the
susceptibilities are all of greater than fourth order in~$\omega$.
Instead, we examine the dielectric functions in the limits of large
and small~$K$, and in the case of~$\chi_{J}$ we can consider the
case~$J\gg1$. In the long-wavelength $K\ll1$ limit, to fourth order,
for the squared Cauchy distribution the regular plasmon mode is 
\begin{equation}
\Omega_{C2}=1+\frac{3}{2}K^{2}-4{\rm i}K^{3}+\left(\frac{H^{2}}{2}-\frac{105}{8}\right)K^{4},\label{eq:dispcau2}
\end{equation}
and for the $J=4$ inverse-quartic distribution it is 
\begin{equation}
\Omega_{4}=1+\frac{3}{2}K^{2}-2\sqrt{2}{\rm iK^{3}+}\left(\frac{H^{2}}{2}-\frac{65}{8}\right)K^{4}.\label{eq:disp2}
\end{equation}
Note that for both of these cases the imaginary part now only appears
to third order in~$K$, and that quantum effects only appear at fourth
order, and only in the real part of the frequency. The exact, numerically-computed
dispersion relations are plotted in figures \ref{fig:1popcau2} (squared
Cauchy) and \ref{fig:1popcaudub} (inverse-quartic).

Dispersion relations are not shown for the cases with non-meromorphic
distribution functions $f_{D}\left(v\right)$ and $f_{FD}\left(v\right)$
in equations \ref{eq:distDeg} and \ref{eq:distFD} as the focus of
this work is to ascertain the influence of individual poles in the
complex distribution function, but they have been studied by \citet{Rightley2016}.
Additionally, further discussion of the distribution functions~$f_{CJ}$
(equation \ref{eq:distcauJ}) and~$f_{J}$ (equation \ref{eq:distJ})
is reserved for a future work.

\section{Dispersion Relations for Two-Population Plasmas\label{sec:twocomp}}

\begin{figure}
\begin{center}%
\begin{tabular}{c}
\includegraphics[width=0.5\columnwidth]{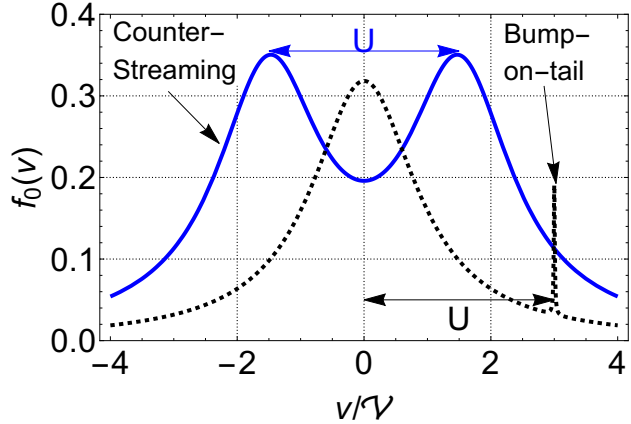}\tabularnewline
\end{tabular}\end{center}

\caption{Two-population distribution functions for symmetrical case (solid,
blue) and for bump-on-tail case (dashed, black). The separation velocity
between the two populations is denoted by~$U$. \label{fig:distributionsUnstable}}
\end{figure}

If two populations are present, we define one to be the primary population
and one the secondary. The non-dimensional variables in section \ref{sec:parameters}
are defined in terms of the primary population. In this paper, for
each type of distribution function we consider two cases: Case (1)
\textit{the symmetrical case}: that of two identical counter-drifting
populations and Case (2) \textit{the bump-on-tail case}: that of a
primary distribution and a drifting secondary delta-function beam.
The issue of frame of reference should be addressed, as for multiple
populations there is not necessarily a natural choice for this frame.
In Case 1 we choose the centre-of-momentum frame, so that each population
moves past the observer with speed $U/2$ in opposite directions,
and in Case 2 we choose the reference frame of the primary (finite-width)
population, with the low-density population streaming by at speed~$U$
in the positive direction. The difference between Cases 1 and 2 is
shown in figure~\ref{fig:distributionsUnstable}.

The dielectric function for Case 1 is 
\begin{equation}
\epsilon_{1}=1+\frac{1}{2}\left[\chi_{s}\left(\omega+kU/2,k\right)+\chi_{s}\left(\omega-kU/2,k\right)\right],
\end{equation}
with the factor $1/2$ ensuring the total density is equal to unity.
The dielectric function for Case 2 is 
\begin{equation}
\epsilon_{2}=1+\left[\left(1-n\right)\chi_{s}\left(\omega,k\right)+n\chi_{\delta}\left(\omega-kU,k\right)\right],
\end{equation}
where the quantity~$n$ is the fraction of particles present in the
beam.

\begin{figure}
\begin{center}%
\begin{tabular}{c}
\includegraphics[width=0.5\columnwidth]{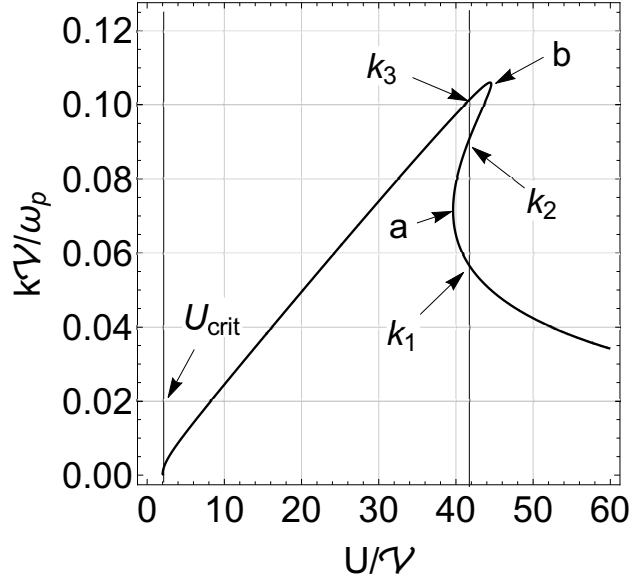}\tabularnewline
\end{tabular}\end{center}

\caption{Example boundary of instability region, for symmetrical counter-propagating
Cauchy distributions with $H=150$. The Cauchy distribution is used for illustration but the other distributions considered in this paper result in the same general features.  The unstable region is under the
curve. For given~$U$, instability exists for $0<k<k_{1}$ and $k_{2}<k<k_{3}$.
Point a refers to the drift speed~$U_{a}$ beyond which the second
region of instability defined by~$k_{2}$ and~$k_{3}$ ceases to
exist when~$k_{2}$ and~$k_{3}$ merge. Point b refers to the minimum
drift speed~$U_{b}$ needed for the existence of the second region
of instability, at which point~$k_{1}$ and~$k_{2}$ merge. $U_{{\rm crit}}$
refers to the minimum drift speed required for the existence of any
instability. \label{fig:defintions_contour}}
\end{figure}

As in the classical case, two-population plasmas may allow for unstable
modes, at least for large enough velocity separations~$U$. For each
distribution function and case, we will be mapping the boundaries
of the region(s) of instability in the $\left(k,U\right)$ parameter
space. We denote the critical value of drift velocity required for
the onset of instability as~$U_{{\rm crit}}$. If a configuration
with given~$U$ allows for an unstable mode, we name the maximum
growth rate of this mode $\gamma_{{\rm max}}\left(U\right)$ and the
wavenumber at which this occurs $k_{{\rm max}}\left(U\right)$. As
will be seen, for fixed~$U$ there exist up to three critical values
of~$k$ which define the boundaries ($\gamma=0$) of the unstable
region, and we label these (in order of increasing value of~$k$)~$k_{1}$,~$k_{2}$,
and~$k_{3}$. In the classical case instability exists only for long
enough wavelengths $k<k_{1}.$ In contrast, in the quantum case for
large enough values of the quantum recoil parameter there are two
instability windows $k<k_{1}$ and $k_{2}<k<k_{3}$. In addition,
there are two further special points denoted a and b that define the
extent of the instability region. These points are demonstrated in
figure \ref{fig:defintions_contour}.

\subsection{Delta-Function Distribution: Most Simple Case}

To begin, we consider a two-population plasma in which both populations
have zero velocity spread; i.e. Dirac delta-function distributions.
In Case 1 both populations are identical, and in Case 2 the populations
have unequal total particle densities, with one beam being substantially
less dense.

\subsubsection{Case 1: Symmetrically counter-propagating distributions}

\begin{figure}
\begin{center}\subfloat[]{\includegraphics[width=0.40\columnwidth]{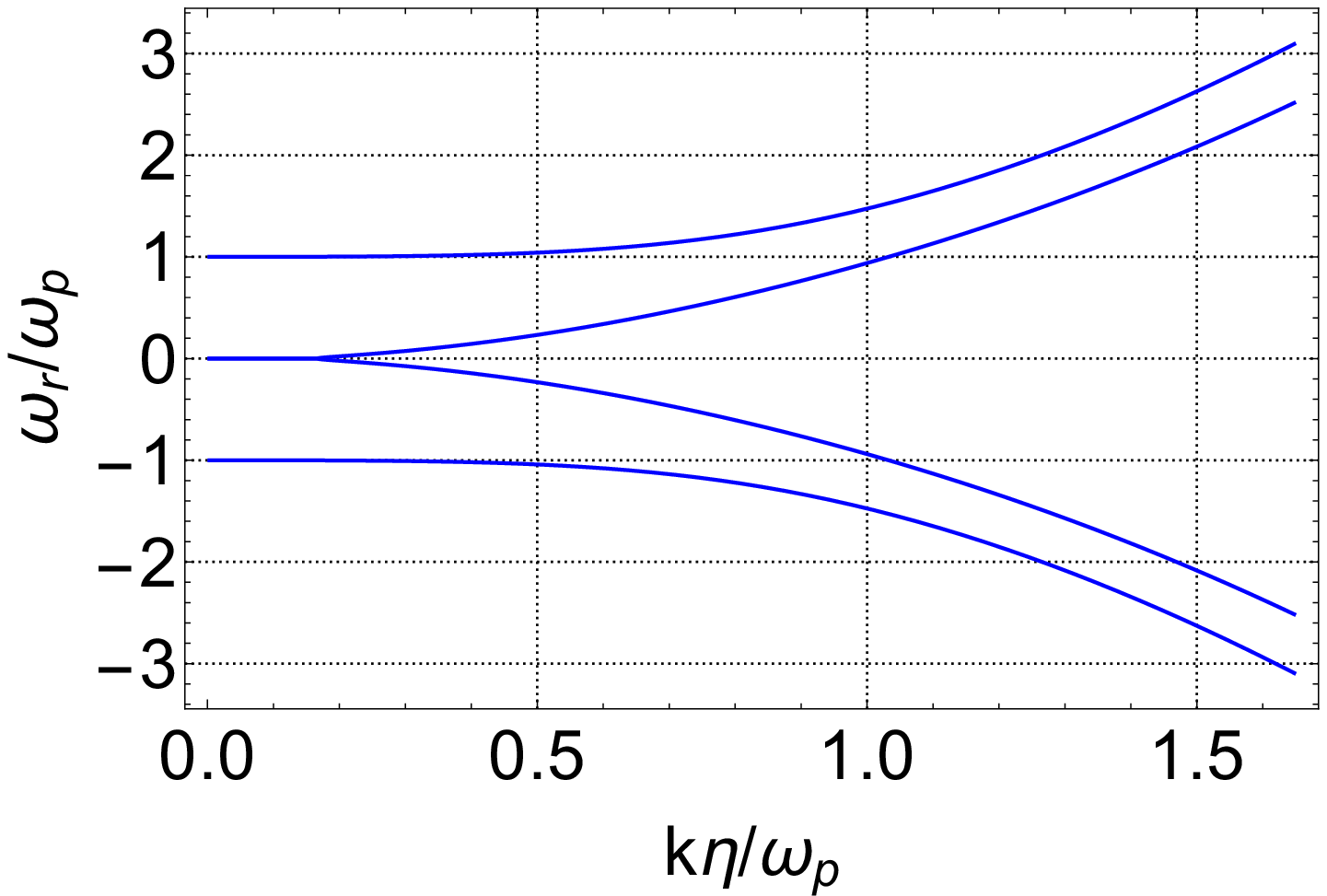}}\hspace{0.02\columnwidth}\subfloat[]{\includegraphics[width=0.40\columnwidth]{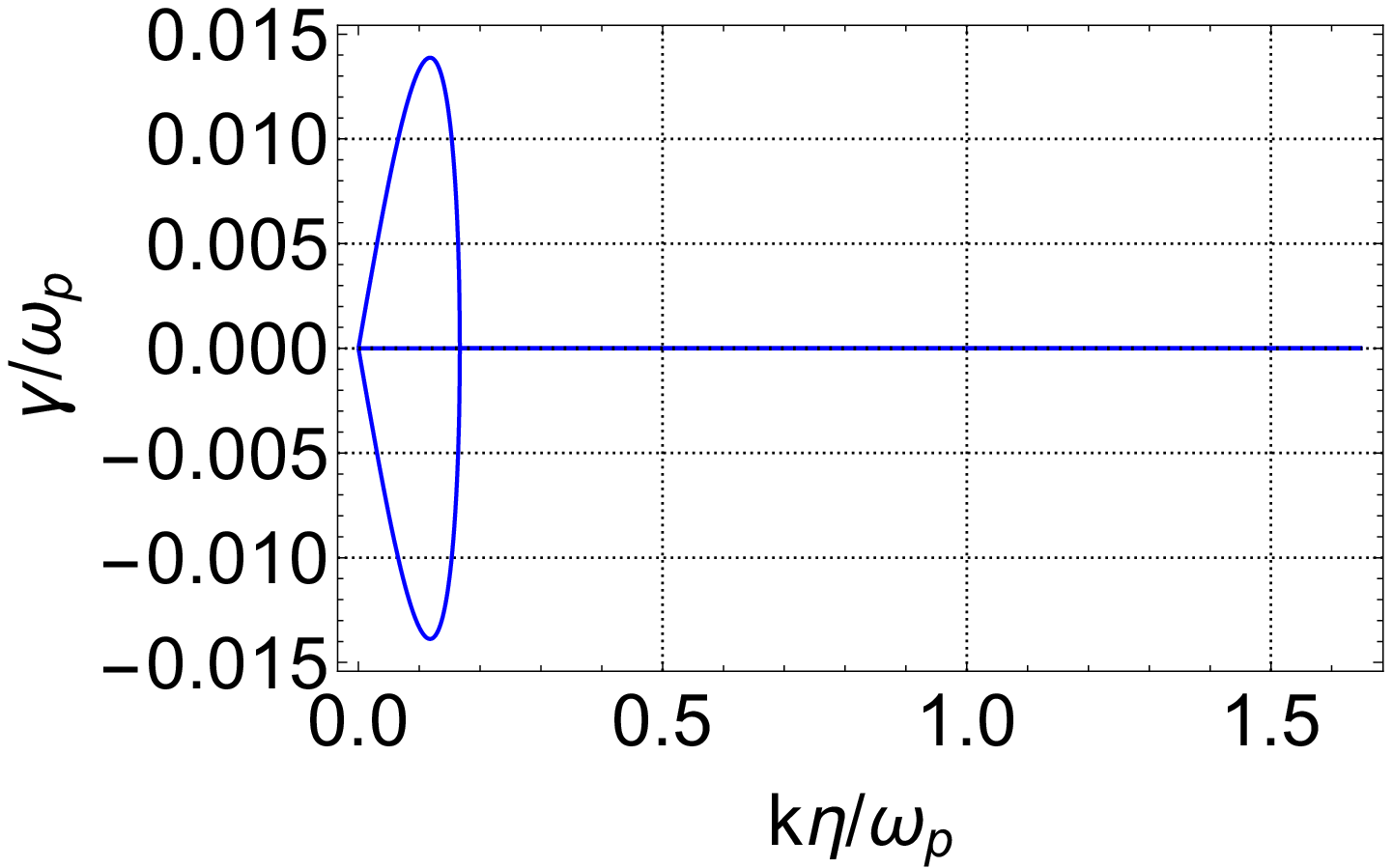}}

\vspace{1mm}
\subfloat[]{\includegraphics[width=0.40\columnwidth]{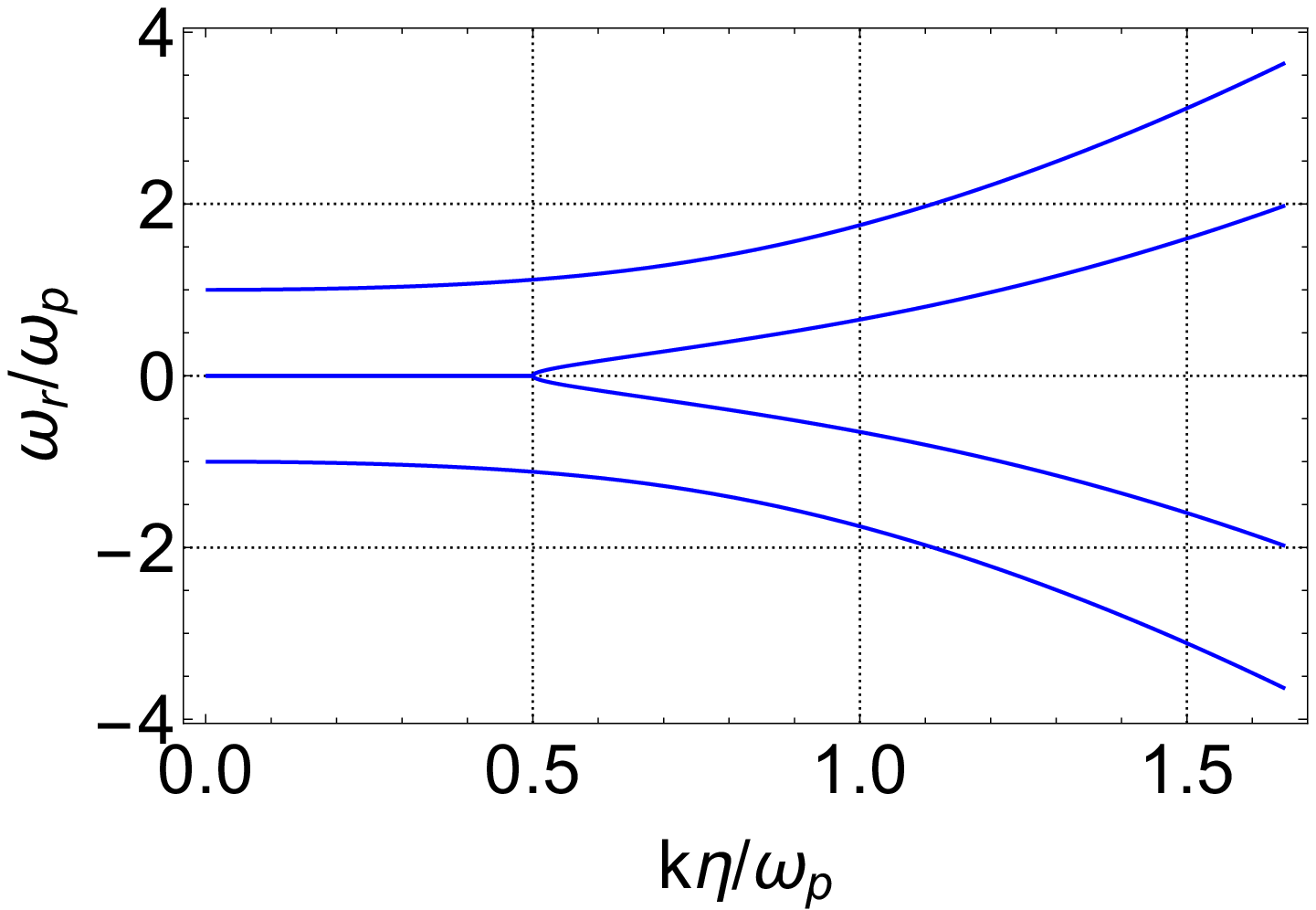}}\hspace{0.02\columnwidth}\subfloat[]{\includegraphics[width=0.40\columnwidth]{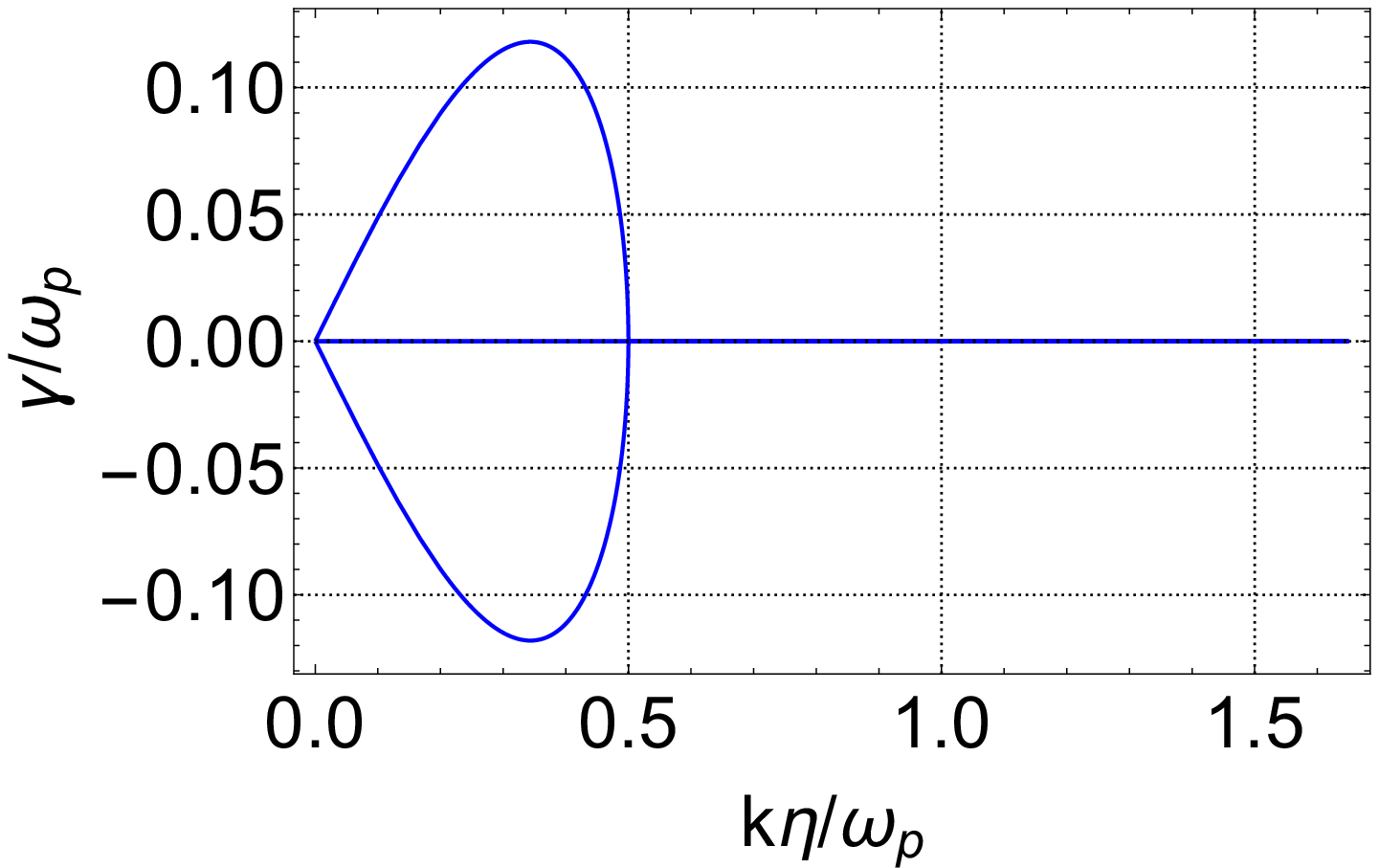}}

\vspace{1mm}
\subfloat[]{\includegraphics[width=0.40\columnwidth]{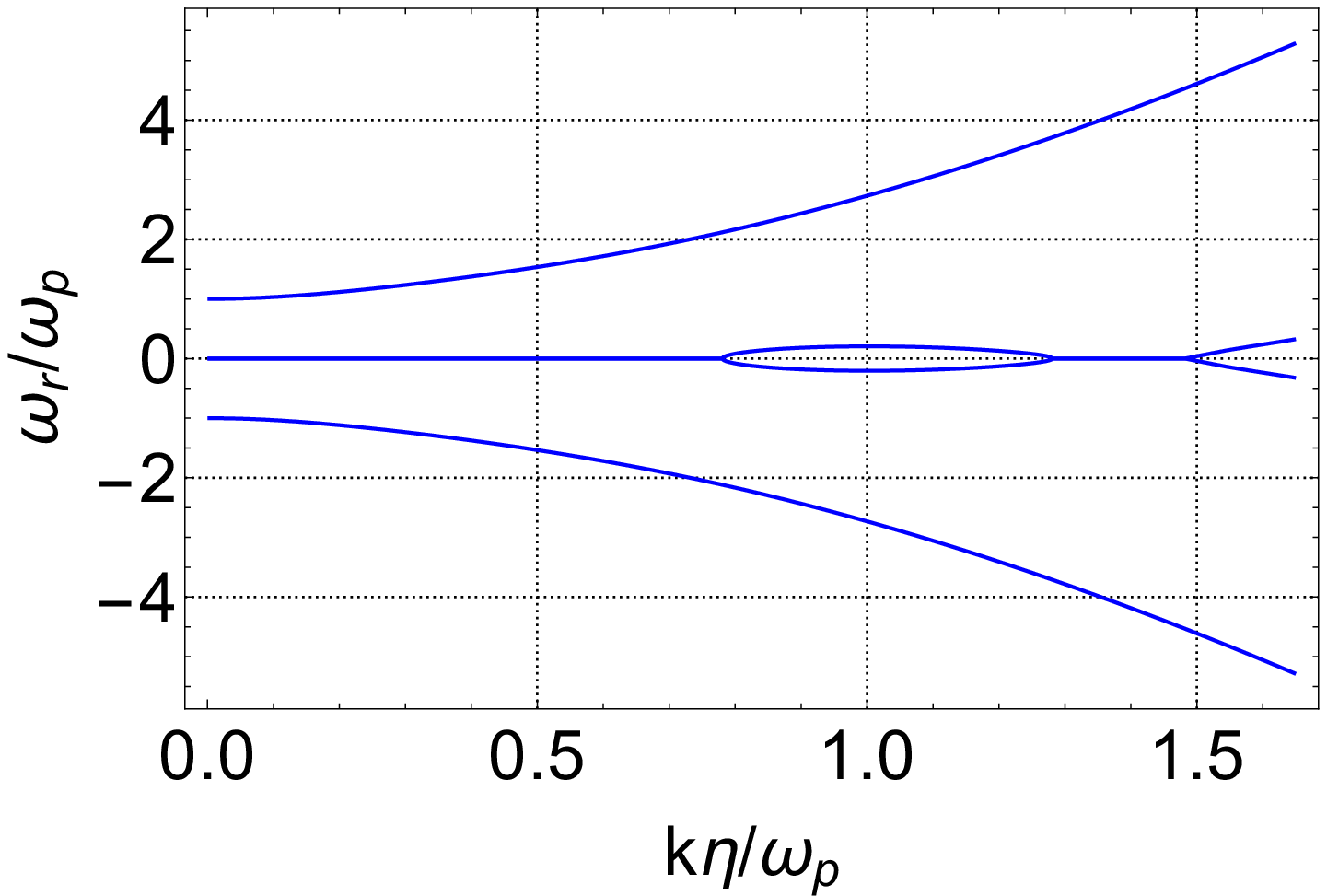}}\hspace{0.02\columnwidth}\subfloat[]{\includegraphics[width=0.40\columnwidth]{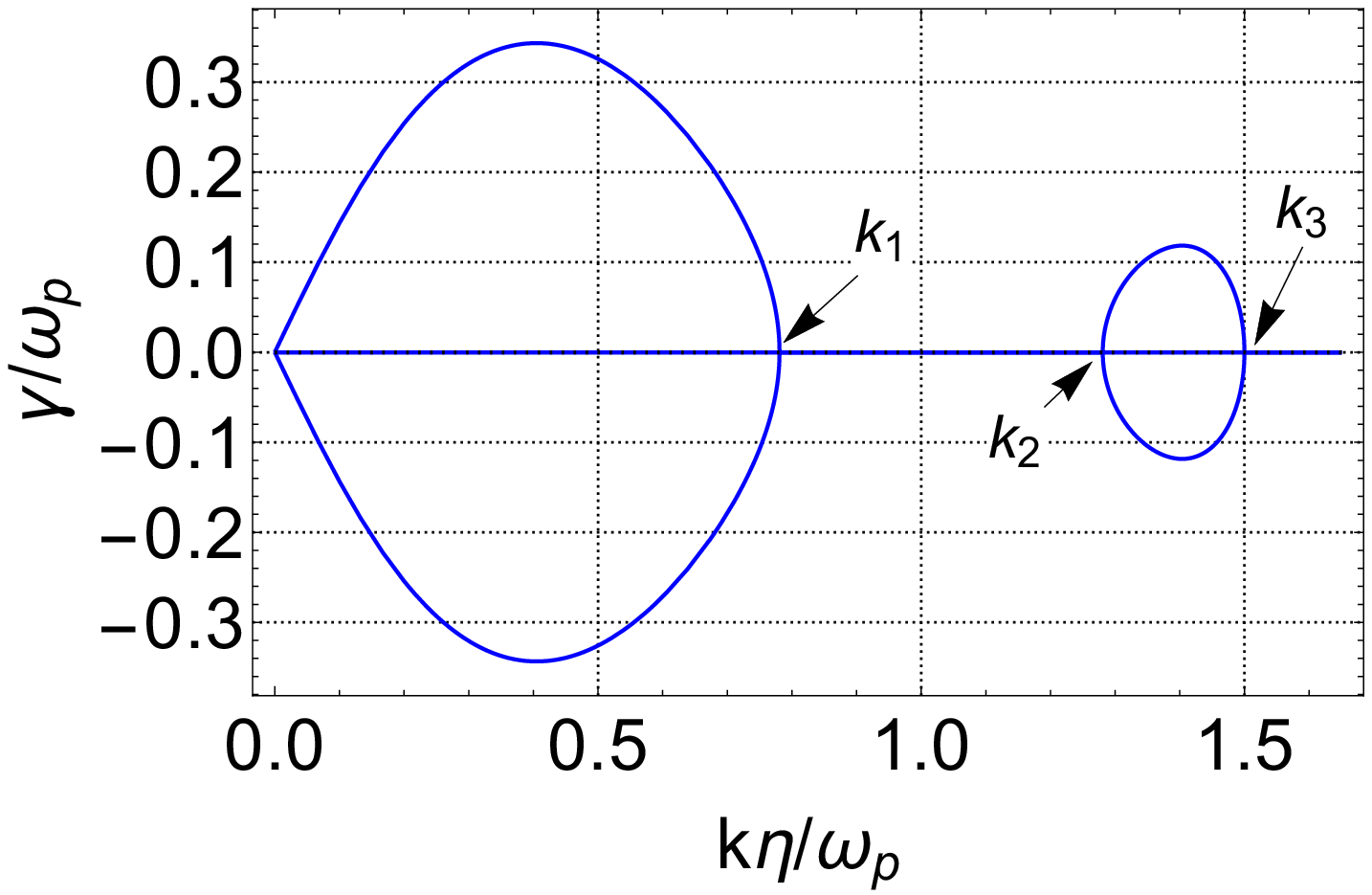}}\end{center}

\caption{Dispersion relations for symmetrical counter-drifting delta-function
distributions with for three different values of the drift velocity~$U$
relative to the characteristic quantum speed $\eta\equiv\sqrt{\hbar\omega_{p}/m_{e}}$:
$U=\eta/3$ (panels a and b), $U=\eta$ (panels c and d), and $U=3\eta$
(panels e and f). The second region of instability appears for $U=3\eta$
(panel f). This region corresponds with the crossing of modes in the
plot of the real part of the frequency (panel e). \label{fig:delta_instabilities}}
\end{figure}

For Dirac delta-function distributions, in the centre-of-mass frame,
utilising equation \ref{eq:chidel} we have

\begin{equation}
\boldsymbol{\epsilon=}1-\frac{1}{2}\left\{ \frac{\omega_{p}^{2}}{\left(\omega-kU/2\right)^{2}-k^{4}\hbar^{2}/m^{2}}+\frac{\omega_{p}^{2}}{\left(\omega+kU/2\right)^{2}-k^{4}\hbar^{2}/m^{2}}\right\} .
\end{equation}
In this case the dispersion relation $\epsilon=0$ can be solved exactly
and is 
\begin{equation}
\frac{\omega_{\pm\pm}}{\omega_{p}}=\frac{1}{2}\left\{ \left(\frac{kU}{\omega_{p}}\right)\pm\sqrt{2+4\left(\frac{\eta k}{\omega_{p}}\right)^{4}+\left(\frac{kU}{\omega_{p}}\right)^{2}\pm2\sqrt{1+2\left(\frac{kU}{\omega_{p}}\right)^{2}\left[2\left(\frac{\eta k}{\omega_{p}}\right)^{4}+1\right]}}\right\} .\label{eq:delta_unstable}
\end{equation}
This solution is plotted in figure \ref{fig:delta_instabilities}
using the second normalisation scheme from section \ref{sec:parameters}.
Expression \ref{eq:delta_unstable} contains four branches: two pairs
of two. Modes~$\omega_{++}$ and~$\omega_{-+}$ are purely real
for all values of~$U$ and become the regular plasmon modes when~$U\rightarrow0$.
Modes~$\omega_{+-}$ and~$\omega_{--}$ are purely imaginary for
small enough values of~$k$ for any~$U$ and purely real for large~$k$,
with mode~$\omega_{+-}$ being positive and unstable and~$\omega_{--}$
being negative and stable.

\begin{figure}
\begin{center}%
\begin{tabular}{c}
\includegraphics[width=0.5\columnwidth]{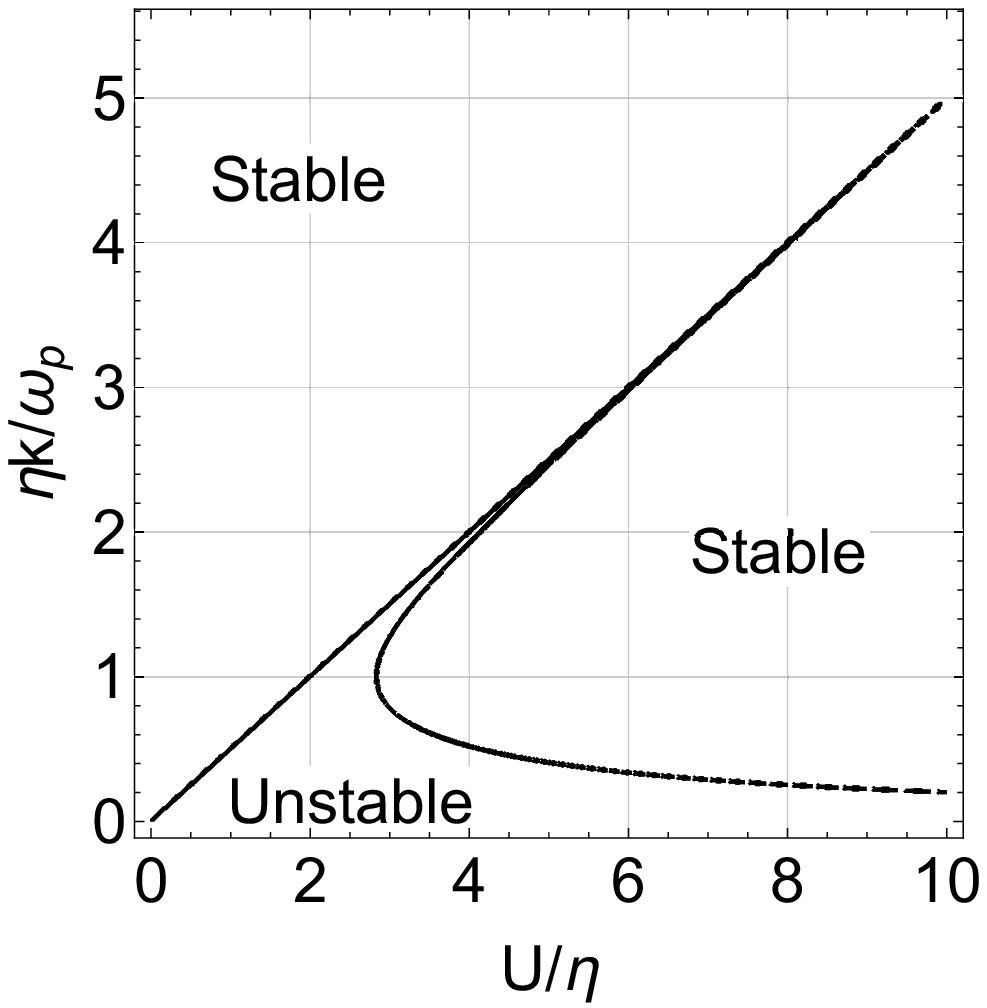}\tabularnewline
\end{tabular}\end{center}

\caption{Instability region for counter-propagating identical cold delta-function
beams. The unstable region fills the region under the plotted curve.
\label{fig:instregion_delta}}
\end{figure}

Focusing on the potentially unstable mode~$\omega_{+-}$, the values
of~$k$ for which the radical becomes zero, 
\begin{equation}
2+4\left(\frac{\eta k}{\omega_{p}}\right)^{4}+\left(\frac{kU}{\omega_{p}}\right)^{2}=2\sqrt{1+2\left(\frac{kU}{\omega_{p}}\right)^{2}\left[2\left(\frac{\eta k}{\omega_{p}}\right)^{4}+1\right]},
\end{equation}
are 
\begin{eqnarray}
k_{1}=\frac{2\omega_{p}}{U}\\
k_{2}=\frac{\omega_{p}}{2\sqrt{2}\eta}\sqrt{\left(U/\eta\right)^{2}-\sqrt{\left(U/\eta\right)^{4}-64}}\label{eq:delta}\\
k_{3}=\frac{\omega_{p}}{2\sqrt{2}\eta}\sqrt{\left(U/\eta\right)^{2}+\sqrt{\left(U/\eta\right)^{4}-64}}
\end{eqnarray}
The dependence of~$k_{1}$ on~$U$ shows that an instability exists
for all non-zero values of ~$U$, with the range of unstable wavelengths
between $k=0$ and $k=k_{1}$ decreasing as~$U$ increases. The additional
region of instability exists for $U/\eta>\sqrt{8}$. Referring to
figure \ref{fig:defintions_contour}, this is point b, at which $U_{b}=\sqrt{8}\eta$
and $k_{b}\eta/\omega_{p}=(1/2\sqrt{2})U_{b}/\eta=1$. The presence
of the velocity~$\eta$ means that this region is explicitly dependent
on quantum phenomena. In the classical limit~$\eta$ vanishes, and
thus~$k_{b}$ approaches infinity (the second region of instability
exists only for shorter and shorter wavelengths and ultimately vanishes).
For counter-streaming delta functions, point a moves out to infinity
along a quantum ray of instability described by $k=U\omega_{p}/2\eta^{2}$.
In the limit $U\gg\eta$, the region of instability is bounded by
$k\eta/\omega_{p}<2\eta/U$ and $U/2\eta-4\eta^{3}/U^{3}<k\eta/\omega_{p}<U/2\eta$.
Furthermore, for $U\gg\eta$ the wavenumbers of maximum growth rate
are $k_{{\rm max}}\eta/\omega_{p}=\sqrt{2}\eta/U$ with maximum growth
rate $\omega_{p}\left(\sqrt{5}/2-1\right)^{1/2}$, which is independent
of~$\hbar$, and $k_{{\rm max}}\eta/\omega_{p}=U/2\eta-2\eta^{3}/U^{3}$
with maximum growth rate $\omega_{p}\eta^{2}/U^{2}$, which is zero
for $\hbar\rightarrow0$, as this second region of instability is
a purely quantum effect. The region of instability is plotted in figure
\ref{fig:instregion_delta}, in which one can see the $\propto1/U$
and $\propto U$ dependence of~$k_{1}$ and~$k_{2,}\thinspace k_{3}$,
respectively.

\subsubsection{Case 2: Primary delta-function population with delta-function beam
of arbitrary density}

\begin{figure}
\begin{center}\subfloat[48]{\includegraphics[width=0.40\columnwidth]{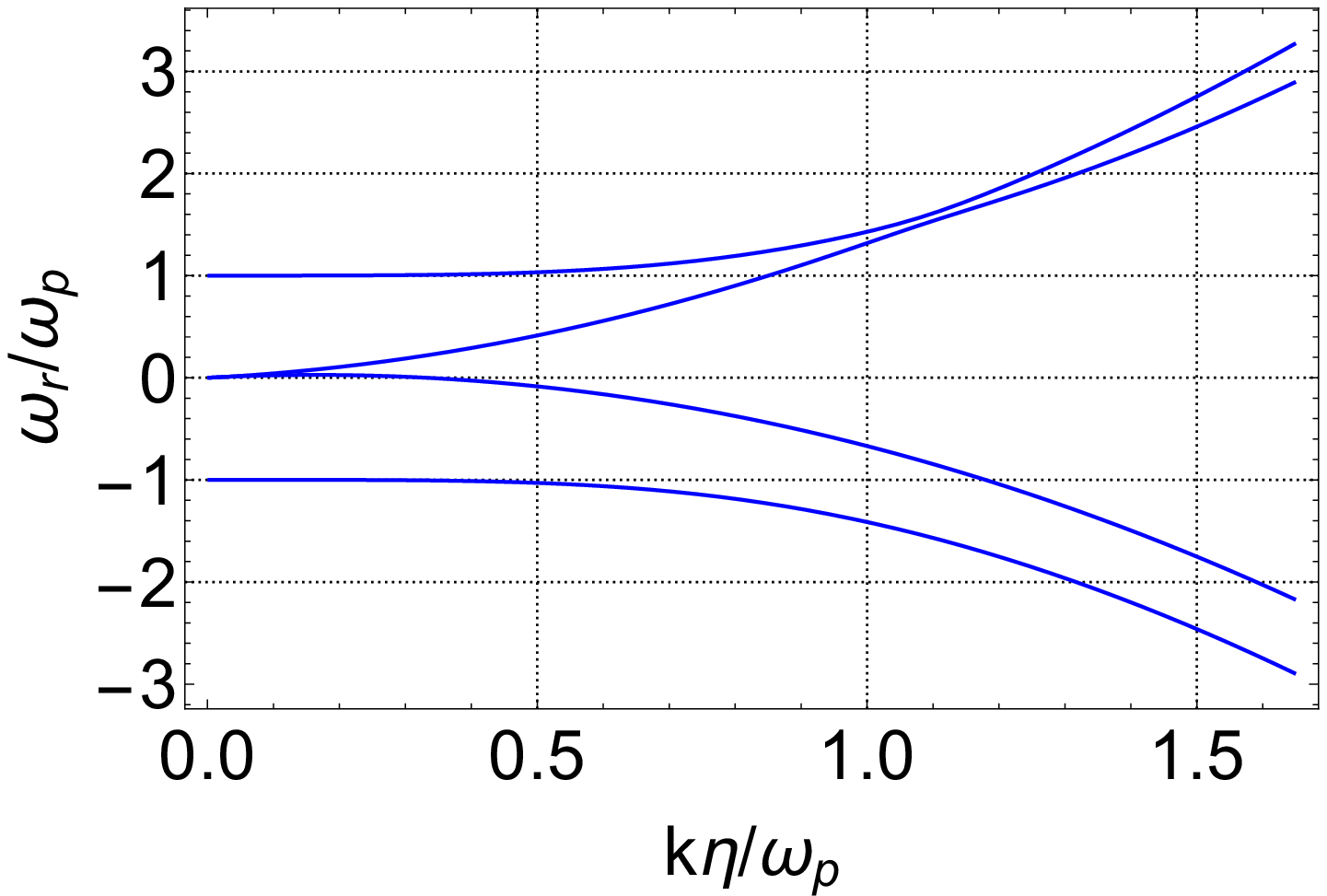}}\hspace{0.02\columnwidth}\subfloat[]{\includegraphics[width=0.40\columnwidth]{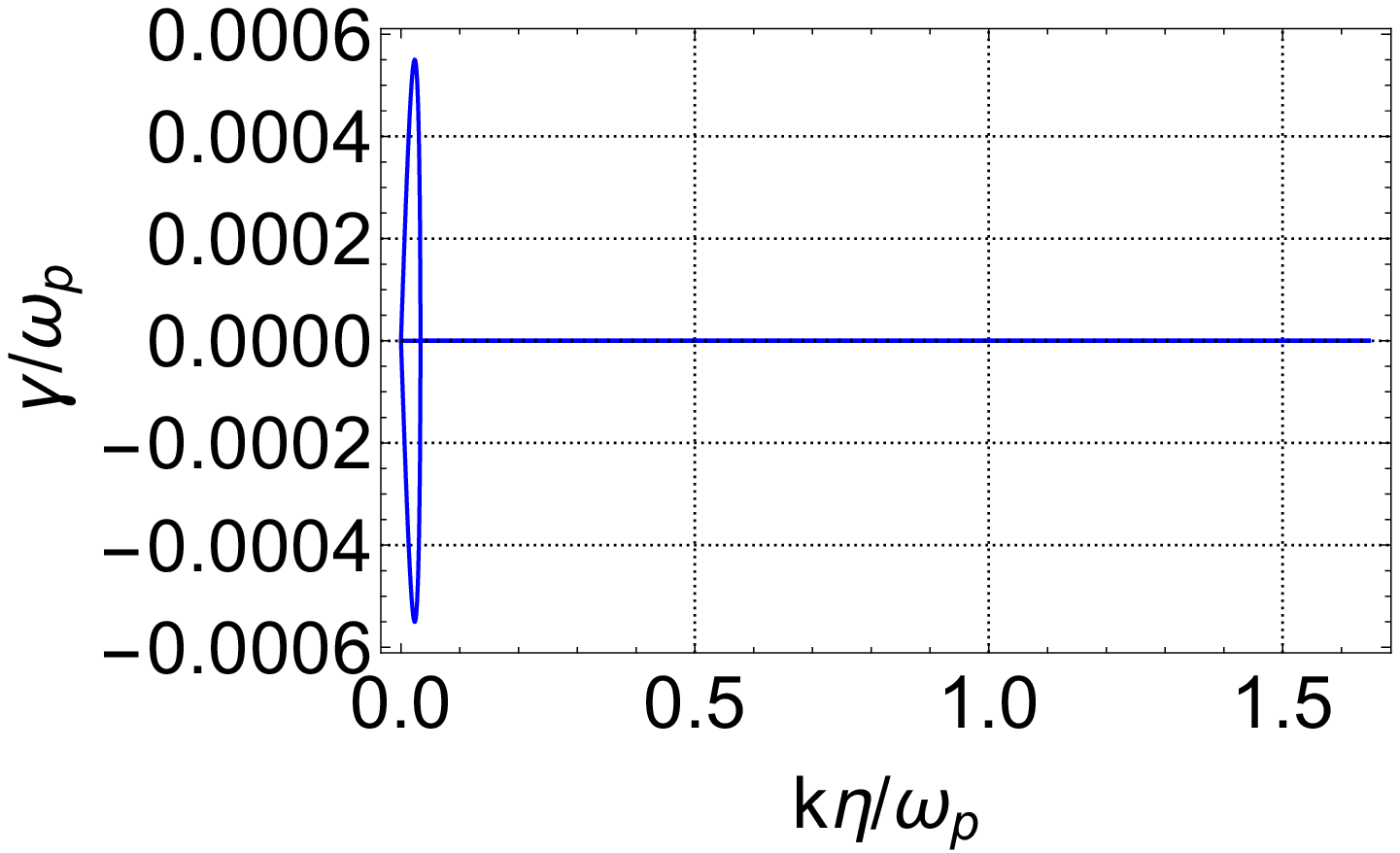}}

\vspace{1mm}
\subfloat[]{\includegraphics[width=0.40\columnwidth]{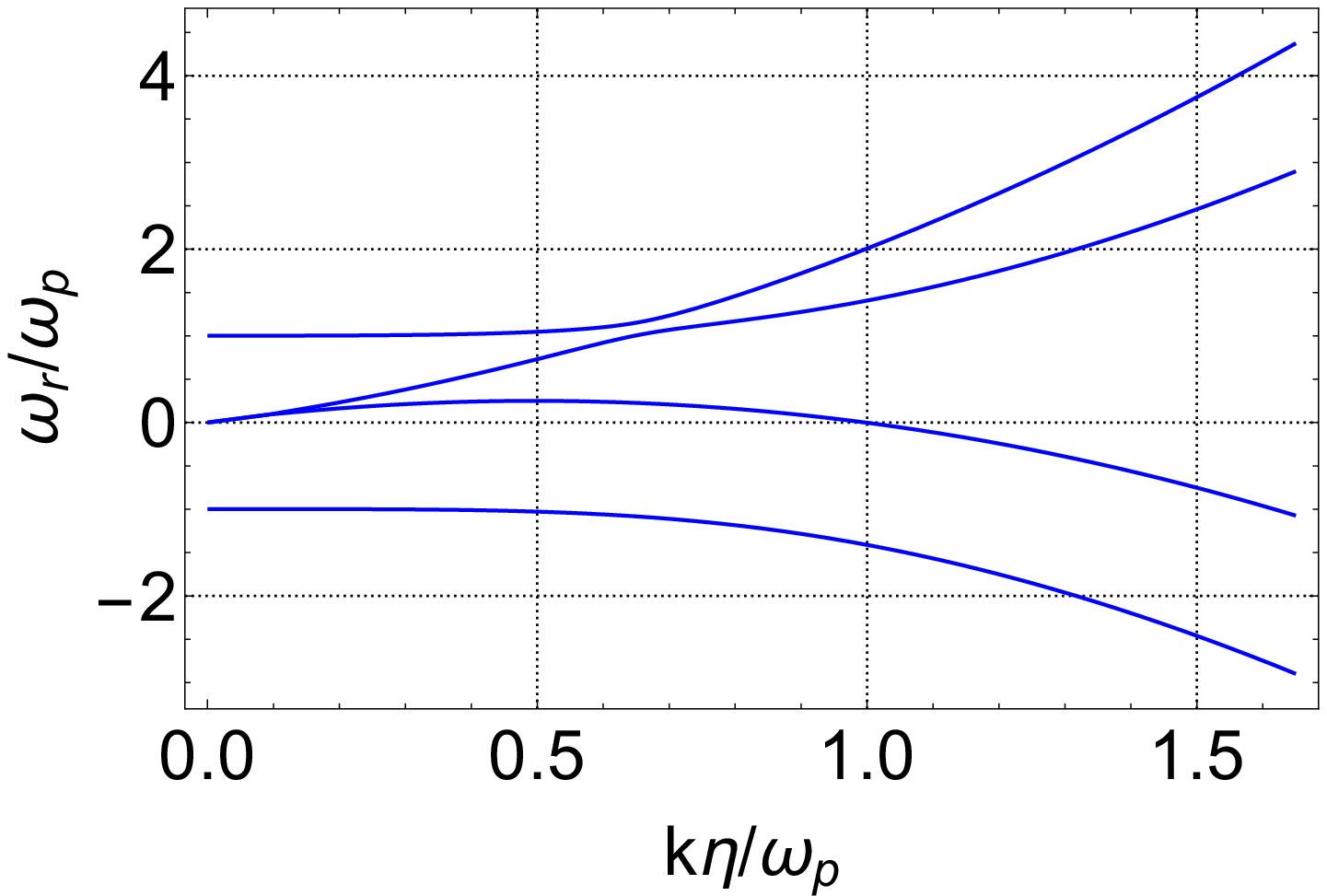}}\hspace{0.02\columnwidth}\subfloat[]{\includegraphics[width=0.40\columnwidth]{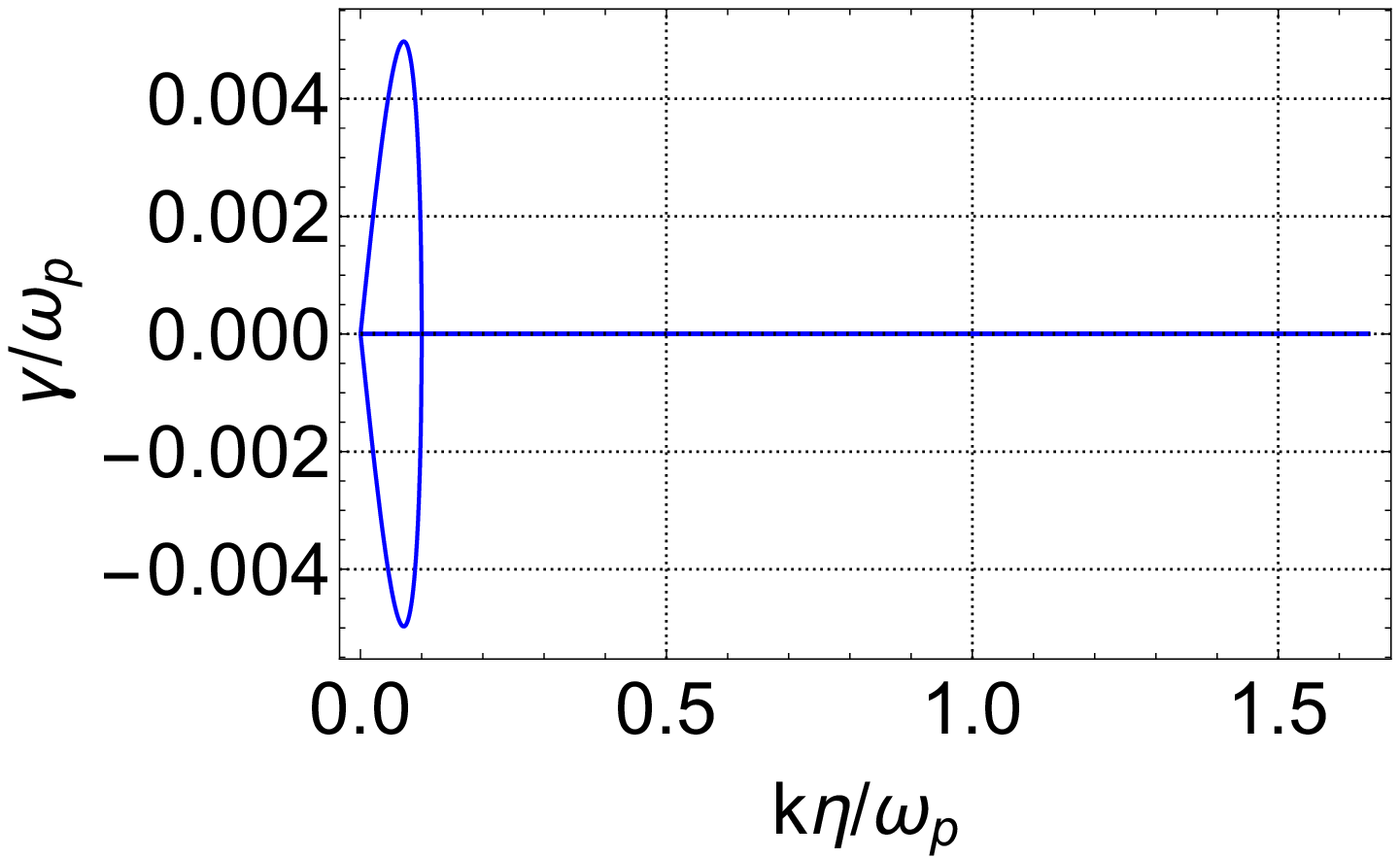}}

\vspace{1mm}
\subfloat[]{\includegraphics[width=0.40\columnwidth]{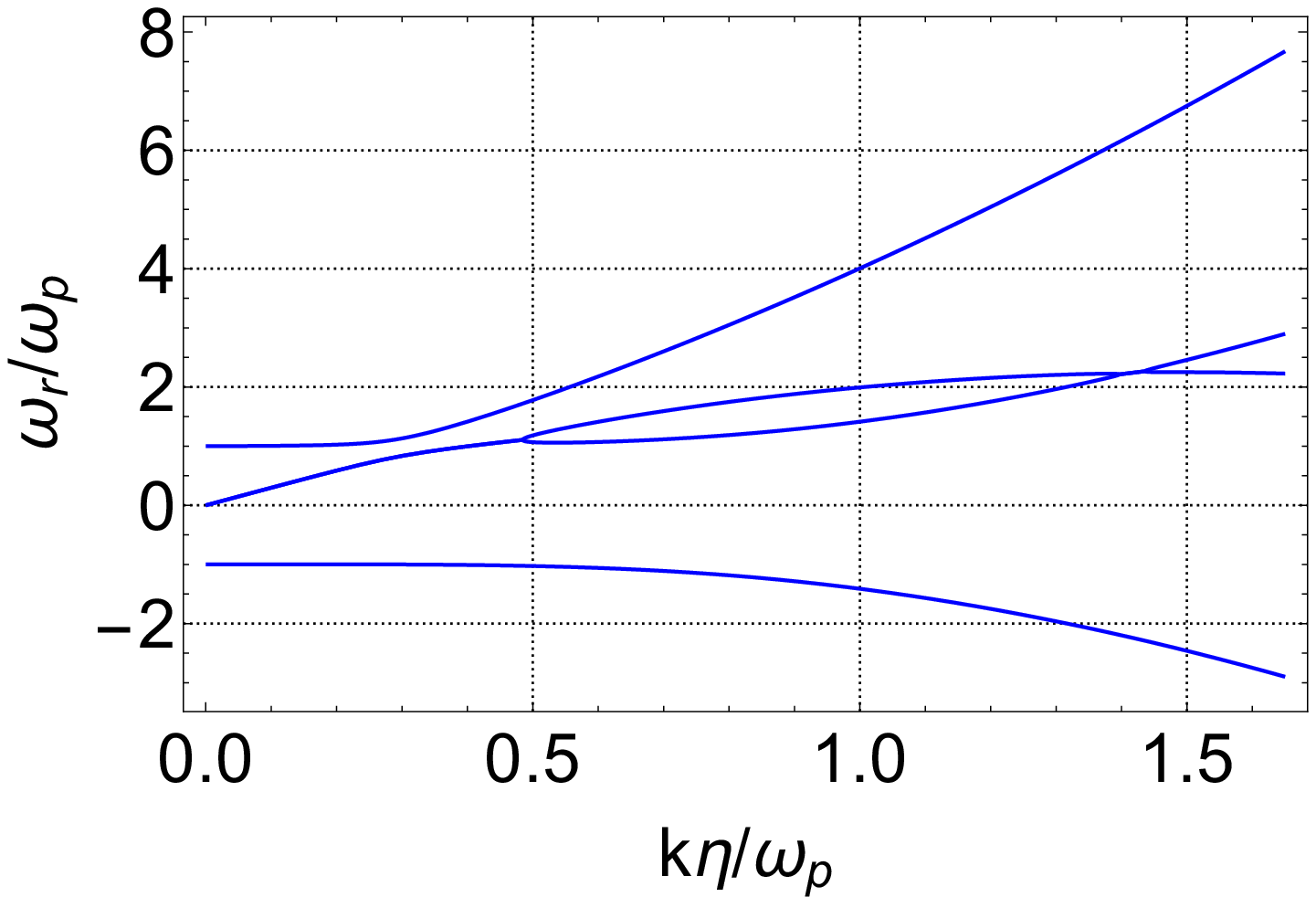}}\hspace{0.02\columnwidth}\subfloat[]{\includegraphics[width=0.40\columnwidth]{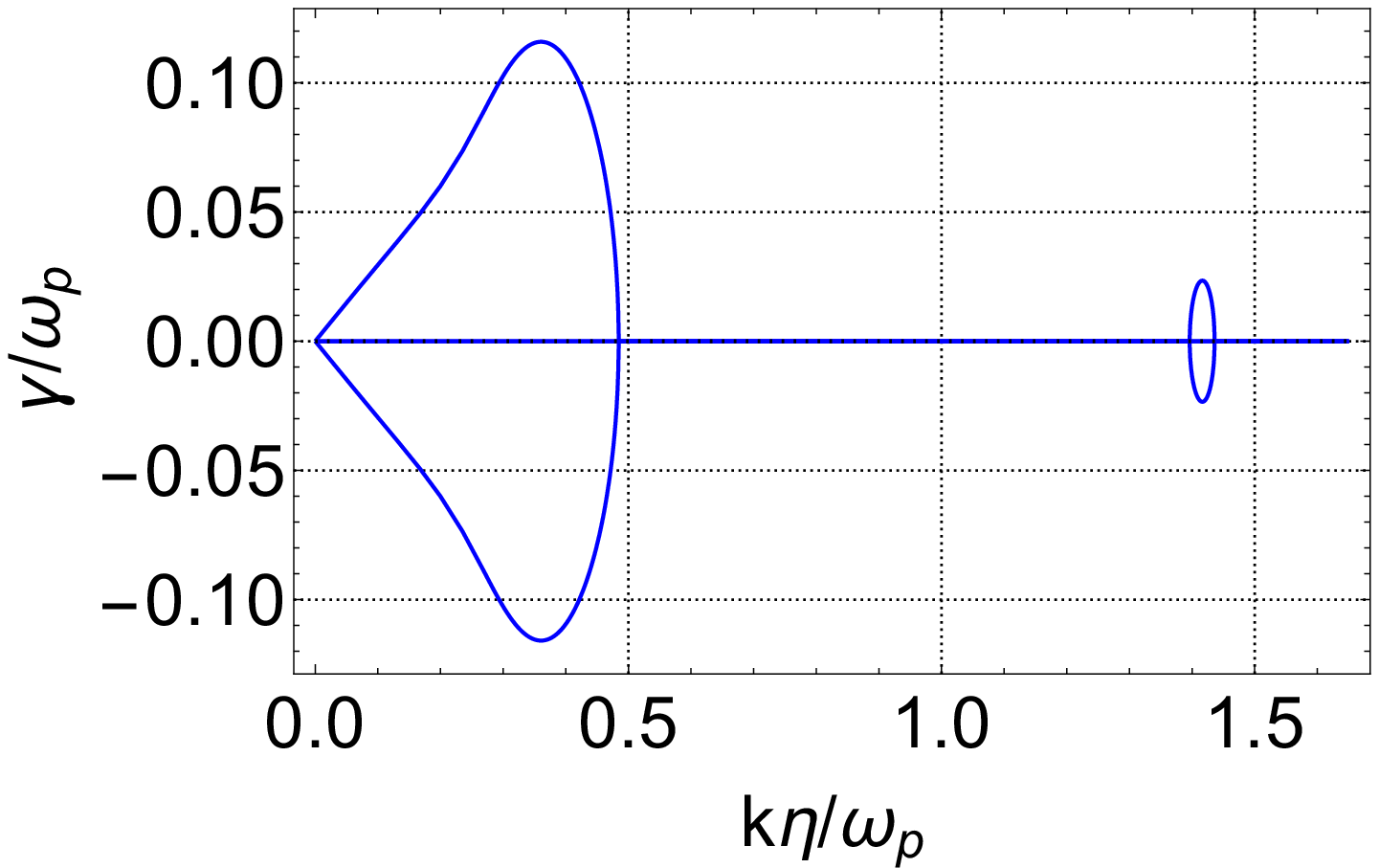}}\end{center}

\caption{Dispersion relations for asymmetrical counter-drifting delta-function
distributions with $n=1/100$ with: $U=\eta/3$ (panels a and b),
$U=\eta$ (panels c and d), and $U=3\eta$ (panels (e) and (f)). Again,
the second region of instability exists near $k\approx1.4$ for $U=3\eta$
(panel f). The effect of the decreased density of the second beam
leads to a decrease in the values of~$k_{1}$, and the difference
between~$k_{2}$ and~$k_{3}$ decreases.\label{fig:delta_instabilities-1}}
\end{figure}

In Case 2 the two delta-functions do not have equal density, the symmetry
is broken and the general solution for the dispersion relation is
quite complicated and there is little to be gained by looking at the
full solution. Instead, here we focus on the small~$k$ (long wavelength)
limit of the dispersion relation, and on numerically-obtained roots.
The exact dispersion relation for this case is plotted in figure \ref{fig:delta_instabilities-1},
where it is seen that both the first and second regions of instability
are diminished in domain and range, and there is dispersion due to
the Doppler shift into the frame of the primary beam. However, the
behaviour is qualitatively similar to Case~1. In the limit $\eta k/\omega_{p}\ll1$
the unstable root is

\begin{equation}
\frac{\omega}{\omega_{p}}={\rm i}\frac{k\eta}{\omega_{p}}\frac{U}{\eta}\left(\sqrt{(3-n)n}+{\rm i}n\right)+O\left(k^{2}\right)
\end{equation}
which is always unstable since $n$ is constrained to be less than~$1$.
This is independent of $\hbar$ up to this order. For a weak beam
($n\ll1$) the growth rate is
\begin{equation}
\frac{\gamma}{\omega_{p}}\approx\frac{kU}{\omega_{p}}\sqrt{3n},
\end{equation}
which is linear in $k$ and grows as the square root of~$n$.

\begin{figure} \begin{center}\subfloat[]{\includegraphics[width=0.40\columnwidth]{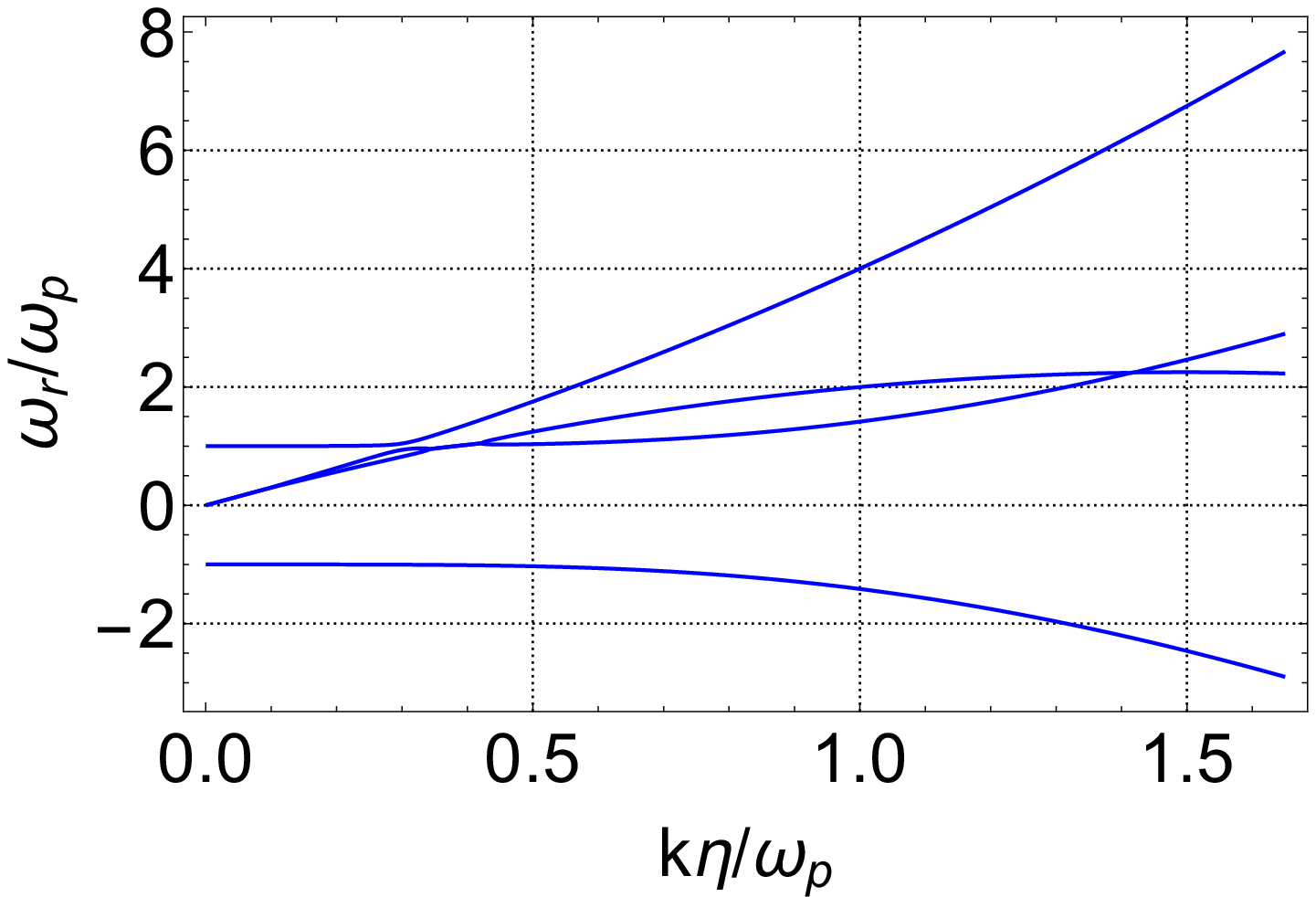}}\hspace{0.02\columnwidth}\subfloat[]{\includegraphics[width=0.40\columnwidth]{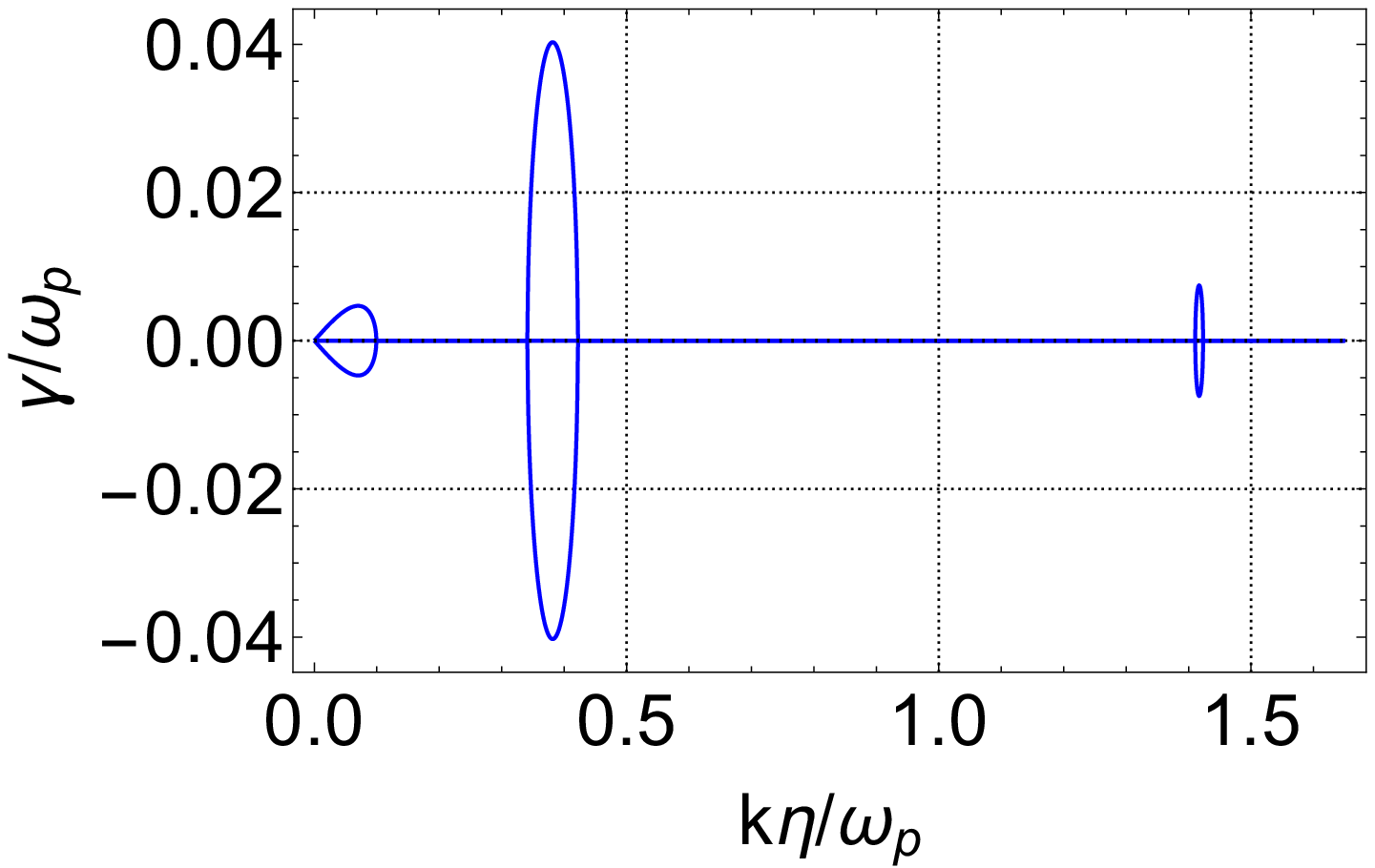}}\end{center}
\caption{Dispersion relation for asymmetrical counter-drifting delta-function distributions with $n=1/1000$ and $U=\eta/3$. The third region of instability is seen in the right panel, and is associated with an additional crossing between modes in the left panel. \label{fig:delta_instabilities-2}} \end{figure}

\begin{figure} \begin{center}
\begin{tabular}{c} \includegraphics[width=0.5\columnwidth]{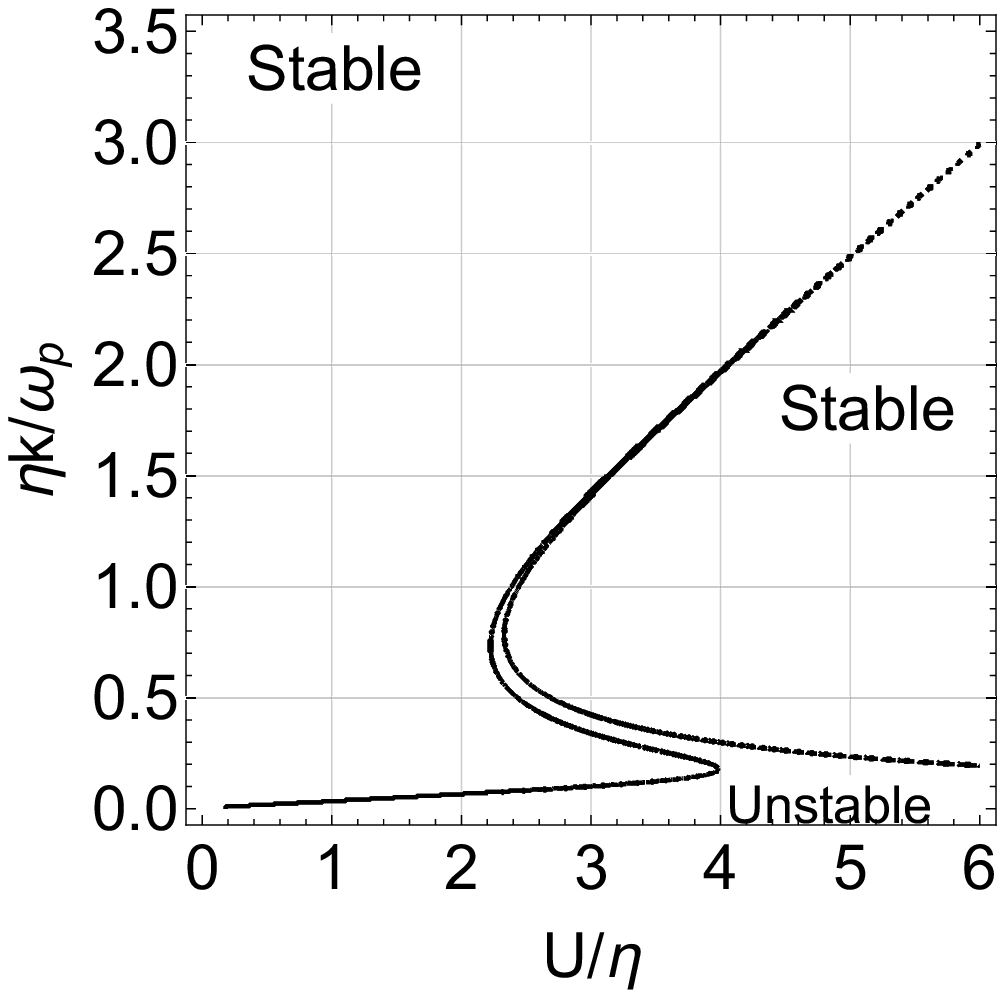}\tabularnewline \end{tabular}
\caption{Instability region for counter-propagating asymmetrical cold delta-function beams,with $n=1/1000$. The unstable region fills the region under the plotted curve, and demonstrates the existence of a third region of unstable values for $k$ when $U$ is between $~2.2$ and $~4$. This is due to the curving backwards of the window of instability for intermediate values of $k$ as $n$ is diminished. \label{fig:instregion_delta-1}} \end{center} \end{figure}

There is an additional new behaviour when $n$ is sufficiently small. For $n$ smaller than approximately $1/500$, a third window of unstable wavenumbers appears for certain values of $U$. This behavior is demonstrated in figure \ref{fig:delta_instabilities-2}, where the right panel shows the behaviour of $\gamma$ as a function of $k$ for fixed $U$. Further information can be gained from figure \ref{fig:instregion_delta-1}, where the region of instability is plotted as a function of $k$ and $U$ for fixed $n=1/1000$. In this figure, it is seen that for smaller $n$ the region of instability curves towards smaller $U$ as $k$ increases, and then curves back and follows the behaviour of the $n=100$ case for sufficiently large $k$. The critical value of $n$ for the onset of this phenomenon could not be determined in this work and remains an open question.

\subsection{Cauchy Distribution: Most Simple Case With Landau Damping}

For the Cauchy distribution the resulting susceptibility is given
by equation \ref{eq:chicau}. In this and the following subsections
we will employ our normalisation scheme 1 from section \ref{sec:parameters},
and correspondingly will measure the separation velocity~$U$ in
units of the width~$\mathcal{V}$ and encapsulate quantum effects
in the parameter $H=\hbar\omega_{p}/m\mathcal{V}^{2}$. The finite
width of the Cauchy distribution allows for Landau damping, and we
will see that, as in the single population case discussed in section
\ref{sec:onecomp}, the only modification to the dispersion relation
is the addition of the Landau damping term. However, this reduces
the region of instability by providing an additional negative component
to the imaginary part of~$\omega$.

\subsubsection{Case 1: Symmetrically counter-propagating distributions}

\begin{figure}
\begin{center}\subfloat[]{\includegraphics[width=0.40\columnwidth]{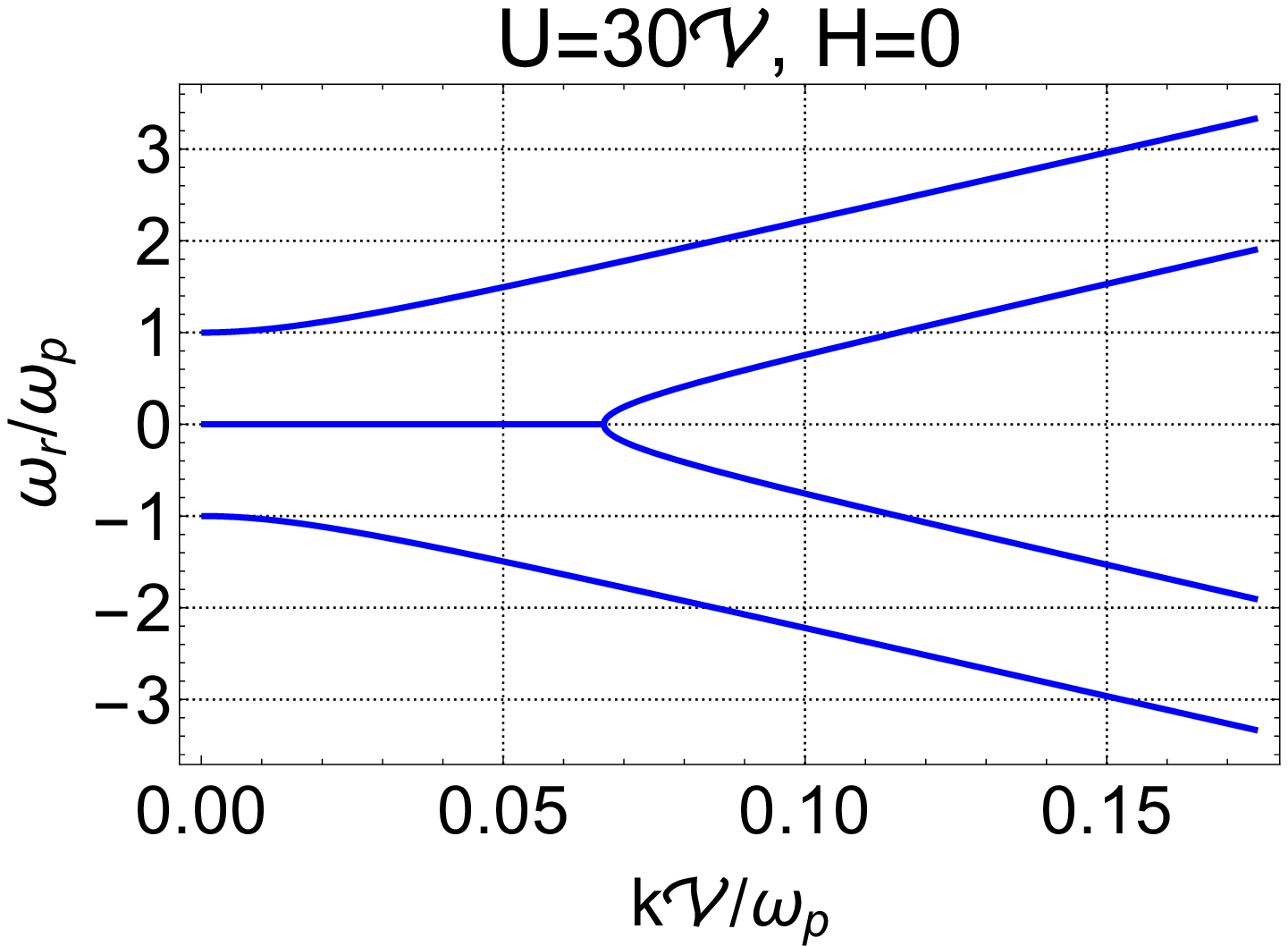}}\hspace{0.02\columnwidth}\subfloat[]{\includegraphics[width=0.40\columnwidth]{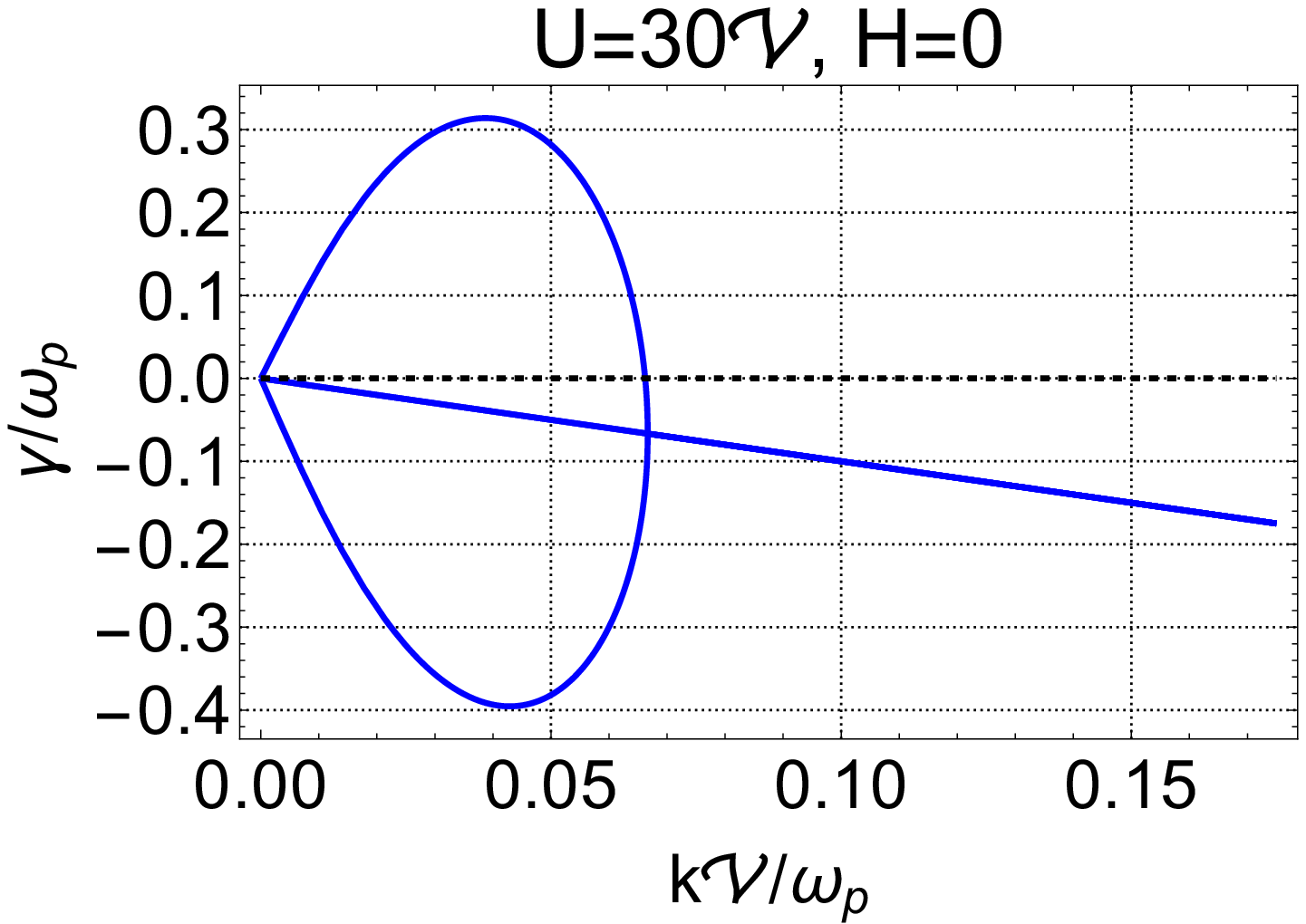}}

\vspace{1mm}
\subfloat[]{\includegraphics[width=0.40\columnwidth]{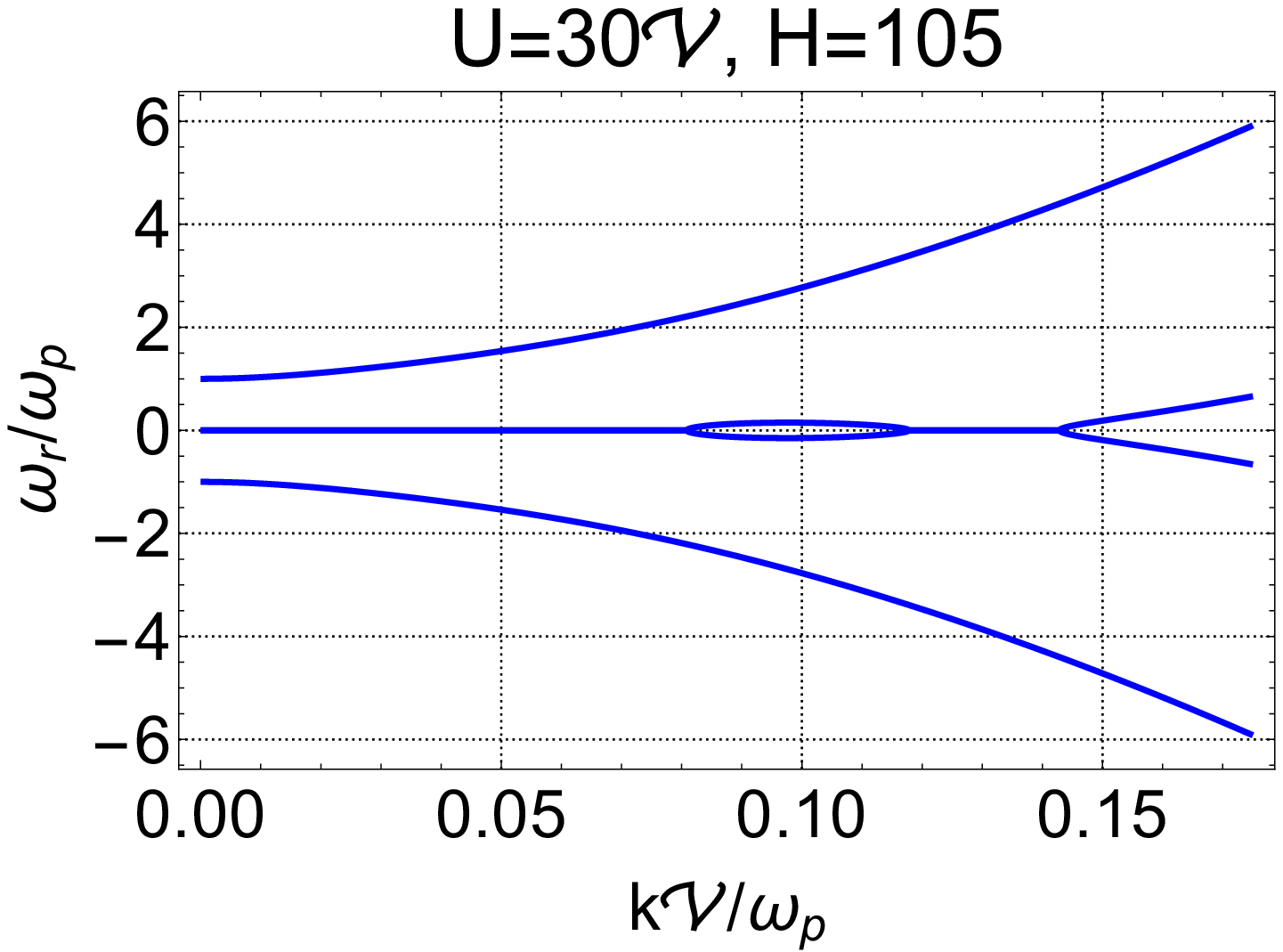}}\hspace{0.02\columnwidth}\subfloat[]{\includegraphics[width=0.40\columnwidth]{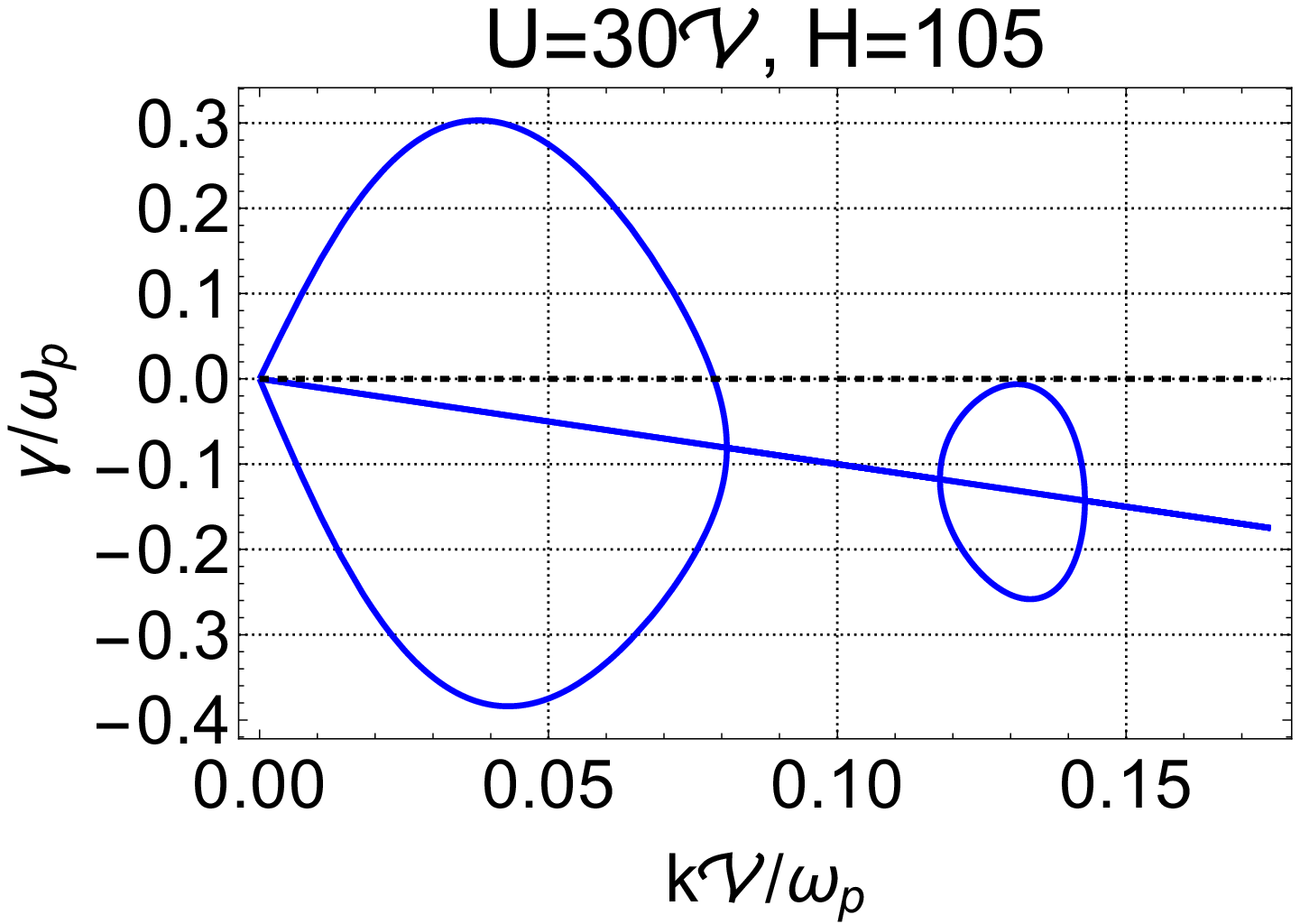}}

\vspace{1mm}
\subfloat[]{\includegraphics[width=0.40\columnwidth]{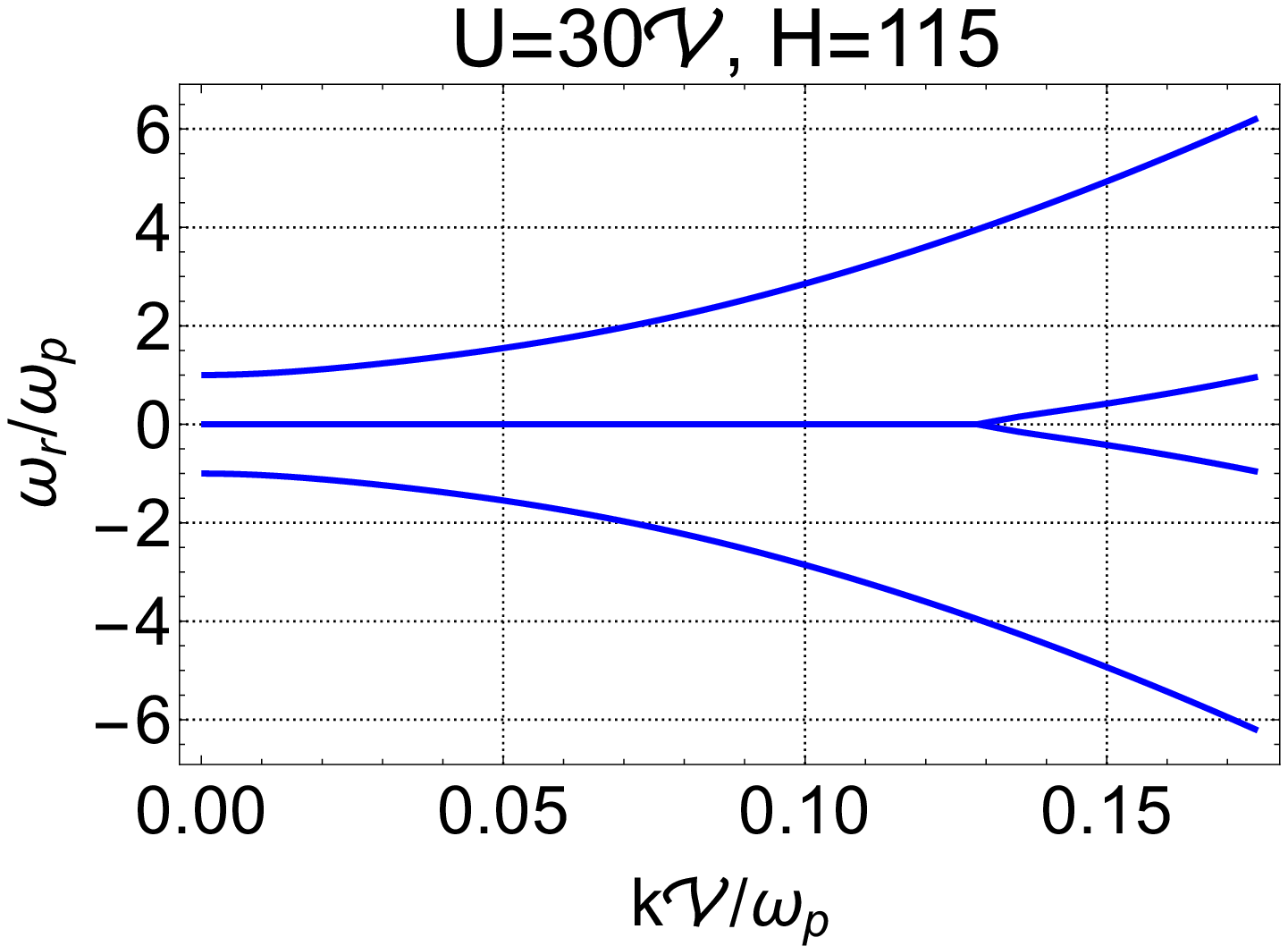}}\hspace{0.02\columnwidth}\subfloat[]{\includegraphics[width=0.40\columnwidth]{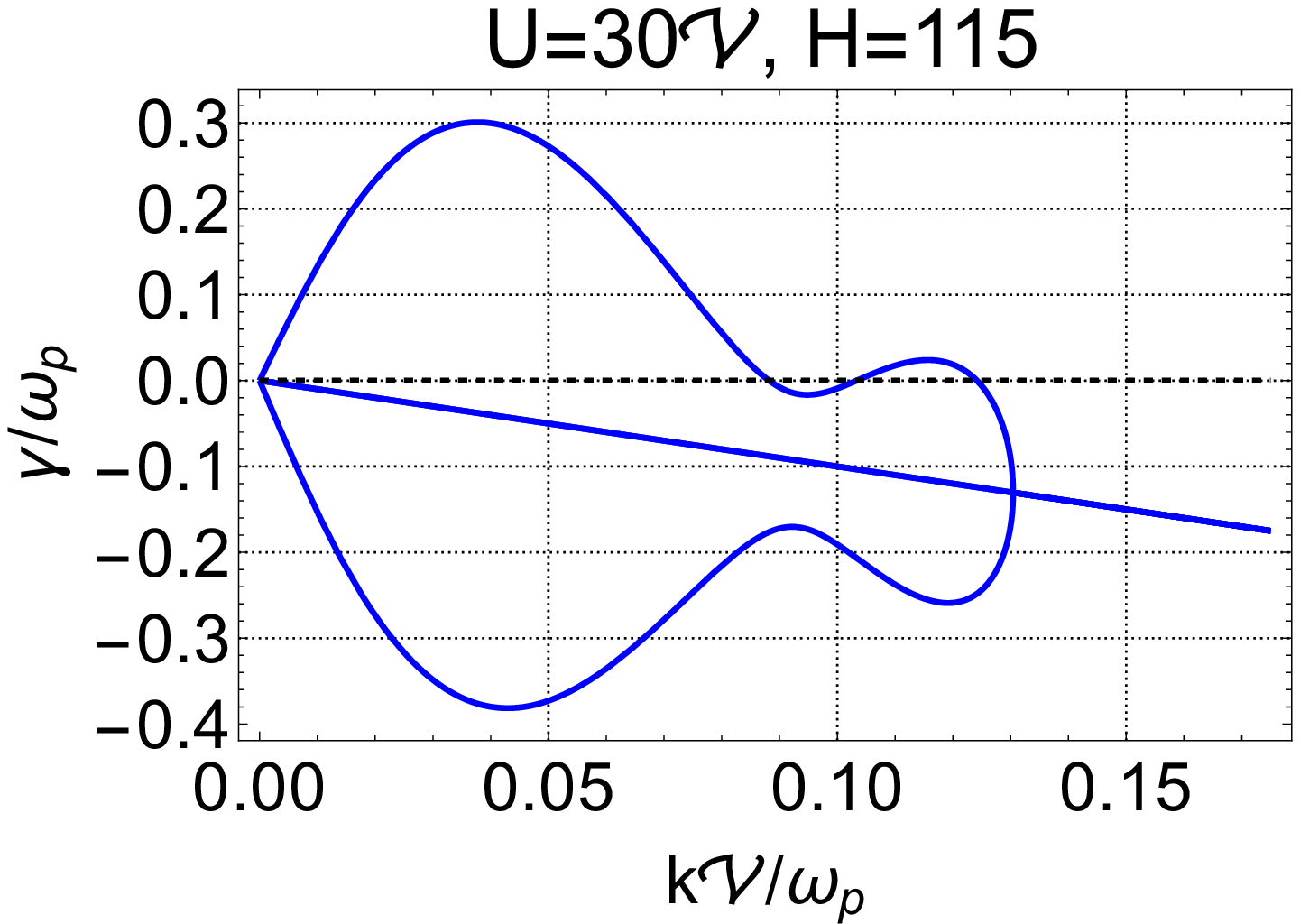}}\end{center}

\caption{Dispersion relations for symmetrical counter-drifting Cauchy distribution
functions (Case 1).  The behaviour is similar to that of the delta-function
distributions as seen in figure \ref{fig:delta_instabilities}. The
presence of Landau damping in the Cauchy distribution case decreases
the maximum growth rates and diminishes the ranges of~$k$ for which
instability exists. The ``bubble'' in panel d moves towards smaller~$k$
as~$H$ increases and is responsible for the second region of instability
defined by~$k_{2}$ and~$k_{3}$ in figure \ref{fig:defintions_contour}.
In panel f this ``bubble'' merges with the primary instability region.
\label{fig:cauchy_dispersion}}
\end{figure}

In Case 1 there is again a simple closed form solution for the dispersion
relation,

\begin{equation}
\Omega=\pm\frac{1}{2}\sqrt{2+4H^{2}K^{4}+U^{2}K^{2}\pm2\sqrt{1+2K^{2}U^{2}\left(2H^{2}K^{4}+1\right)}}-{\rm i}K.\label{eq:disp_cauchy}
\end{equation}
This is plotted in figure \ref{fig:cauchy_dispersion}. equation \ref{eq:disp_cauchy}
is nearly identical to the dispersion relation equation \ref{eq:delta_unstable}
in the previously considered scenario, but with the additional term
$-{\rm i}k\mathcal{V}/\omega_{p}$ reflecting that every mode experiences
Landau damping. The similarity to the delta-function case can be attributed
to the lack of dispersion in a Cauchy plasma due to the lack of a
finite pressure. Additionally, the second region of instability ceases
to exist for large enough~$U$, terminating at point a (see figure
\ref{fig:defintions_contour}). For large~$U$ the instability boundary
is defined by 
\begin{equation}
K_{1}\approx\frac{2}{U}-\frac{12}{U^{3}}+\frac{16H^{2}+44}{U^{5}},
\end{equation}
where quantum effects appear at fourth order in $\mathcal{V}/U$.
The maximum growth rate occurs at 
\begin{equation}
K_{{\rm max}}\approx\frac{\sqrt{2}}{U}+\frac{\sqrt{3}}{U^{2}}-\frac{9}{2\sqrt{2}U^{3}}+\frac{3\sqrt{3}}{U^{4}}+\frac{256H^{2}-225}{16\sqrt{2}U^{5}},
\end{equation}
with maximum growth rate 
\begin{equation}
\frac{\gamma_{{\rm max}}}{\omega_{p}}\approx\frac{1}{2\sqrt{3}}-\frac{\sqrt{2}}{U}-\frac{3\sqrt{3}}{2U^{2}}+\frac{15}{2\sqrt{2}U^{3}}+\frac{\left(16H^{2}-63\right)}{4\sqrt{3}U^{4}}.
\end{equation}
It is apparent that quantum effects appear at fourth order in~$U^{-1}$.

\begin{figure}
\begin{center}%
\begin{tabular}{c}
\includegraphics[width=0.5\columnwidth]{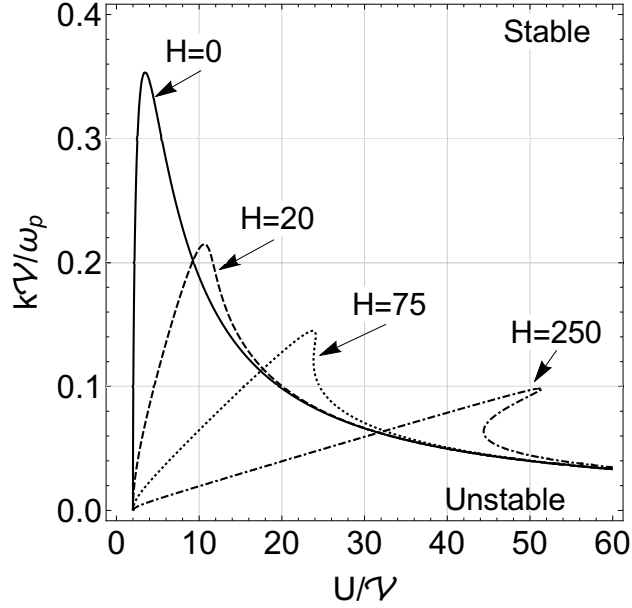}\tabularnewline
\end{tabular}\end{center}

\caption{Region of instability for the dispersion relation for Case 1 (counter-drifting
Cauchy distributions), equation \ref{eq:disp_cauchy} for $H=0$,
$30$, $75$ and $250$ (labelled in figure). The second region of
instability emerges for $H\gtrsim50$, and its existence is thus reliant
on sufficient strong quantum effects. \label{fig:cauchy_inst_region}}
\end{figure}

\begin{figure}
\begin{center}\subfloat[]{\includegraphics[width=0.40\columnwidth]{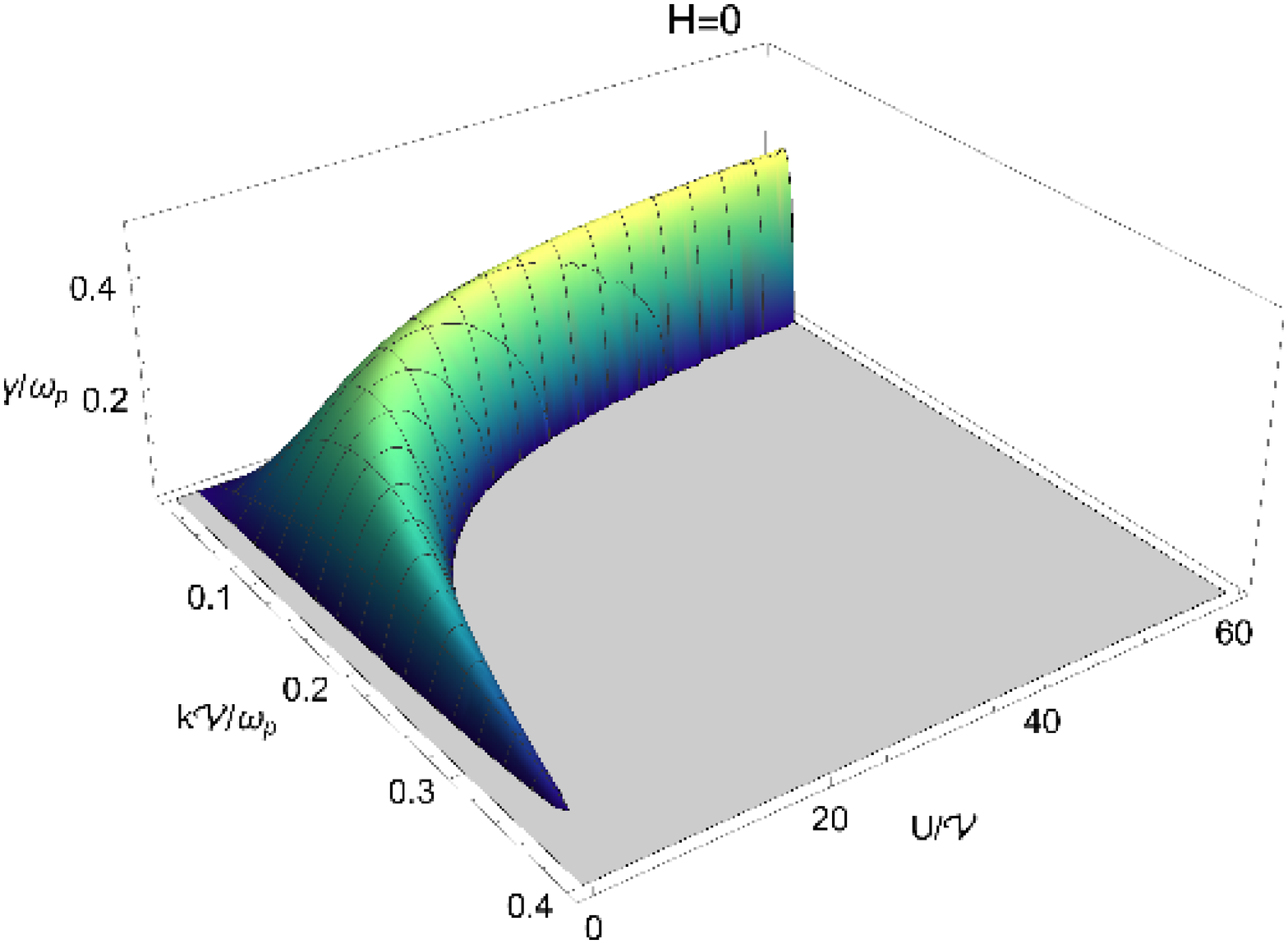}}\hspace{0.02\columnwidth}\subfloat[]{\includegraphics[width=0.40\columnwidth]{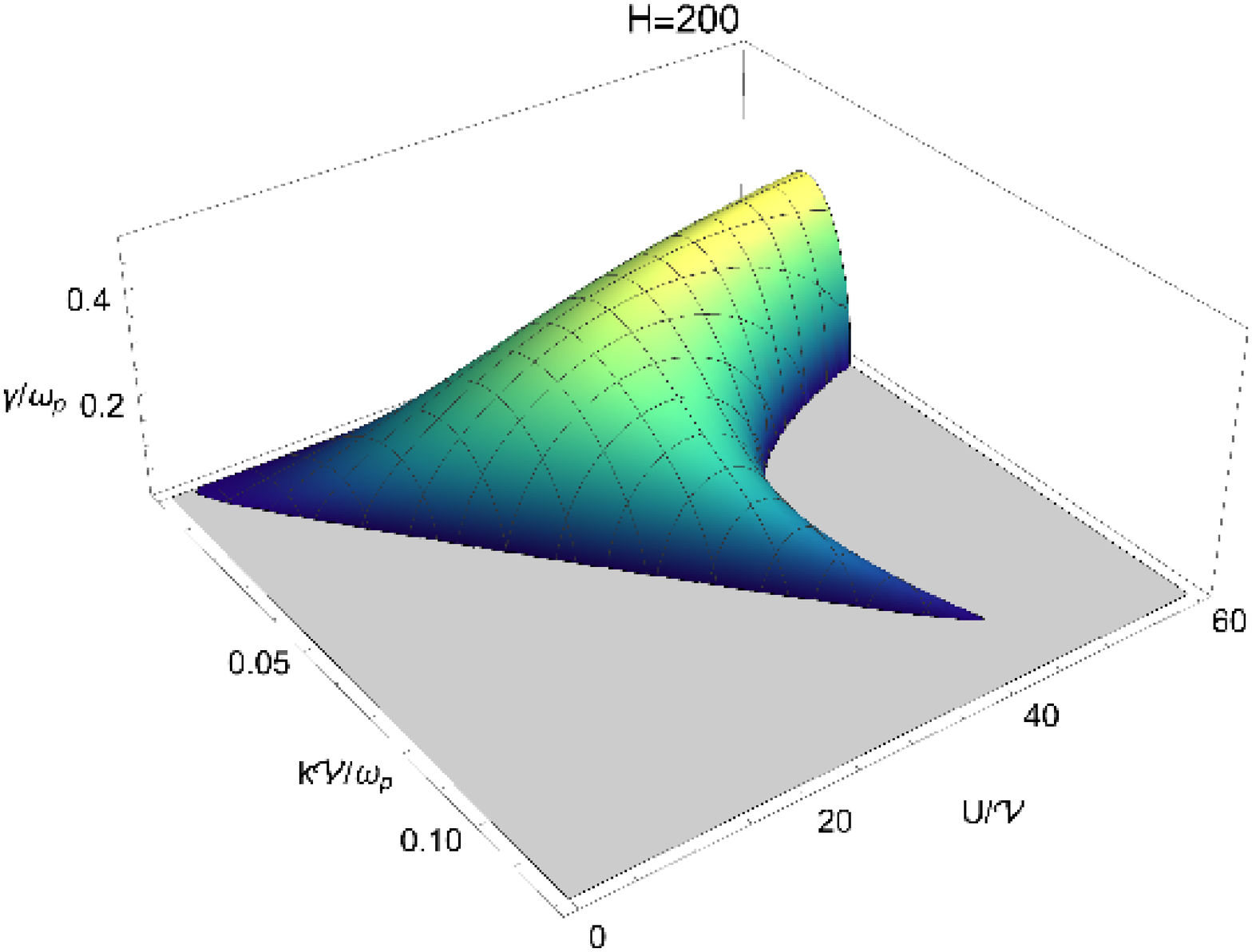}}

\end{center}

\caption{Dispersion relation for the imaginary part of the frequency with counter-streaming
Cauchy distributions in Case 1, equation \ref{eq:disp_cauchy}. The
unstable unstable region is shown, with $H=0$ (panel a) and $H=200$
(panel b). \label{fig:cauchy_inst_region3d}}
\end{figure}

The boundary of the region of instability is shown in figure \ref{fig:cauchy_inst_region}
and the growth rate in the unstable region is shown in figure \ref{fig:cauchy_inst_region3d},
which demonstrate similar behaviour to the case of counter-drifting
delta-functions with the following difference. Importantly, the instability
region does not extend to arbitrarily small~$U$ or arbitrarily large~$K$.
This shows that for finite-width distribution functions instability
only occurs when the populations are separated by sufficiently large
drift velocity. This is also shown by expanding for small~$K.$ The
boundary of the instability region for small~$K$ is given by 
\begin{equation}
K_{`1}\approx\frac{2\sqrt{U^{2}-4}\left(U^{2}+4\right)}{\sqrt{\left(U^{2}+4\right)^{4}-16H^{2}\left(U^{4}-24U^{2}+16\right)}},
\end{equation}
with the classical limit
\begin{equation}
K_{1}=\frac{2\sqrt{U^{2}-4}}{U^{2}+4}.
\end{equation}
from which it can be shown that the instability exists only for $U>2$
in both the classical and quantum cases. In this regime the max growth
rate is found at 
\begin{equation}
K_{{\rm max}}\approx\sqrt{\frac{2}{3}}\sqrt{\frac{(U-2)U}{4H^{2}+U^{4}}}
\end{equation}
with growth rate 
\begin{equation}
\frac{\gamma_{{\rm max}}}{\omega_{p}}\approx\frac{1}{3}\sqrt{\frac{2}{3}}\left(U-2\right)\sqrt{\frac{(U-2)U}{4H^{2}+U^{4}}}.
\end{equation}
Unfortunately, we are unable to obtain analytical expressions for
the behaviour of the second region of instability, as there is no
small parameter in this region, but it is apparent from the full solution
obtained numerically in figure \ref{fig:cauchy_inst_region} that
both~$K_{a}$ and~$K_{b}$ decrease with~$H$, and~$U_{a}$ and~$U_{b}$
increase with~$H$. This means the range from~$K_{1}$ to~$K_{2}$
decreases with~$H$, while the range of velocities $U_{b}-U_{a}$
increases with~$H$. The second region of instability appears for~$H\sim50$.
A determination of the precise critical value of~$H$ could not be
carried out in this work.

\begin{figure}
\begin{center}\subfloat[]{\includegraphics[width=0.40\columnwidth]{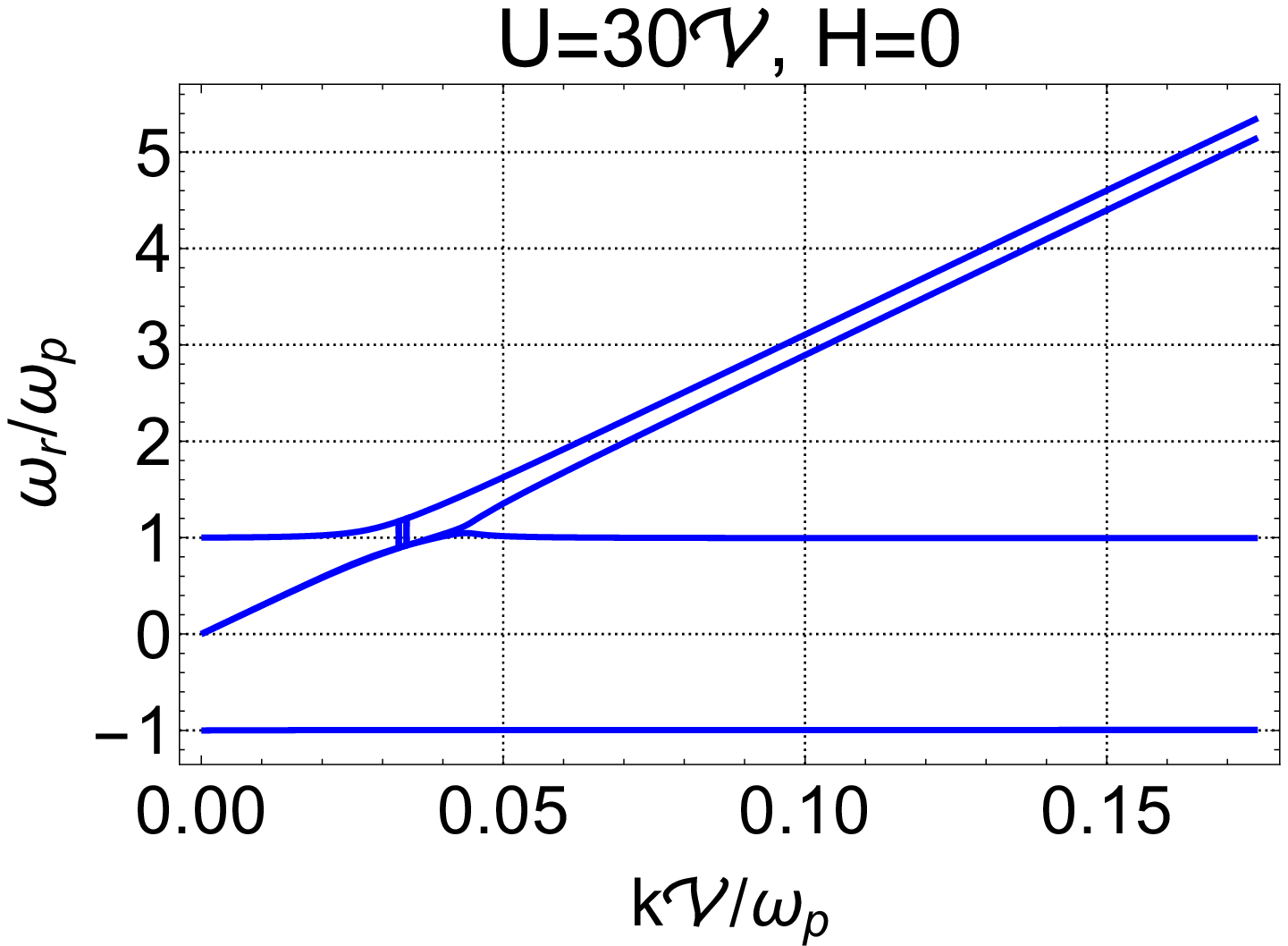}}\hspace{0.02\columnwidth}\subfloat[]{\includegraphics[width=0.40\columnwidth]{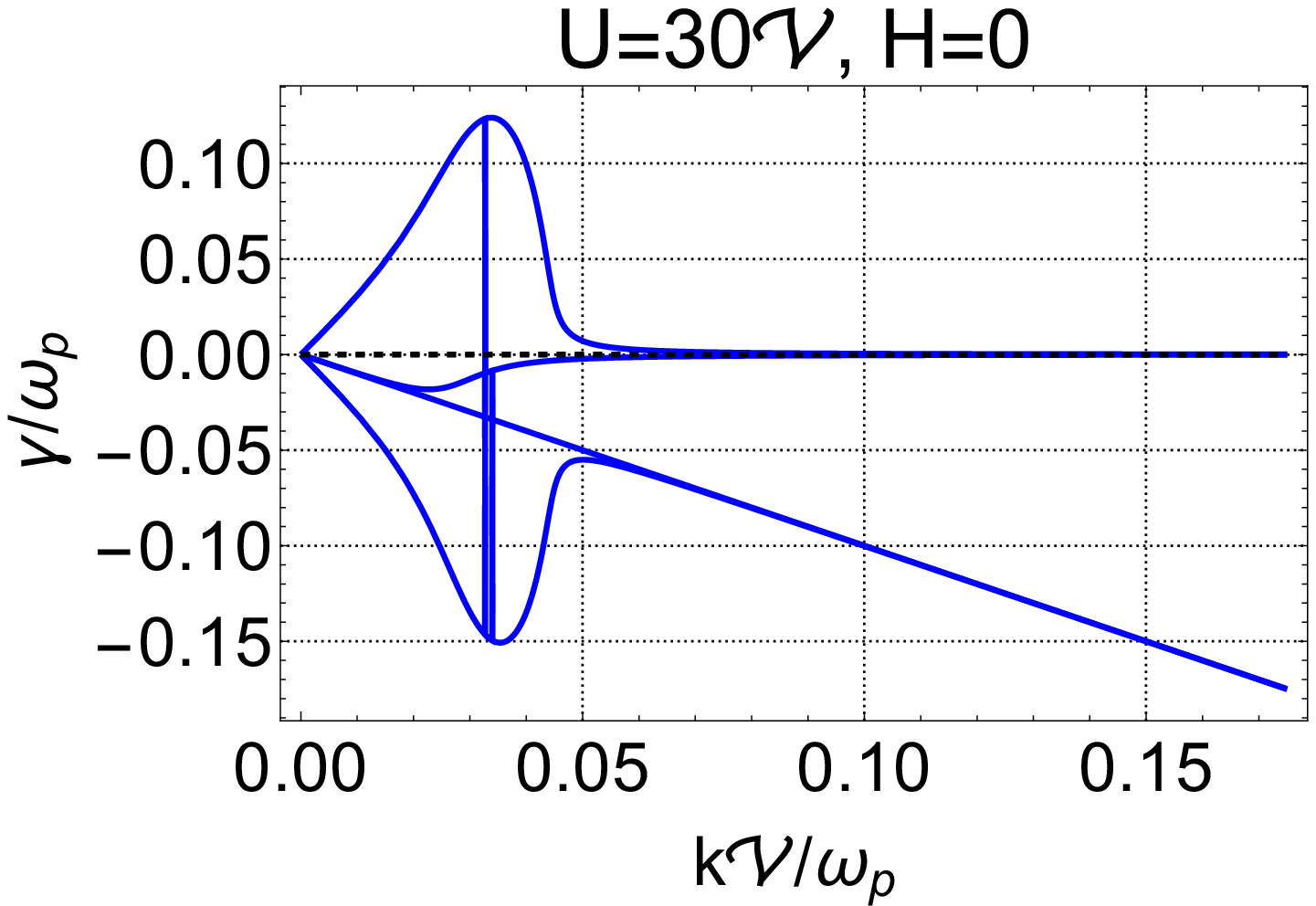}}

\vspace{1mm}
\subfloat[]{\includegraphics[width=0.40\columnwidth]{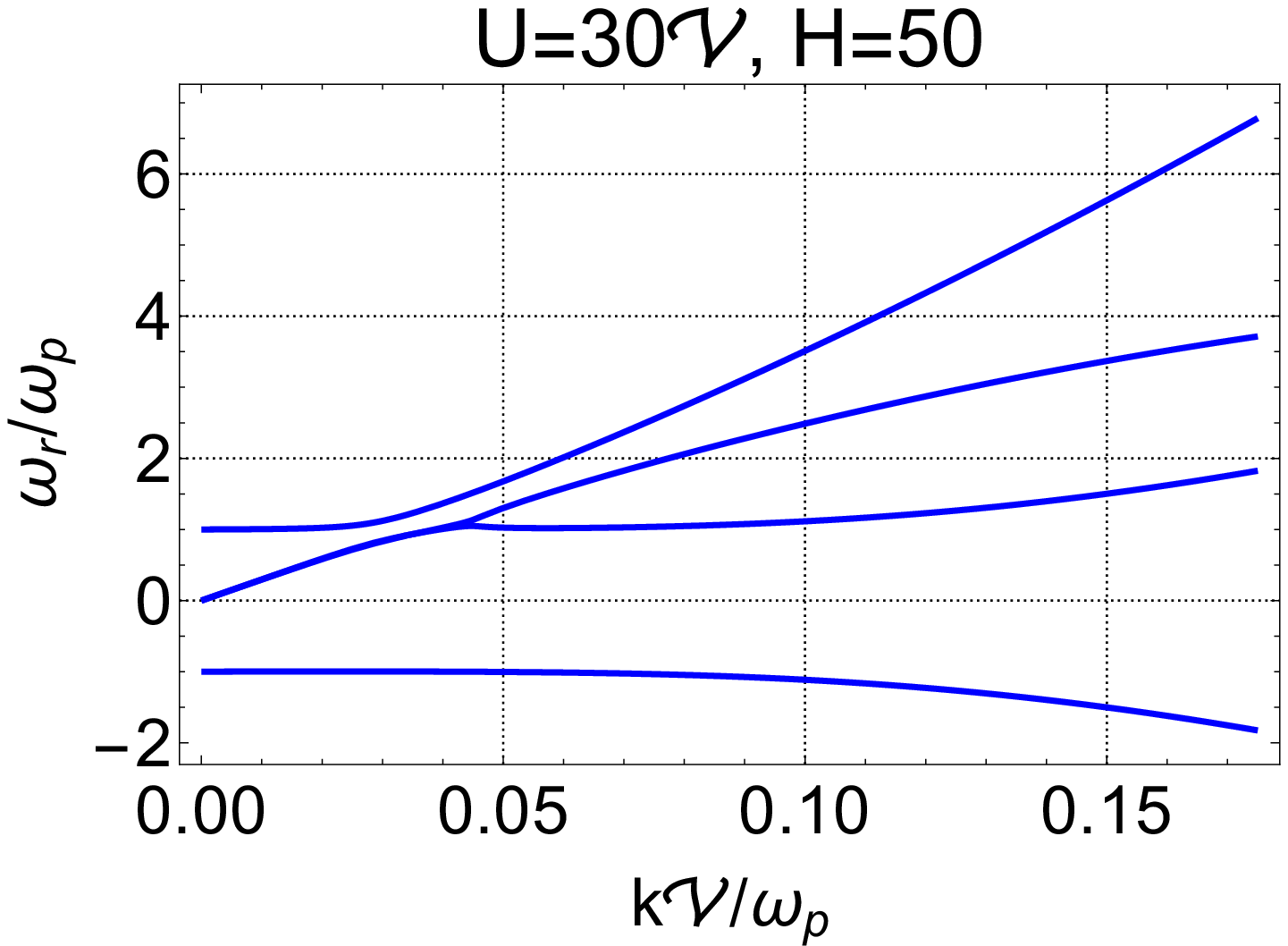}}\hspace{0.02\columnwidth}\subfloat[]{\includegraphics[width=0.40\columnwidth]{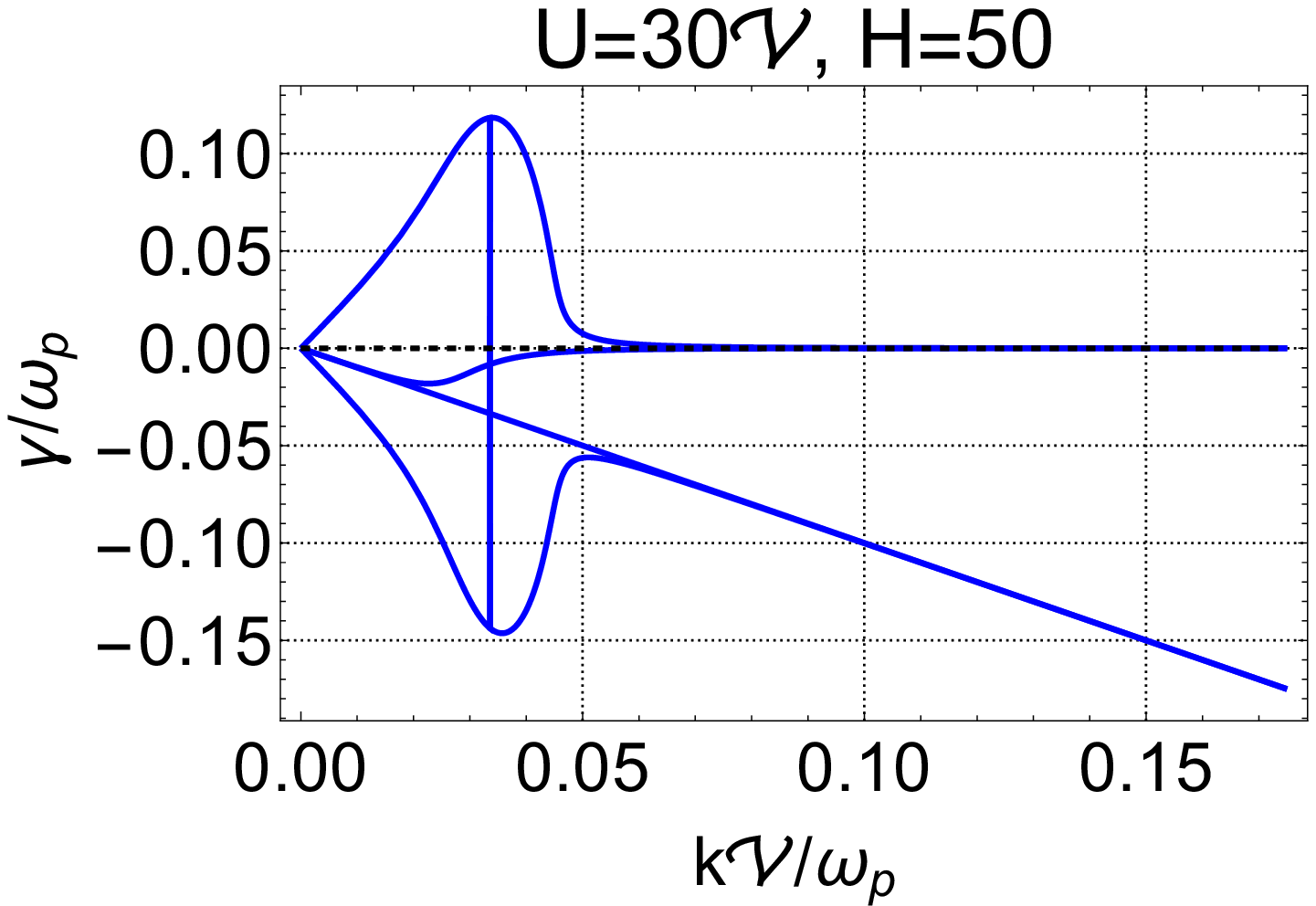}}

\vspace{1mm}
\subfloat[]{\includegraphics[width=0.40\columnwidth]{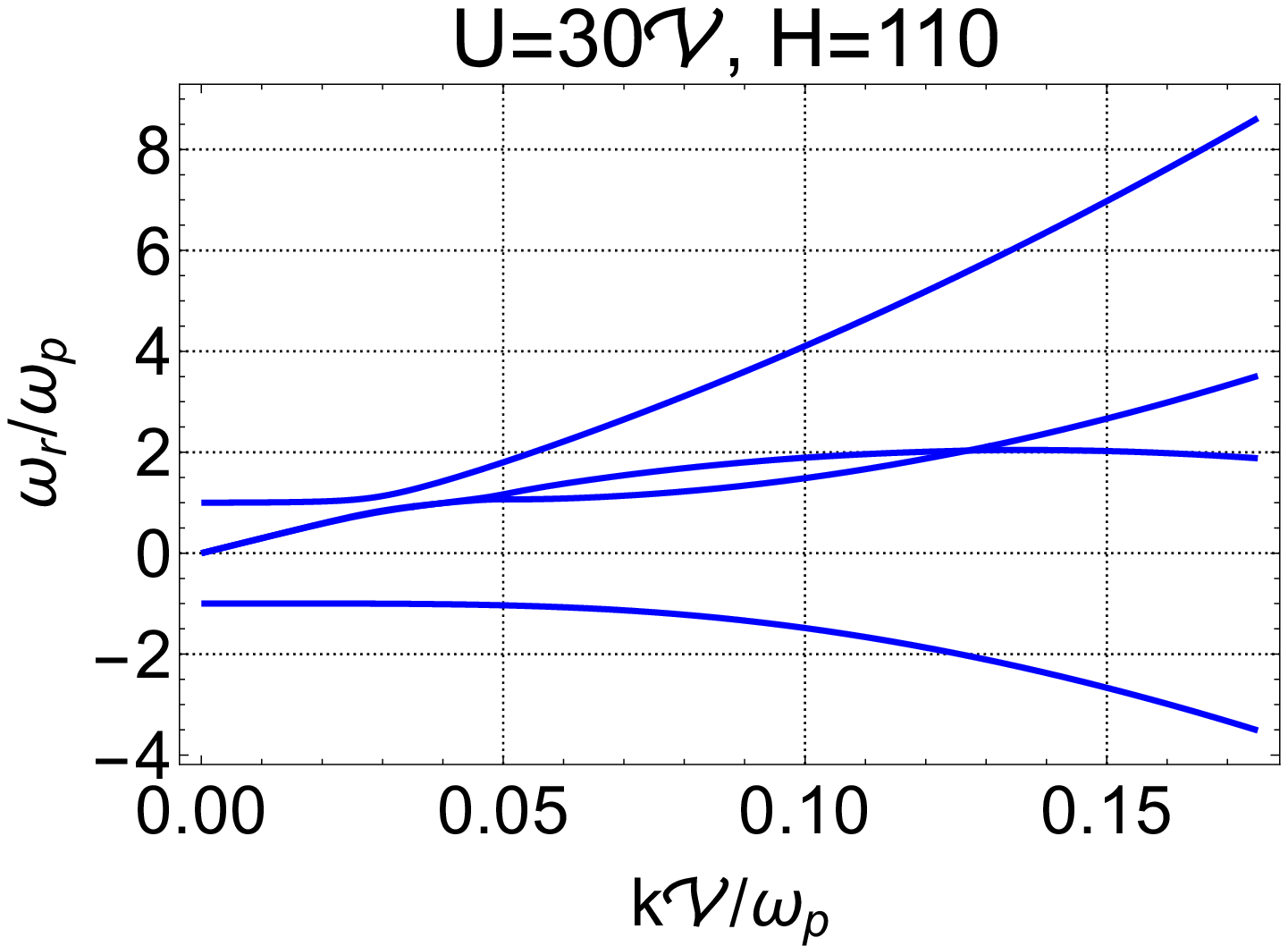}}\hspace{0.02\columnwidth}\subfloat[]{\includegraphics[width=0.40\columnwidth]{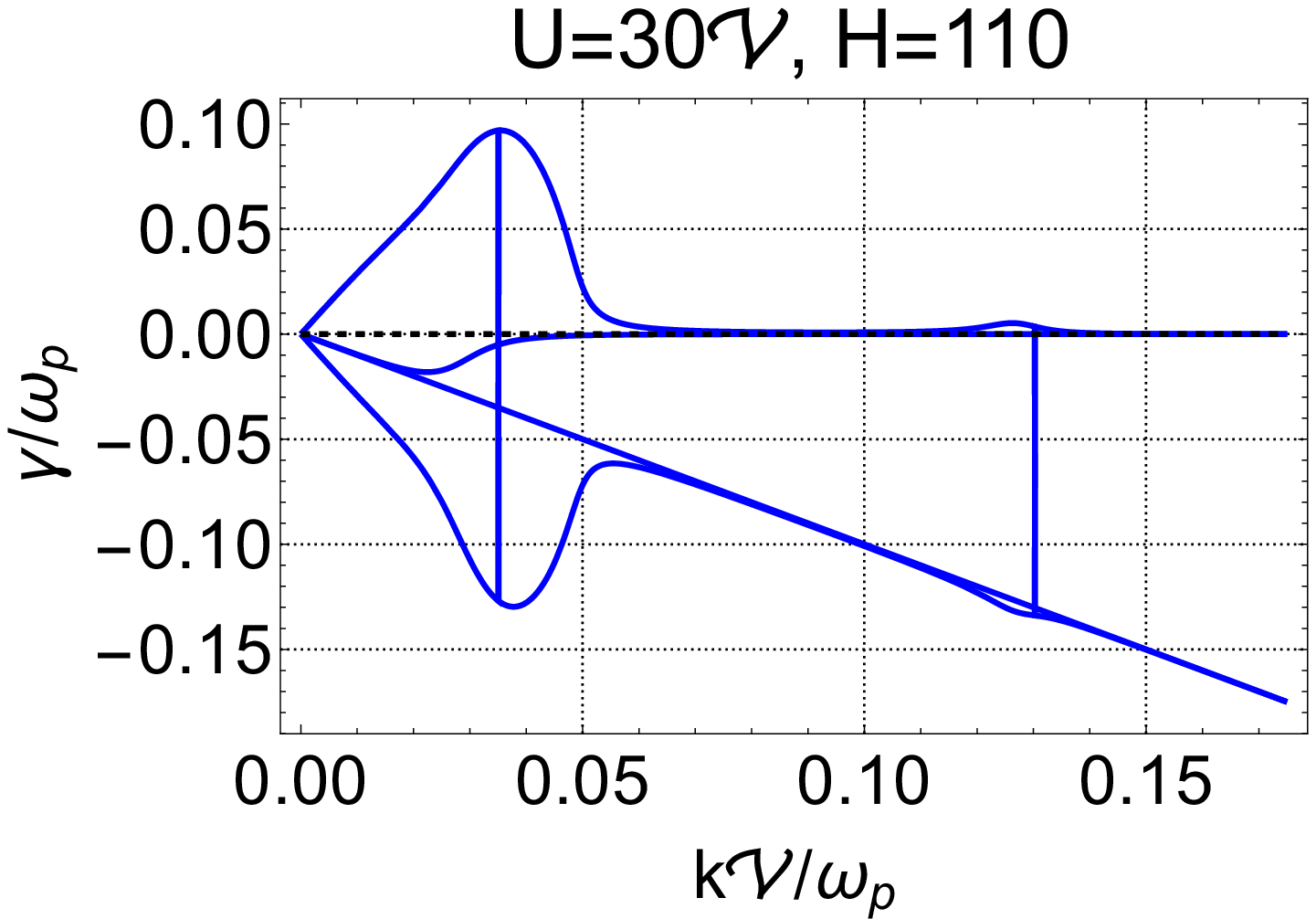}}\end{center}

\caption{Dispersion relations for primary Cauchy distribution function population
with low-density drifting beam with $n=1/100$ (Case 2). The modes
are quantitatively altered from that of the symmetrical case in figure
\ref{fig:cauchy_dispersion}. However, the general behaviour is the
same, with a region of instability for $k<k_{1}$ and a second, quantum,
region of instability for $k_{2}<k<k_{3}$ for large enough~$H$.\label{fig:cauchy_dispersion-1}}
\end{figure}

\subsubsection{Case 2: Primary population with delta-function beam}

Similarly to the delta-function beams, in Case 2 there is no simple
solution for the dispersion relation. However, the asymptotic behaviour
of $\omega$ can still be determined. In the small-$K$ limit the
unstable mode is

\begin{equation}
\omega\approx-kn(U+i)+ik\sqrt{n}(U+i)+kU+O\left(K^{2}\right)
\end{equation}
which is unstable for arbitrarily small values of~$U$. The numerically-obtained
dispersion relation for this asymmetrical case is plotted in figure
\ref{fig:cauchy_dispersion-1}. The distinction in comparison to Case
1 is more notable here than for the delta-functions, in that the damping
of the unstable mode for $K>K_{1}$ disappears. However, the existence
of one region of instability for small~$H$ and the appearance of
a second window for sufficiently large~$H$ remain as important features.

The behaviour of the modes shown in figure \ref{fig:cauchy_dispersion-1}
differs from that in Case 1 in part due to the change in frame of
reference, which Doppler shifts the real part of the frequency and
accounts for the phase velocity of the unstable mode for $k<k_{1}$
being equal to the beam velocity. Additionally, the ``bubble'' in
the plot of~$\gamma$ in figure \ref{fig:cauchy_dispersion} panels
d and f is not present, but the second region of instability still
exists. The ``bubble'' instead is split by the difference in Landau
damping rate between the two modes.

\begin{figure}
\begin{center}\subfloat[]{\includegraphics[width=0.40\columnwidth]{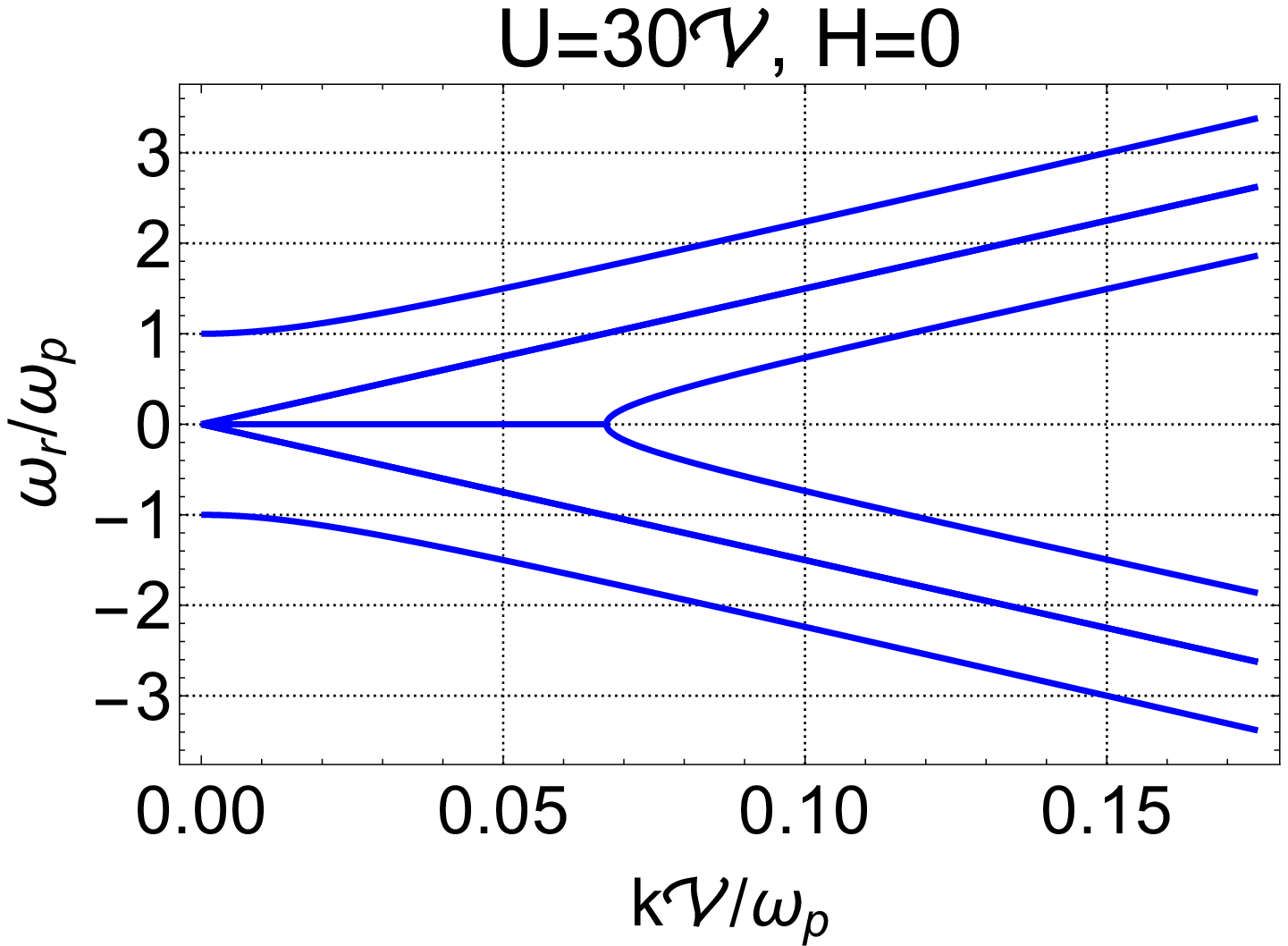}}\hspace{0.02\columnwidth}\subfloat[]{\includegraphics[width=0.40\columnwidth]{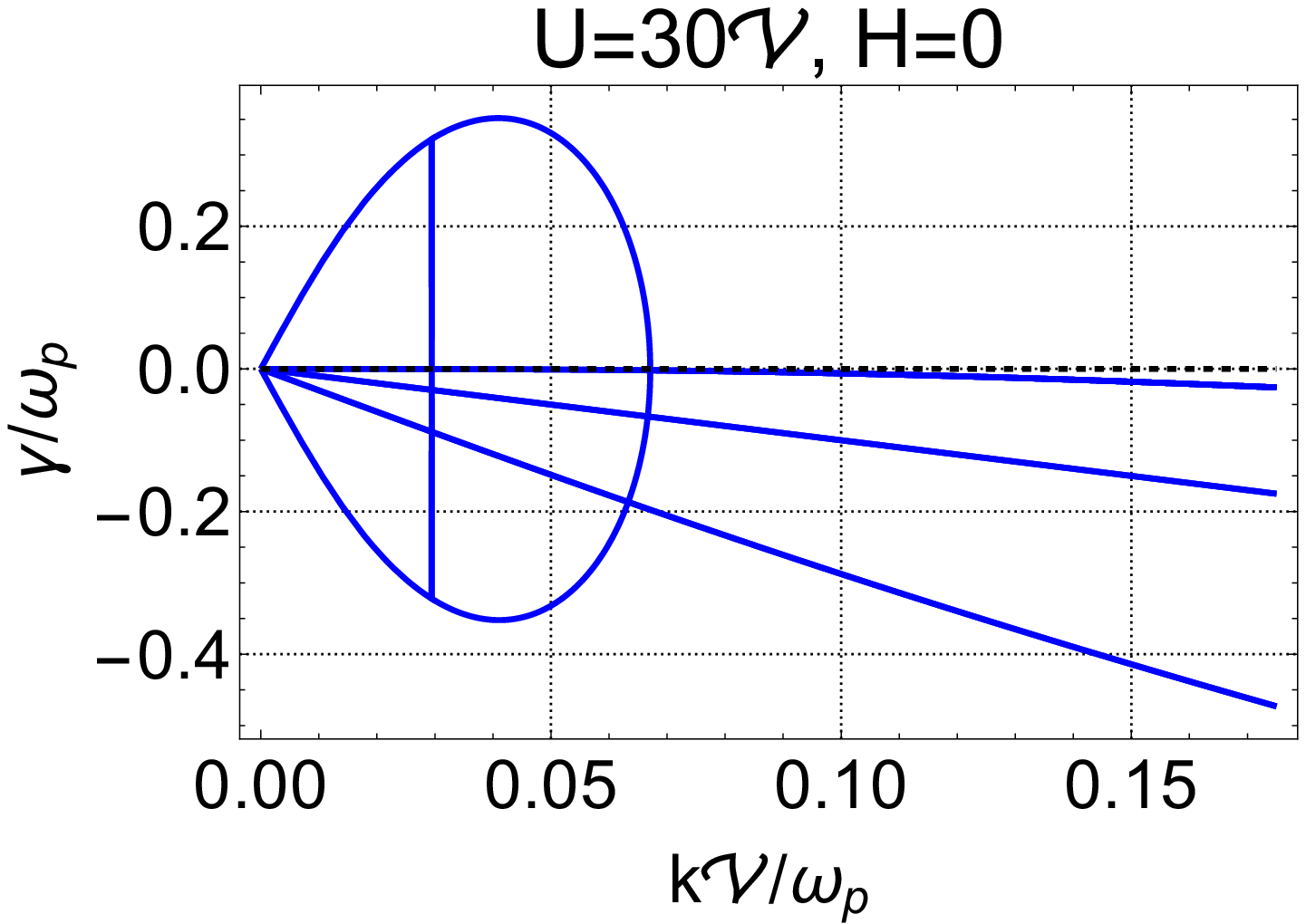}}

\vspace{1mm}
\subfloat[]{\includegraphics[width=0.40\columnwidth]{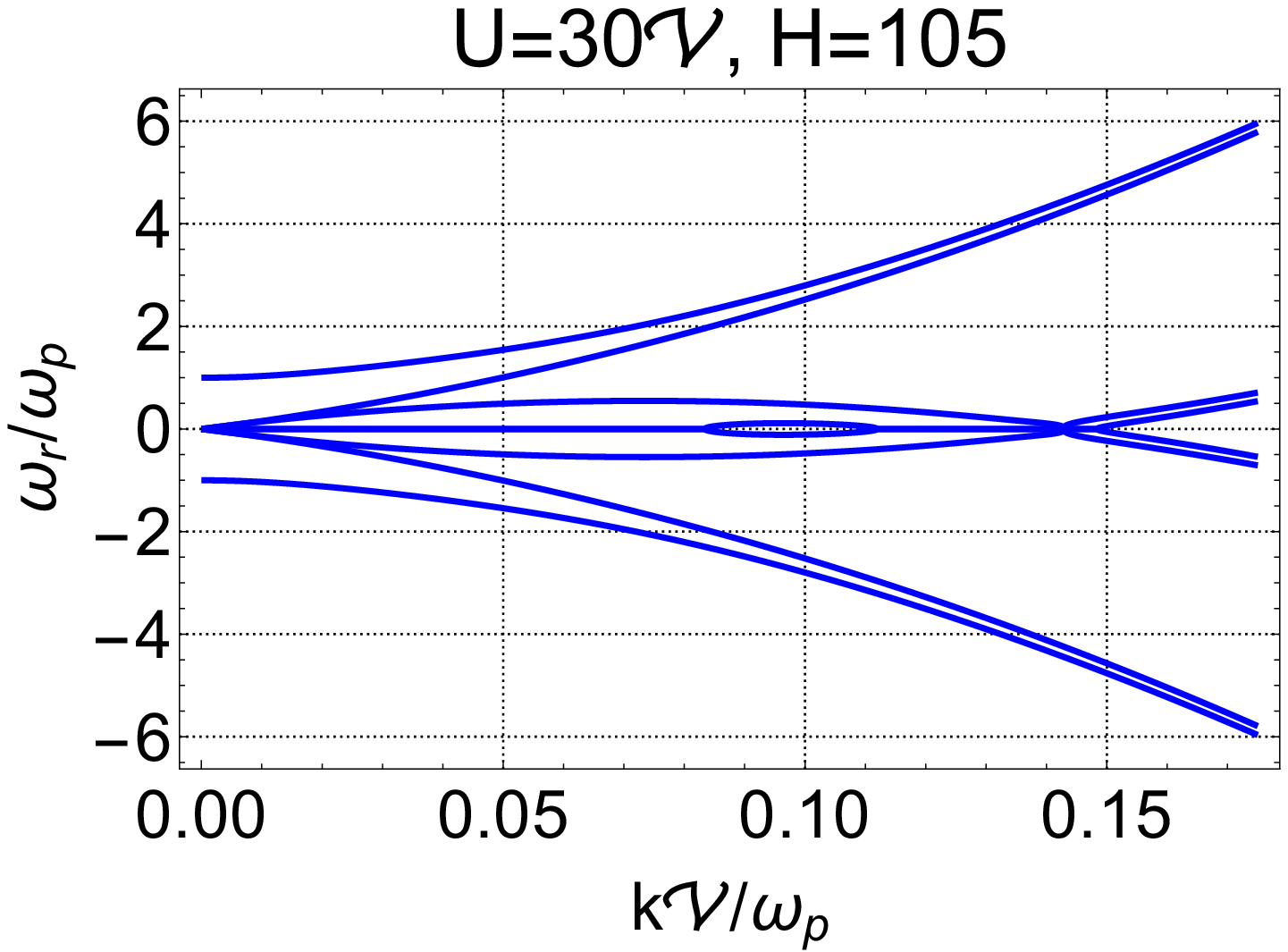}}\hspace{0.02\columnwidth}\subfloat[]{\includegraphics[width=0.40\columnwidth]{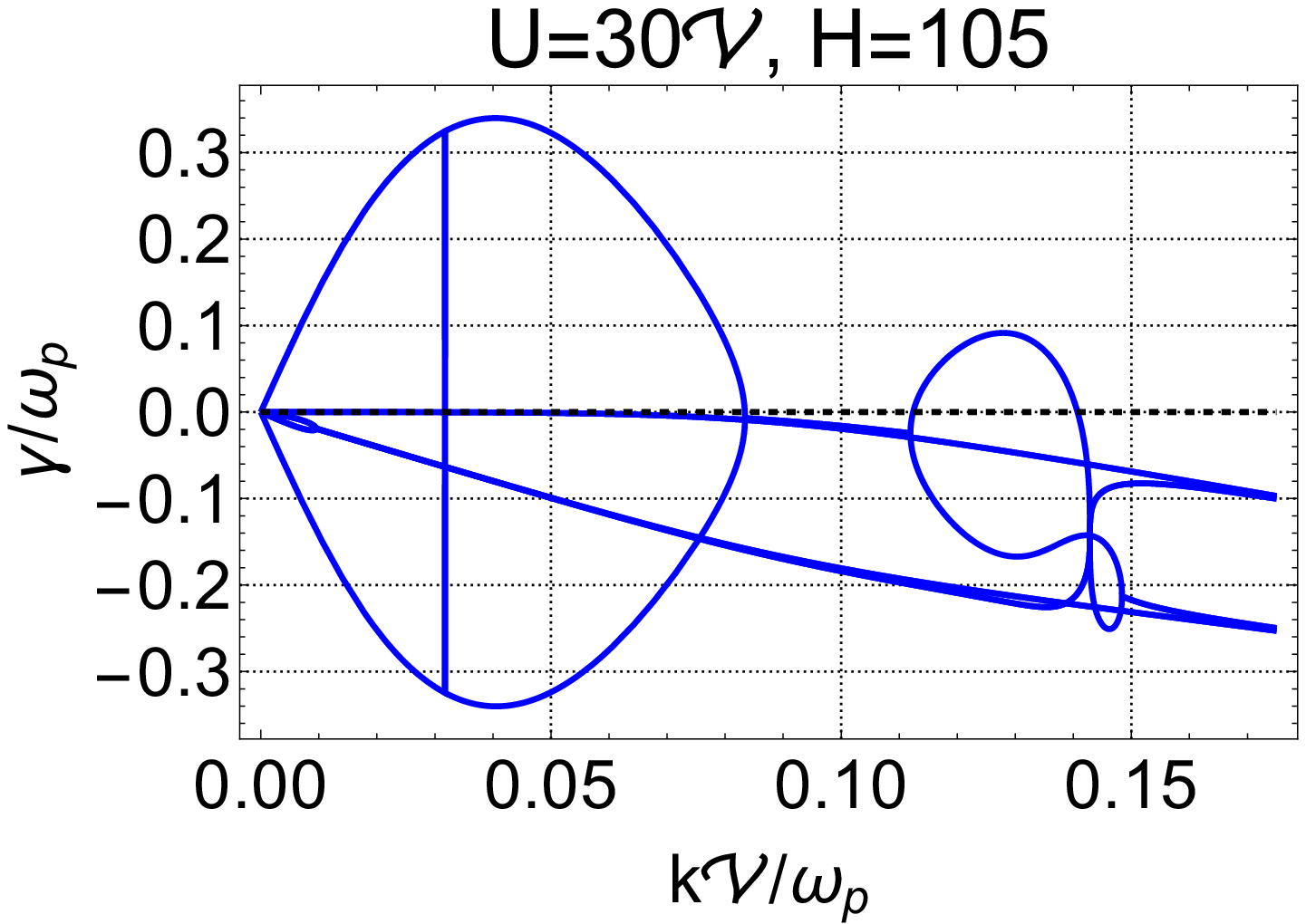}}

\vspace{1mm}
\subfloat[]{\includegraphics[width=0.40\columnwidth]{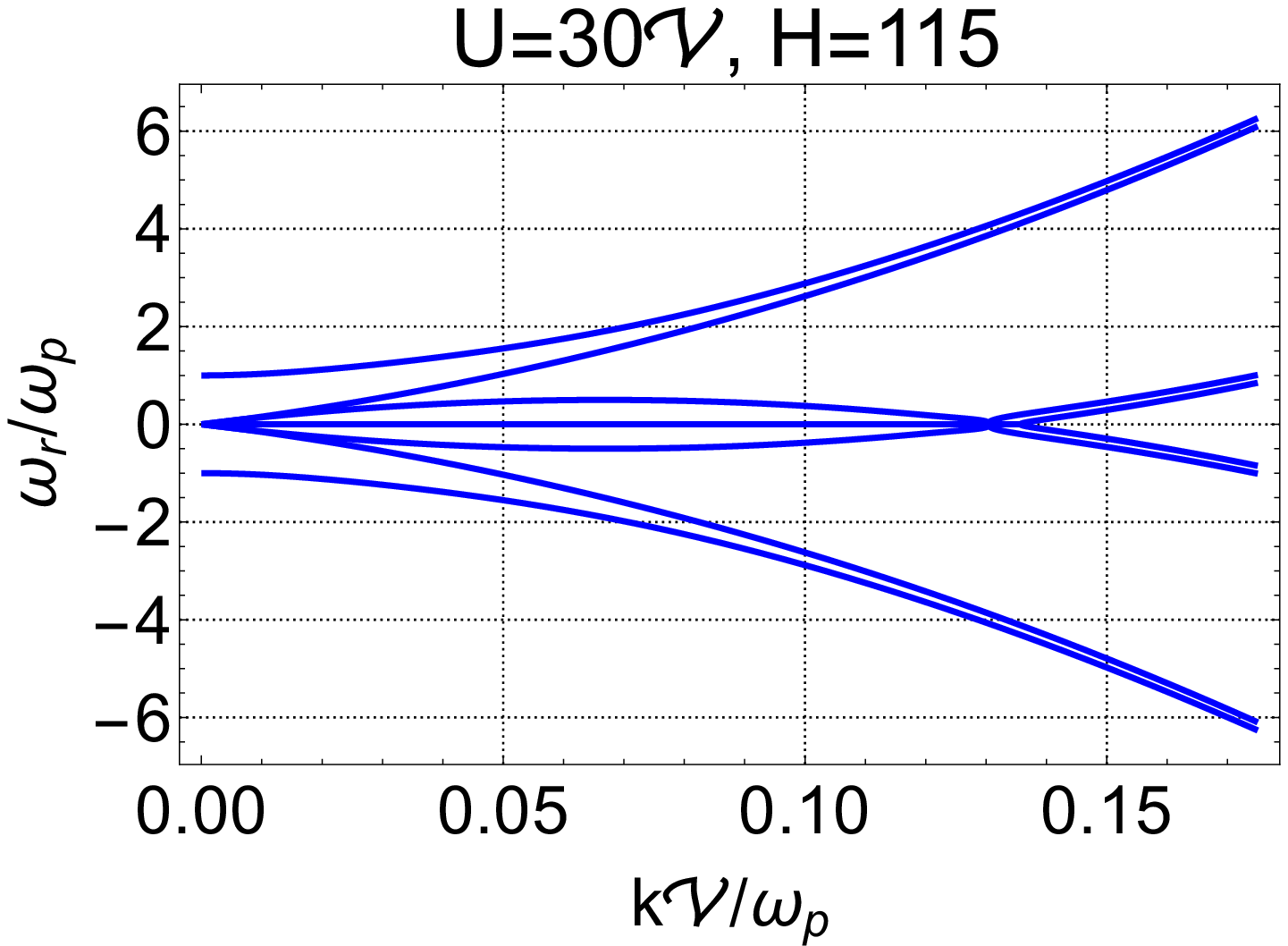}}\hspace{0.02\columnwidth}\subfloat[]{\includegraphics[width=0.40\columnwidth]{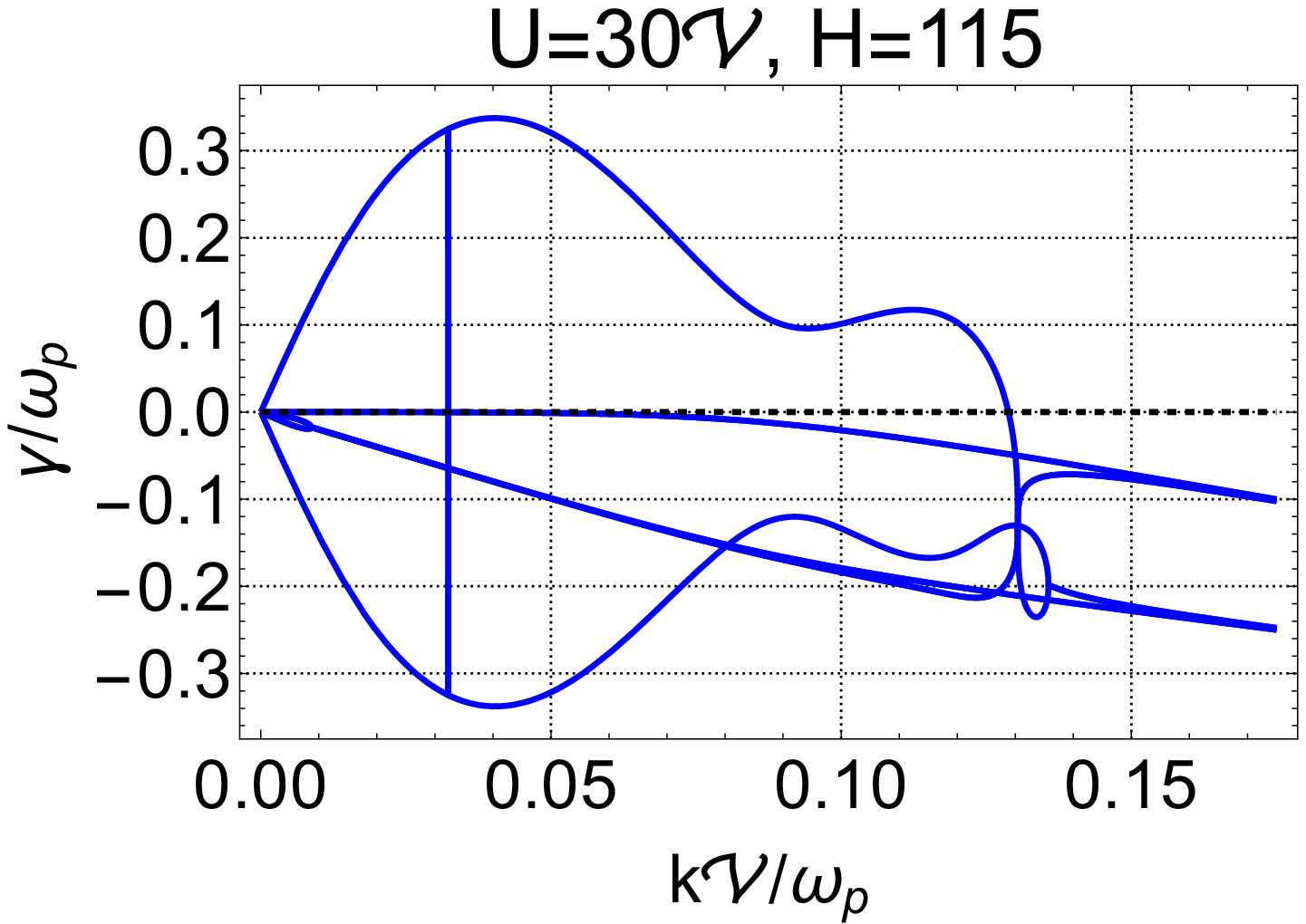}}\end{center}

\caption{Exact dispersion relations for the symmetrical counter-drifting squared
Cauchy distribution functions (Case 1). The behaviour is generally
similar to that of the Cauchy distributions as seen in figure \ref{fig:cauchy_dispersion},
with the existence of a pair of new modes which interact with the
unstable mode at the point where the second region of instability
terminates (i.e.~$k_{3}$). \label{fig:cauchy2_instability}}
\end{figure}

\begin{figure}
\begin{center}\subfloat[]{\includegraphics[width=0.40\columnwidth]{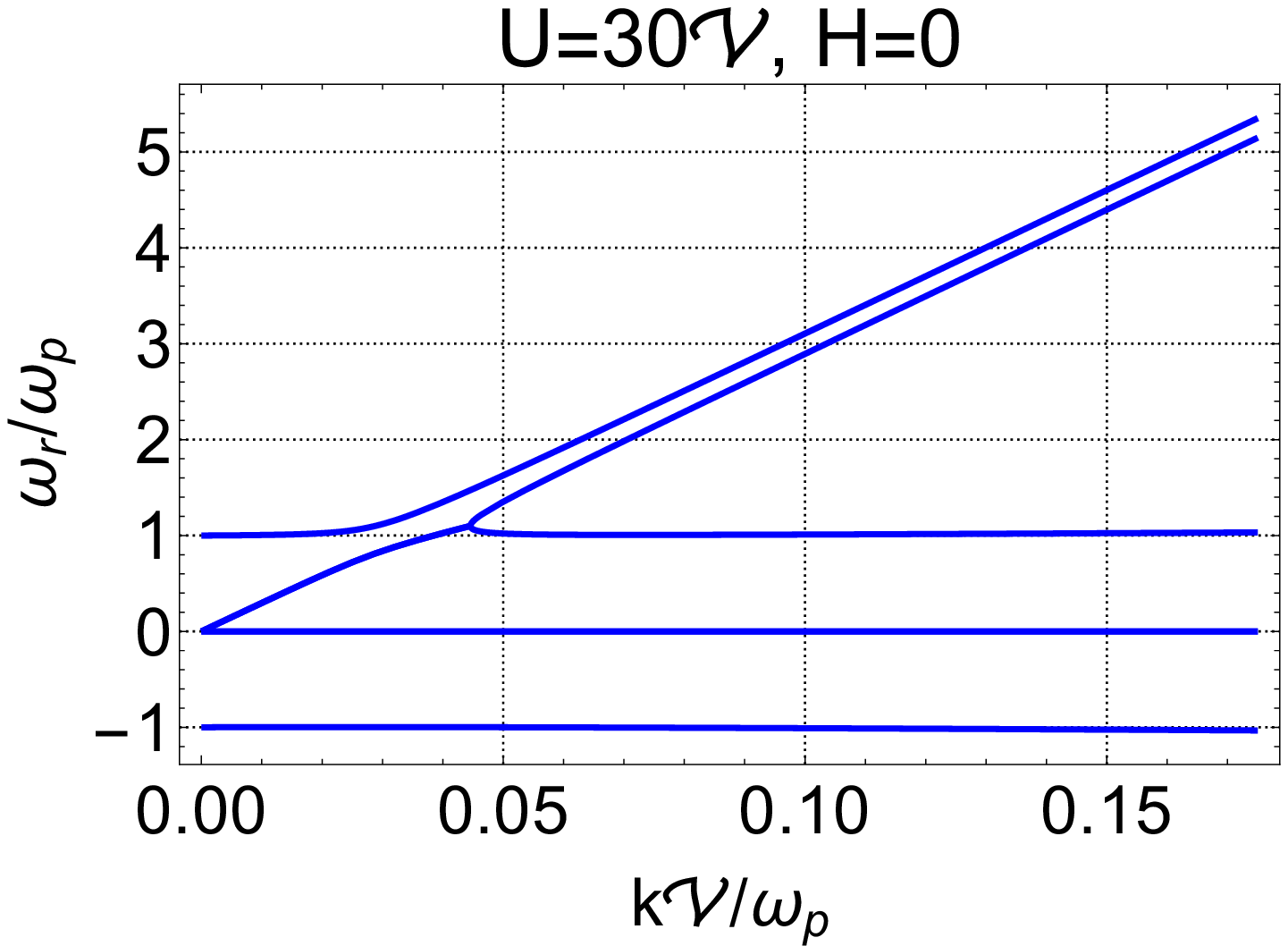}}\hspace{0.02\columnwidth}\subfloat[]{\includegraphics[width=0.40\columnwidth]{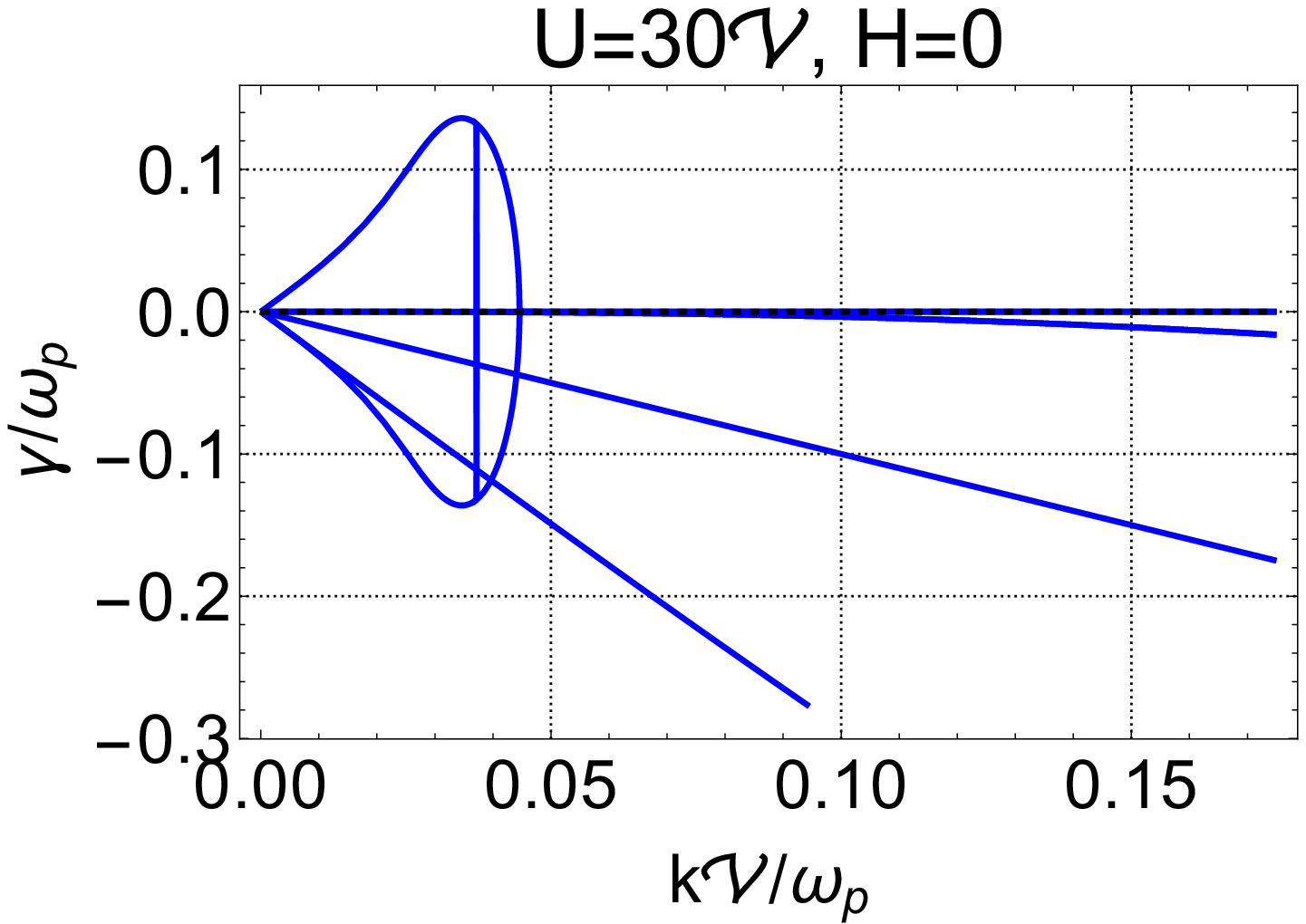}}

\vspace{1mm}
\subfloat[]{\includegraphics[width=0.40\columnwidth]{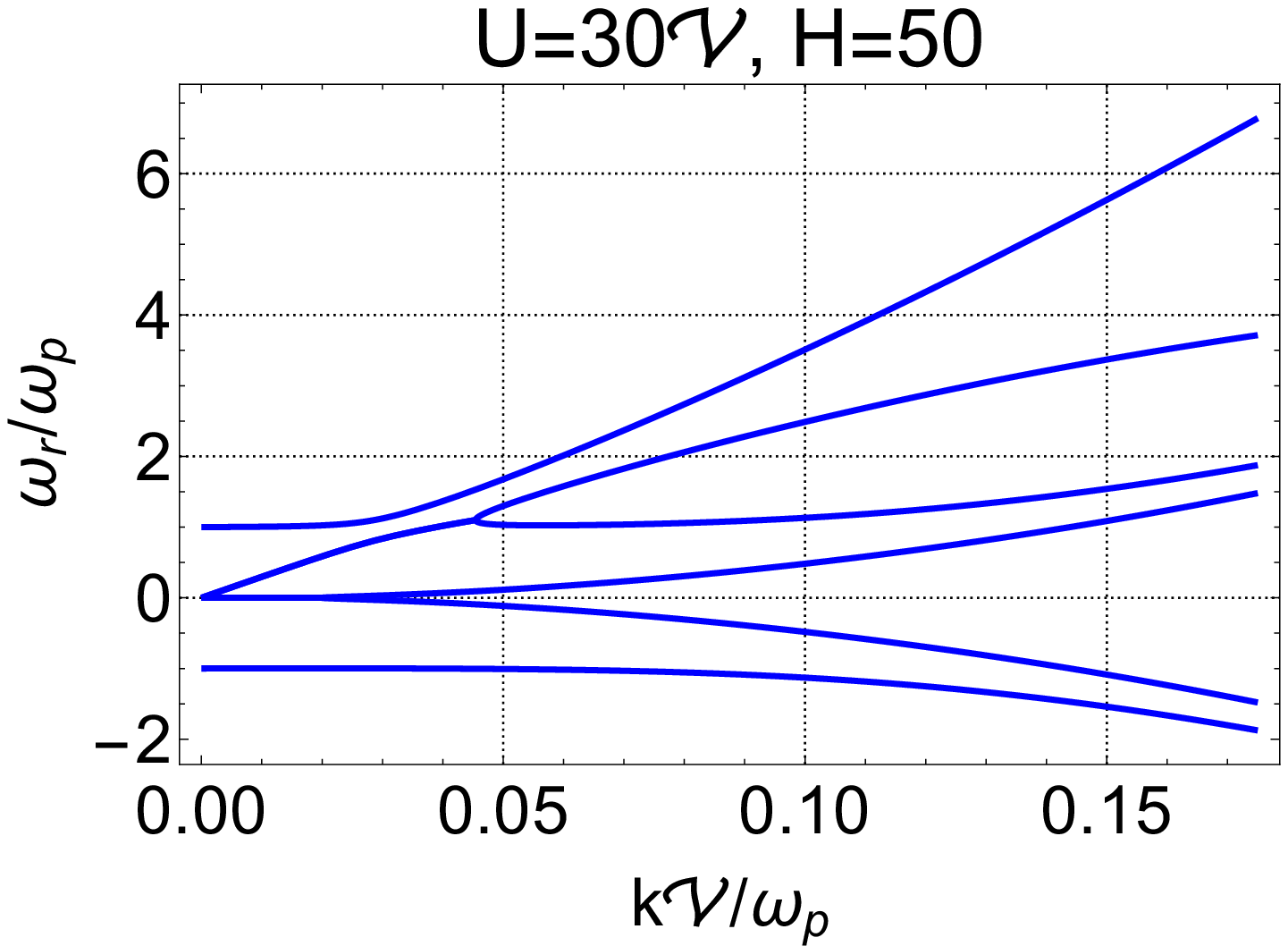}}\hspace{0.02\columnwidth}\subfloat[]{\includegraphics[width=0.40\columnwidth]{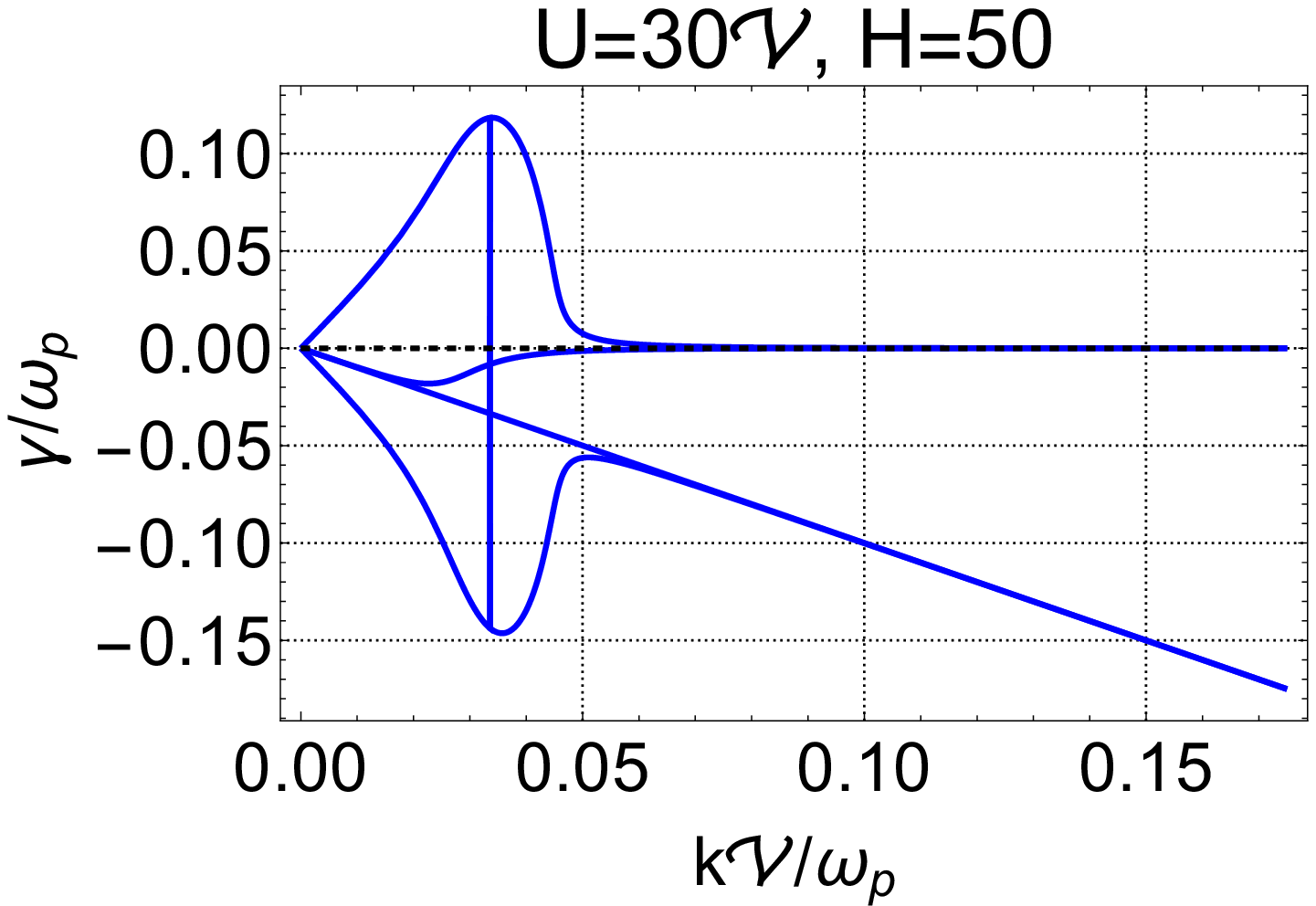}}

\vspace{1mm}
\subfloat[]{\includegraphics[width=0.40\columnwidth]{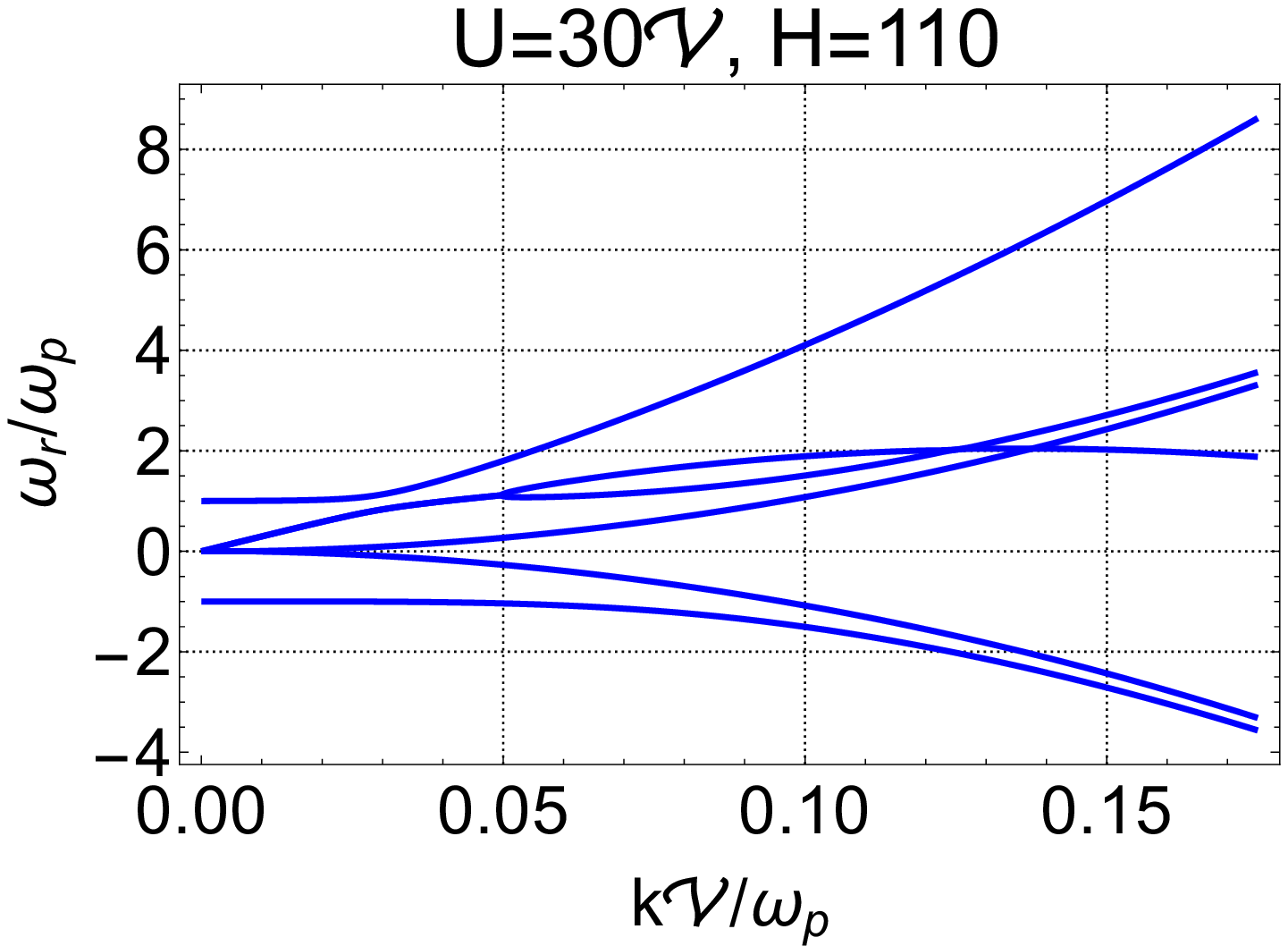}}\hspace{0.02\columnwidth}\subfloat[]{\includegraphics[width=0.40\columnwidth]{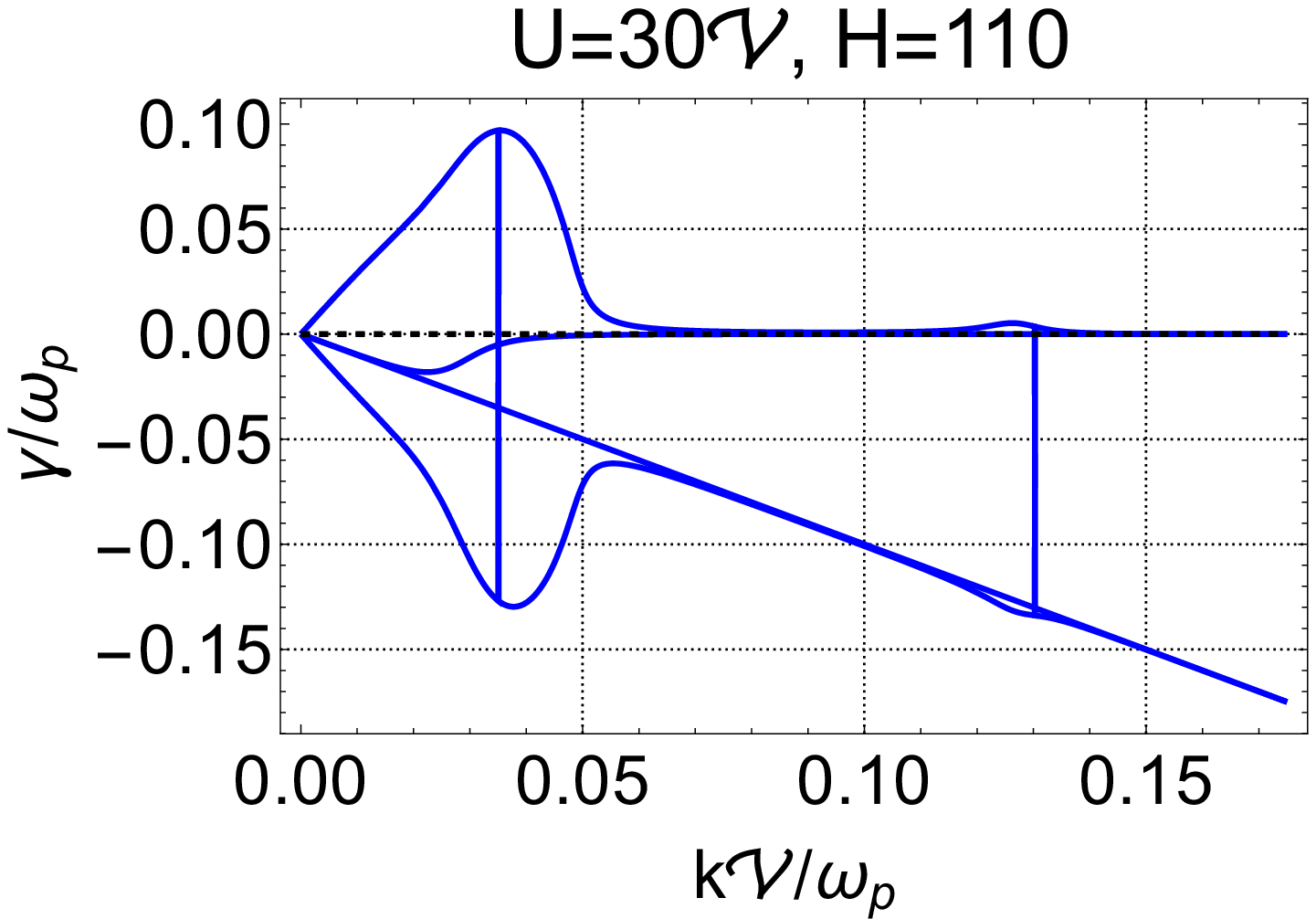}}\end{center}

\caption{Exact dispersion relations for primary squared Cauchy distributed
population with low-density drifting beam with $n=1/100$ (Case 2).
The asymptotic behaviour is different from Case 1 (figure \ref{fig:cauchy2_instability})
in that the the plot of~$\omega$ is Doppler shifted due to the change
reference frames, and the unstable mode becomes an undamped plasma
oscillation for large wavenumbers. This can be interpreted as a stationary
oscillation in the beam, which explains the lack of Landau damping.
\label{fig:cauchy2_instability-1}}
\end{figure}

\begin{figure}
\begin{center}%
\begin{tabular}{c}
\includegraphics[width=0.5\columnwidth]{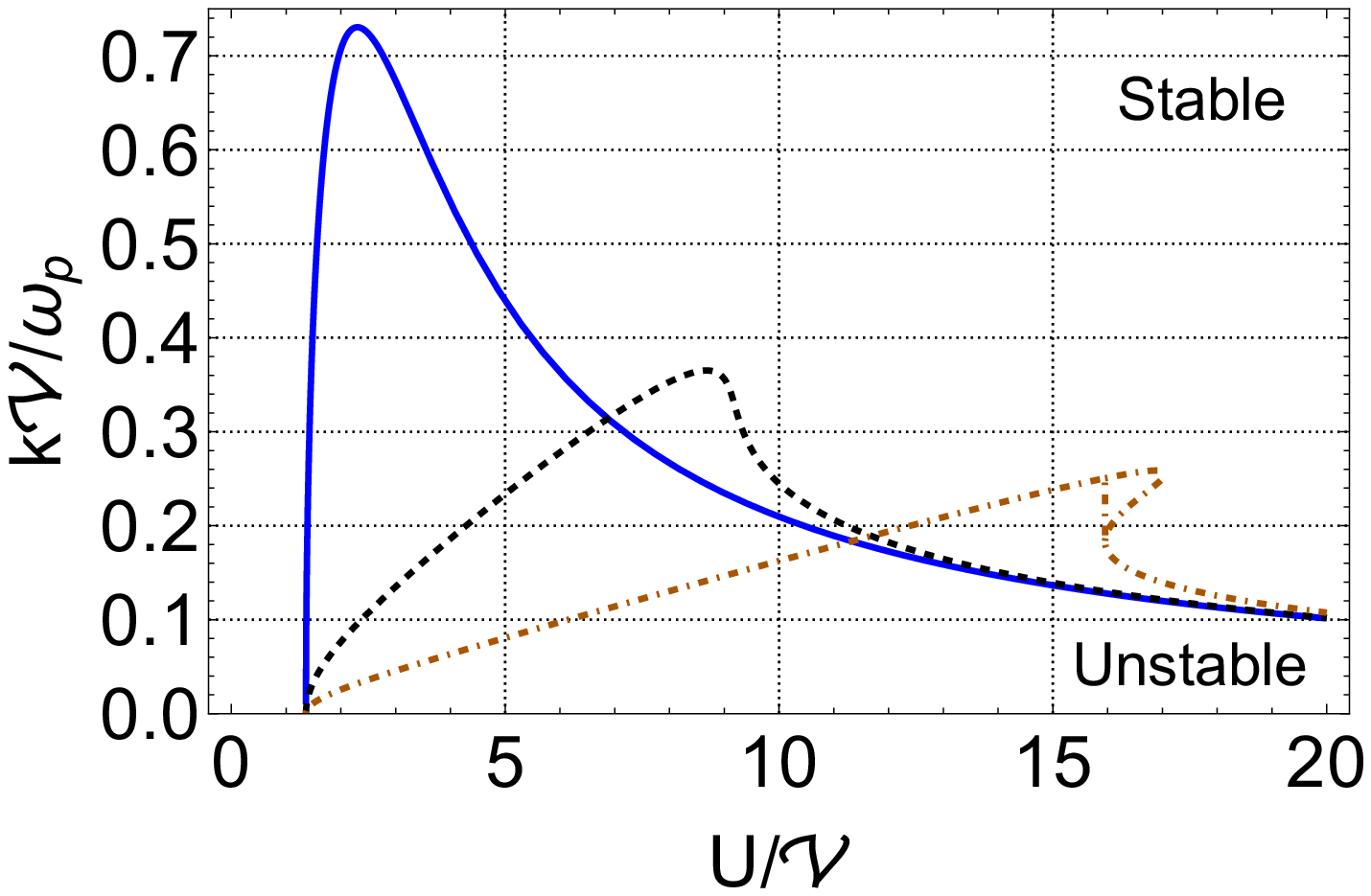}\tabularnewline
\end{tabular}\end{center}

\caption{Region of instability for the dispersion relation for counter-streaming
squared Cauchy distributions with $H=0$, (blue solid), $H=10$ (black
dashed) and $H=30$ (orange dot-dashed). The second region of instability
is seen to emerge for smaller~$H$ than for the counter-drifing Cauchy
distributions. \label{fig:cauchy2_inst_region}}
\end{figure}

\subsection{Squared Cauchy Distribution: Case With Classical Dispersion}

\label{sec:cau2}

The primary physical difference between the squared Cauchy distribution
and the Cauchy distribution is that it has a finite second moment.
This means that there can be thermal dispersion of the electrostatic
waves, which we do not see for the Cauchy case. Due to the presence
of a velocity scale $\mathcal{V}$ associated with the distribution
function, we utilise the first normalisation scheme discussed in section
\ref{sec:parameters}.

\paragraph{Case 1: Symmetrically counter-propagating distributions}

The dispersion relation for Case 1 can only be obtained analytically
under approximations of $U\gg\mathcal{V}$ or $K\ll1$, but the roots
of $\epsilon\left(\Omega,K\right)$ can be found numerically for any
values of~$K$ and~$U$. We first consider the case $U\gg\mathcal{V}$
in which case the dispersion relation is 
\begin{equation}
\Omega\approx\frac{K\sqrt{K^{2}\left(U^{4}-16\right)-4\left(U^{2}+12{}^{2}\right)}}{4\sqrt{3}}.
\end{equation}
In this limit, the region of instability is bounded by 
\begin{equation}
K_{1}\approx\frac{2}{U}+\frac{12}{U^{3}}+\frac{4\left(4H^{2}-69\right)}{U^{5}}+O\left(U^{-7}\right).
\end{equation}
The maximum growth occurs at 
\begin{equation}
K_{max}\approx\frac{\sqrt{2}}{U}+\frac{6\sqrt{2}}{U^{3}}+\frac{2\sqrt{2}\left(4H^{2}-9\right)}{U^{5}}+O\left(U^{-7}\right),
\end{equation}
and is 
\begin{equation}
\frac{\gamma}{\omega_{p}}\approx\frac{1}{2\sqrt{3}}+\frac{2\sqrt{3}}{U^{2}}+\frac{4H^{2}}{\sqrt{3}U^{4}}+O\left(U^{-6}\right).
\end{equation}
To lowest order, the maximum growth rate is purely classical and is
independent of~$U$.

In the large-wavelength approximation, $K\ll1$, the boundary of the
unstable region obeys the expression
\begin{equation}
K_{1}\approx\frac{2\left(U^{2}+4\right)\sqrt{48-U^{2}\left(U^{2}+24\right)}}{\sqrt{16H^{2}\left(U^{6}+60U^{4}-720U^{2}+320\right)-\left(U^{2}+4\right)^{5}}},
\end{equation}
with the classical limit
\begin{equation}
K_{1}\approx\frac{2\sqrt{U^{4}+24U^{2}-48}}{\left(U^{2}+4\right)^{3/2}},
\end{equation}
from which it can be shown that the instability exists only for $U>2\left(2\sqrt{3}-3\right)^{1/2}\approx1.36$
in both the classical and quantum cases. Notably, this differs from
the value in the Cauchy distribution case and, as will be seen, the
inverse-quartic distribution case.

In the presentation of the full, numerically-obtained, dispersion
relation in figure \ref{fig:cauchy2_instability}, it can be seen
that the imaginary part of the frequency becomes quite complicated.
The mode-crossings in the real part of the frequency coincide with
dramatic ``bubbles'' consisting of splitting modes in the graphs
of~$\gamma$. These bubbles produce the second region of instability
defined by~$K_{2}$ and~$K_{3}$, as seen in figure \ref{fig:cauchy2_instability}f.
As in the instance of Cauchy distributions,~$K_{2}$ and~$K_{3}$
do not have simple analytical representations, and despite the outwardly
more complicated behaviour in this example, the essential characteristics
defined by~$K_{1}$,~$K_{2}$, and~$K_{3}$ remain.

\subsubsection{Case 2: Primary population with delta-function beam}

The unstable mode in this situation is 
\begin{equation}
\omega=K\left(-nU+{\rm i}\sqrt{(3-n)nU^{2}-3n+3}\right)+O\left(K^{2}\right)
\end{equation}
which is unstable for all $U$. Again, this is purely classical to
this level of accuracy. The dispersion relation for Case 2 is plotted
in figure \ref{fig:cauchy2_instability-1}, from which the general
similarities to the case with a primary Cauchy distribution are apparent.
The primary difference is the dependence of the Landau damped modes
on~$K$, and the sharp cutoff of the first region of instability
at~$K_{1}$.

\subsection{Inverse-Quartic Distribution: Second Case With Classical Dispersion}

\label{sec:caudub}

While the squared Cauchy susceptibility \ref{eq:chic2} contains terms
due to the second order poles in the distribution function, a similar
case that we consider here is that of the inverse-quartic distribution
function 
\[
f_{4}\left(v\right)=\frac{\sqrt{2}\mathcal{V}^{3}}{\pi\left(v^{4}+\mathcal{V}^{4}\right)},
\]
which has four first order poles, and which has a flatter top and
steeper wings than the distribution functions \ref{eq:distcau} and
\ref{eq:distcau2}.

\paragraph{Case 1: Symmetrically Counter-Propagating Distributions}

The susceptibility equation \ref{eq:chi2} is of greater than fourth
order in~$k$ and the dispersion equation cannot be solved algebraically;
instead we turn to approximations and numerical solutions. We again
consider the cases $U\gg\mathcal{V}$ (separation is much greater
than the thermal widths) or $k\mathcal{V}/\omega_{p}\ll1$ (phase
speed large compared to thermal speed). We additionally plot the numerically-obtained
solution for the full dispersion relation for three values of~$U$
and~$H$ in figure \ref{fig:cauchyD_instability}.

In the limit $U\gg\mathcal{V}$, the dispersion relation is 
\begin{equation}
\Omega=\frac{\sqrt{-16H^{2}K^{4}+K^{4}U^{4}-4K^{2}U^{2}-48K^{2}}}{4\sqrt{3}}.
\end{equation}
The long-wavelength instability region is bounded by 
\begin{equation}
K_{1}=\frac{2}{U}+\frac{12}{U^{3}}+\frac{16H^{2}-116}{U^{5}}+O\left(U^{-7}\right),
\end{equation}
the same as for the squared Cauchy distribution. The wavenumber of
maximum growth rate is 
\begin{equation}
K_{{\rm max}}=\frac{\sqrt{2}}{U}+\frac{6\sqrt{2}}{U^{3}}+\frac{2\sqrt{2}\left(4H^{2}-9\right)}{U^{5}}+O\left(U^{-7}\right)
\end{equation}
and that maximum growth rate is 
\begin{equation}
\frac{\gamma}{\omega_{p}}=\frac{1}{2\sqrt{3}}+\frac{2\sqrt{3}}{U^{2}}+\frac{4H^{2}}{\sqrt{3}U^{4}}+O\left(U^{-6}\right).
\end{equation}
Note that these are identical to the results of section \ref{sec:cau2}.
The asymptotic behaviour only differs beyond fifth order in $1/U$.
However, the detailed behaviour of the modes for moderate values of~$U$
differs quantitatively. This is seen in the difference between the
cases in figures \ref{fig:cauchy2_instability} and \ref{fig:cauchyD_instability},
where the region of instability defined by~$K_{1}$ is slightly larger
for large~$U$ in the present case, and the complicated mode crossing
is not present in the $U=4\mathcal{V}$ and $H=1$ case, but re-appears
for the $U=52\mathcal{V}$ case, which appears nearly identical to
what is seen for the squared Cauchy distribution.

In the $K\ll1$ limit, the instability boundary obeys
\begin{gather}
K_{1}\approx2\left(U^{4}+16\right)\left[(U-2)(U+2)\left(U^{4}+16U^{2}+16\right)\right]^{-1/2}\times\nonumber \\
\left[\left(U^{4}+16\right)^{4}-16H^{2}\left(U^{12}+40U^{10}-496U^{8}-2816U^{6}+7936U^{4}+10240U^{2}-4096\right)\right]^{-1/2},
\end{gather}
with the classical limit
\begin{equation}
K_{1}\approx\frac{2\sqrt{(U-2)(U+2)\left(U^{4}+16U^{2}+16\right)}}{U^{4}+16},
\end{equation}
from which it can be shown that instability exists for $U>2$, which
is the same as for the Cauchy case and greater than the squared Cauchy
distribution.

\subsubsection{Case 2: Primary population with delta-function beam}

\begin{figure}
\begin{center}\subfloat[]{\includegraphics[width=0.40\columnwidth]{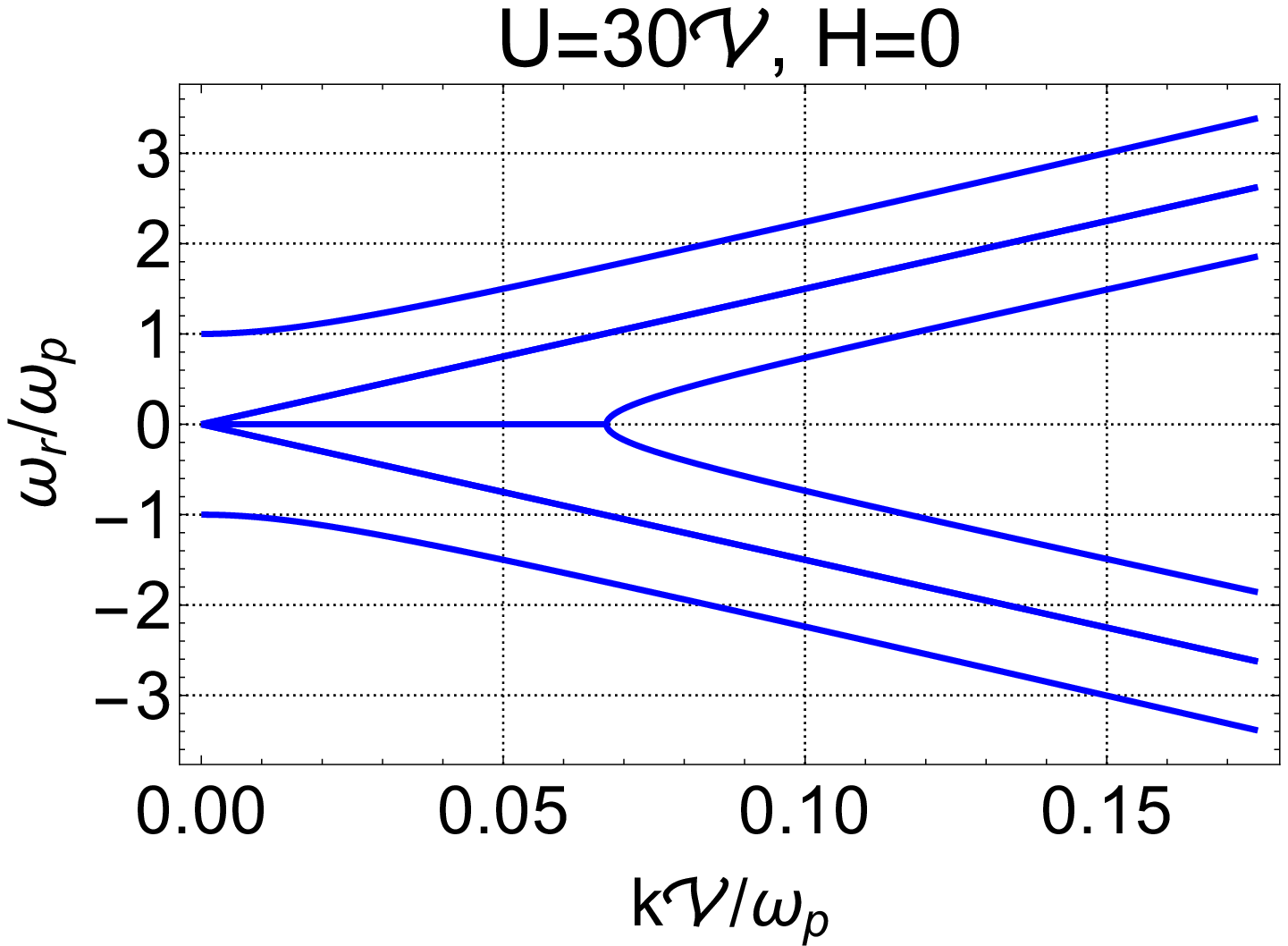}}\hspace{0.02\columnwidth}\subfloat[]{\includegraphics[width=0.40\columnwidth]{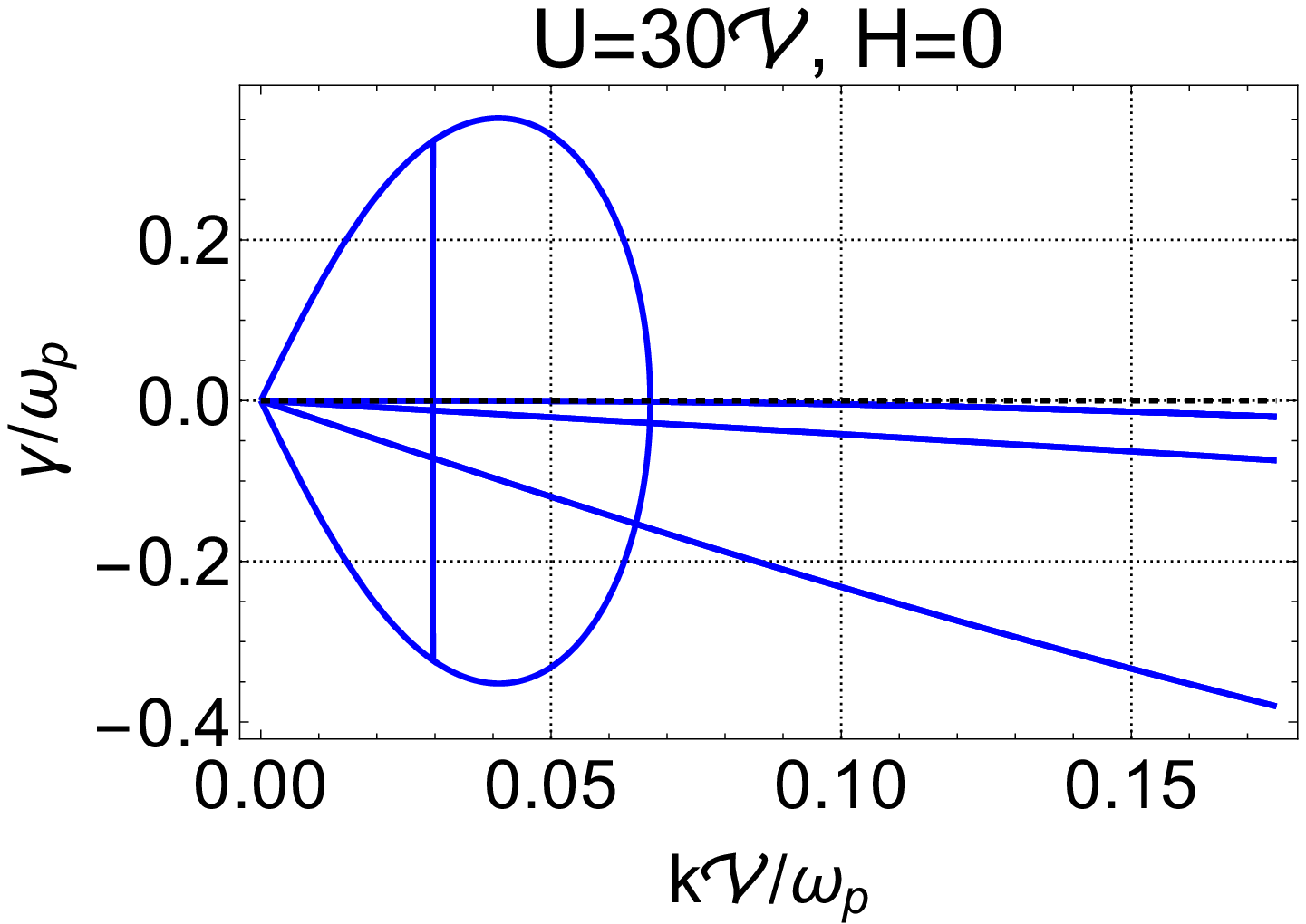}}

\vspace{1mm}
\subfloat[]{\includegraphics[width=0.40\columnwidth]{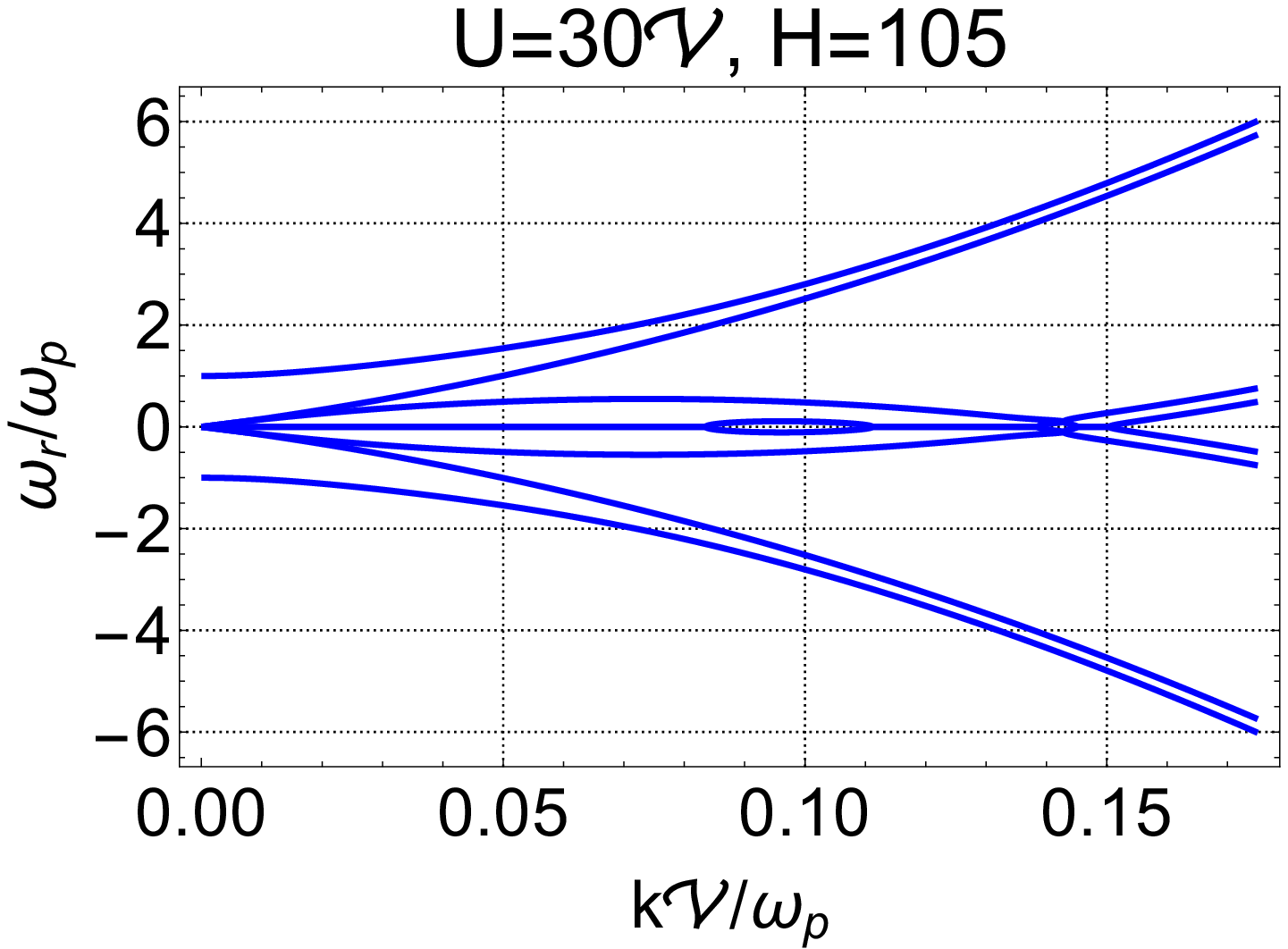}}\hspace{0.02\columnwidth}\subfloat[]{\includegraphics[width=0.40\columnwidth]{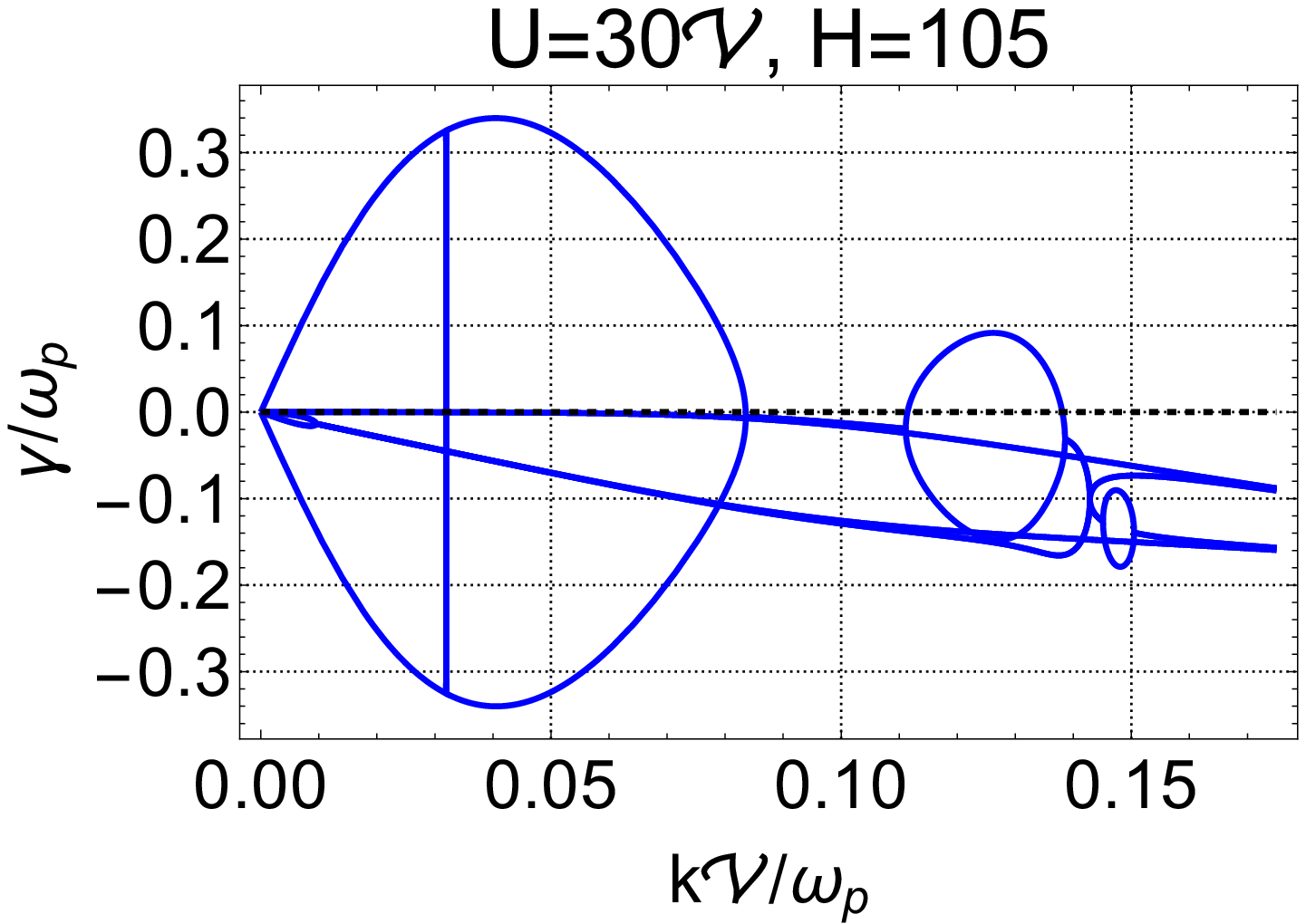}}

\vspace{1mm}
\subfloat[]{\includegraphics[width=0.40\columnwidth]{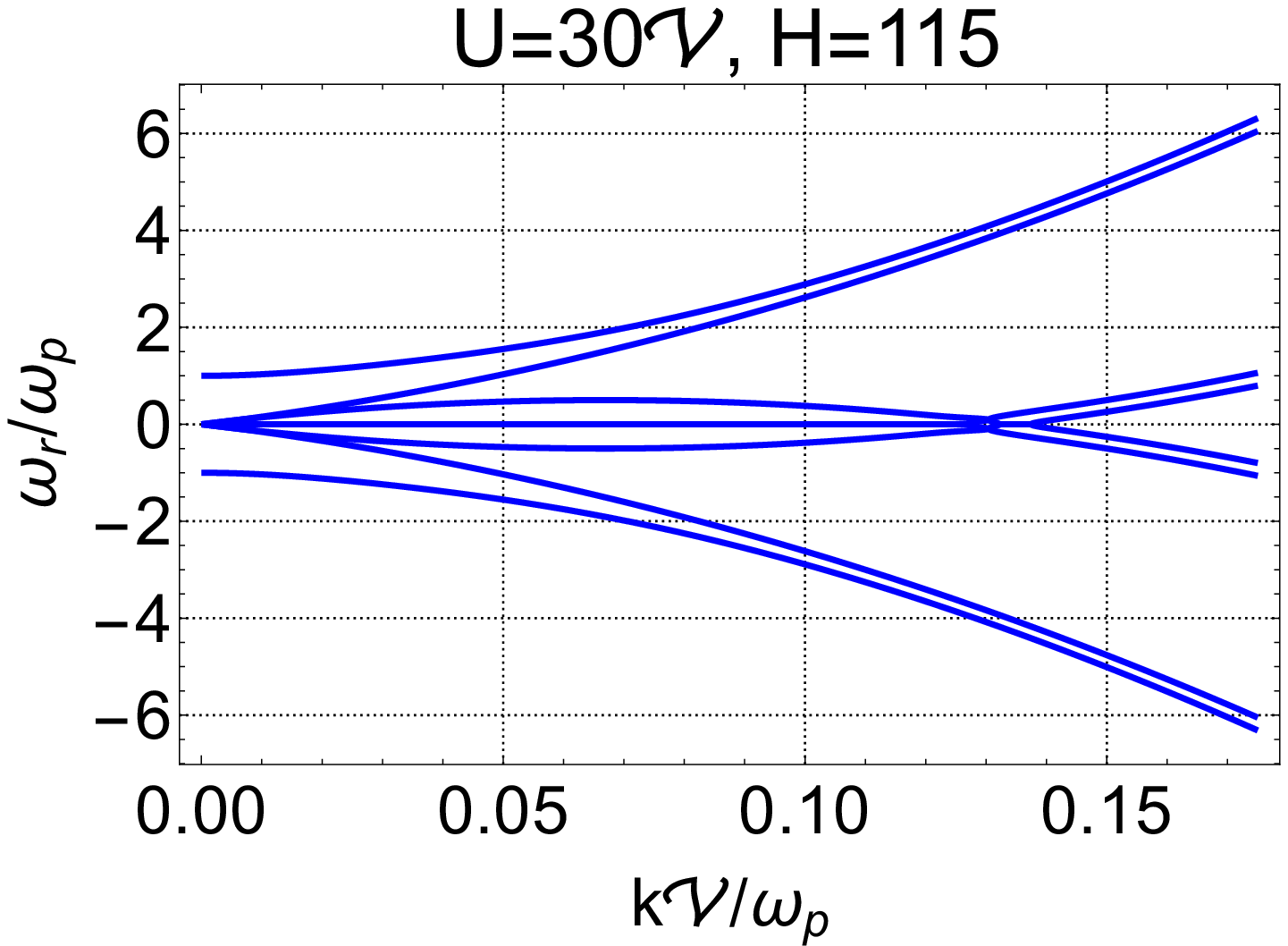}}\hspace{0.02\columnwidth}\subfloat[]{\includegraphics[width=0.40\columnwidth]{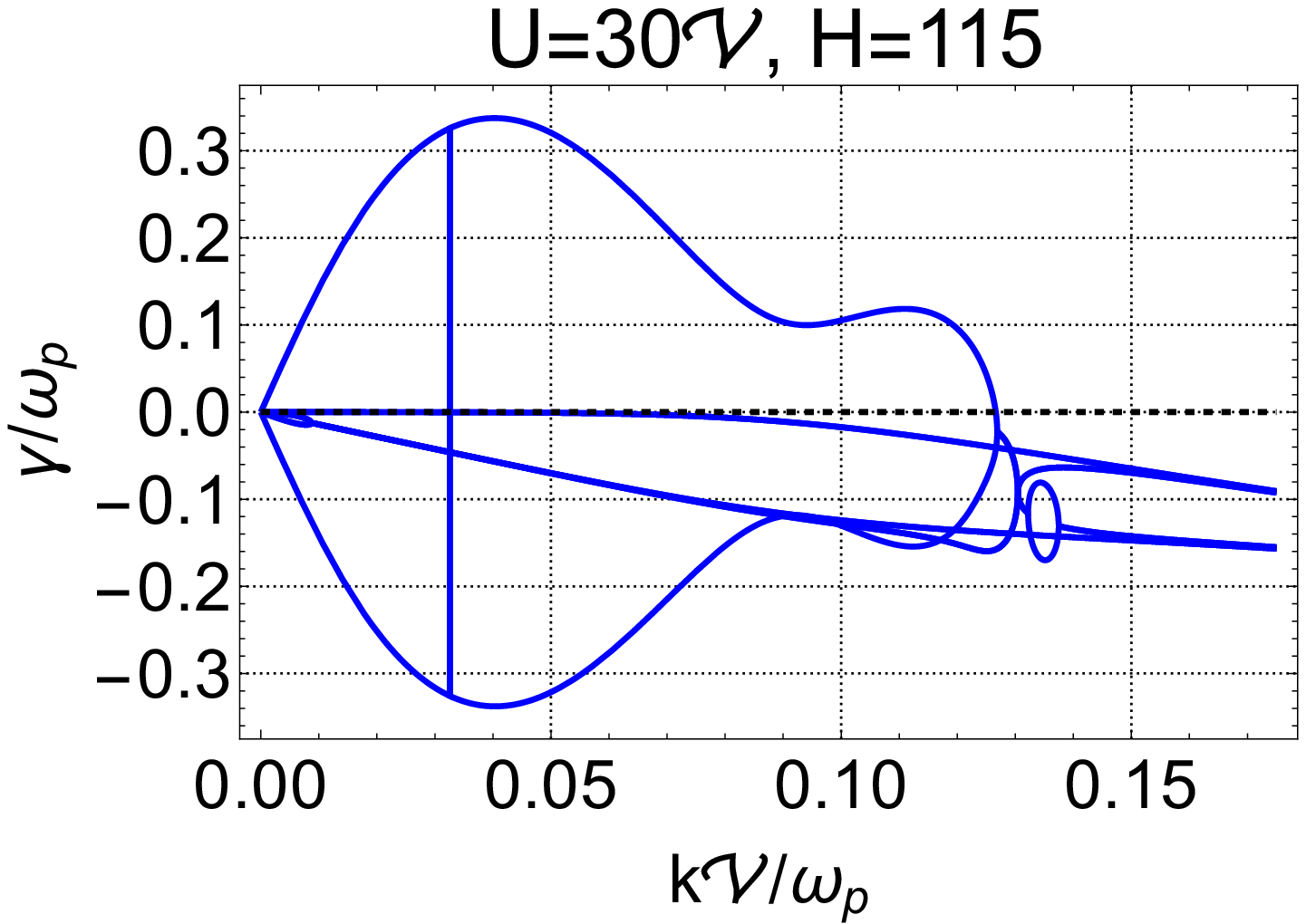}}\end{center}

\caption{Exact dispersion relations for symmetrical counter-drifting inverse-quartic
flat-top~$\chi_{4}$ distribution functions (Case 1). The behaviour
is generally similar to that of the Cauchy distributions as seen in
figures \ref{fig:cauchy_dispersion} and \ref{fig:cauchy2_instability},
with the existence of a pair of new modes which interact with the
unstable mode at the point where the second region of instability
terminates (i.e.~$k_{3}$). \label{fig:cauchyD_instability}}
\end{figure}

\begin{figure}
\begin{center}\subfloat[]{\includegraphics[width=0.40\columnwidth]{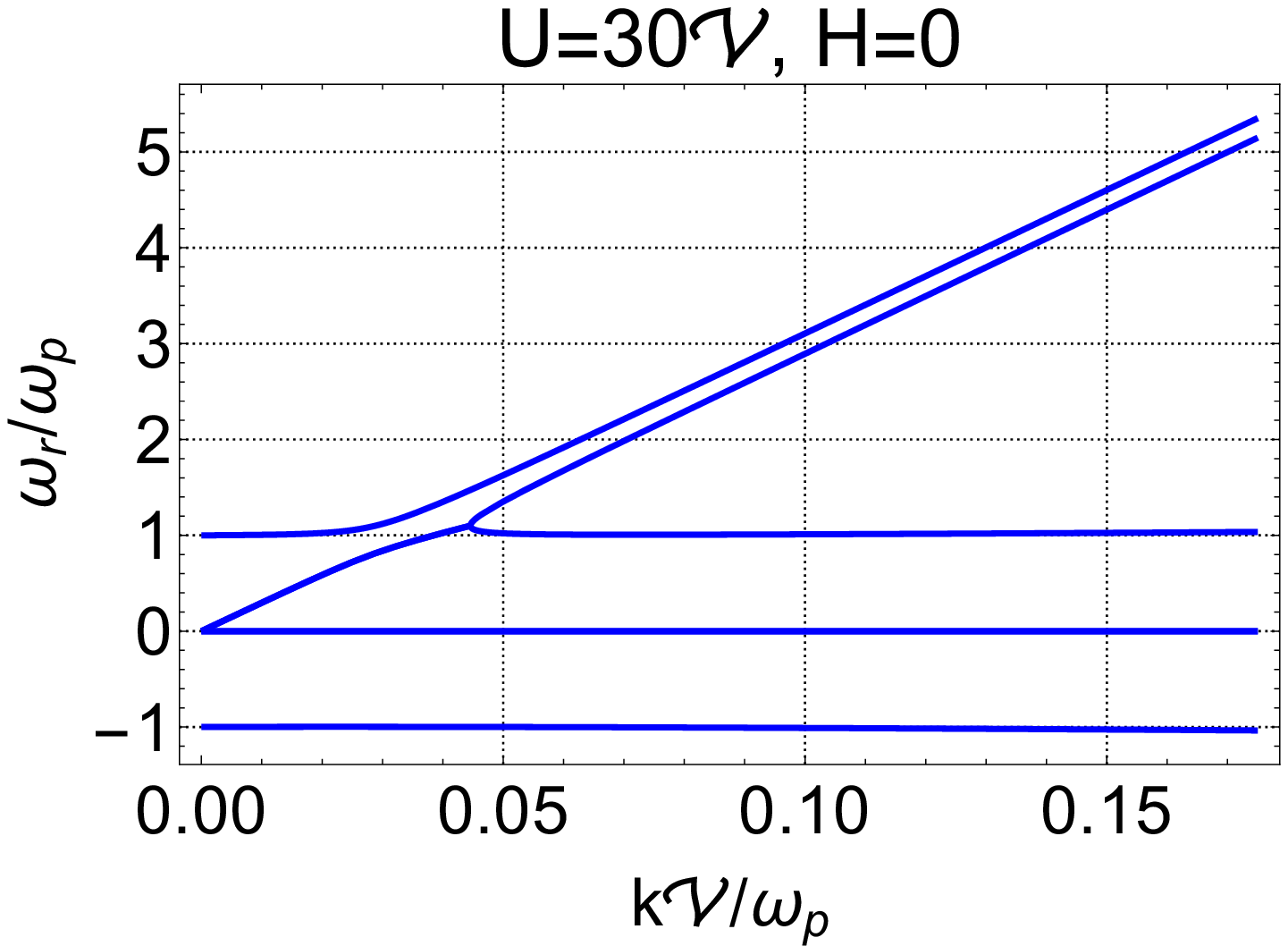}}\hspace{0.02\columnwidth}\subfloat[]{\includegraphics[width=0.40\columnwidth]{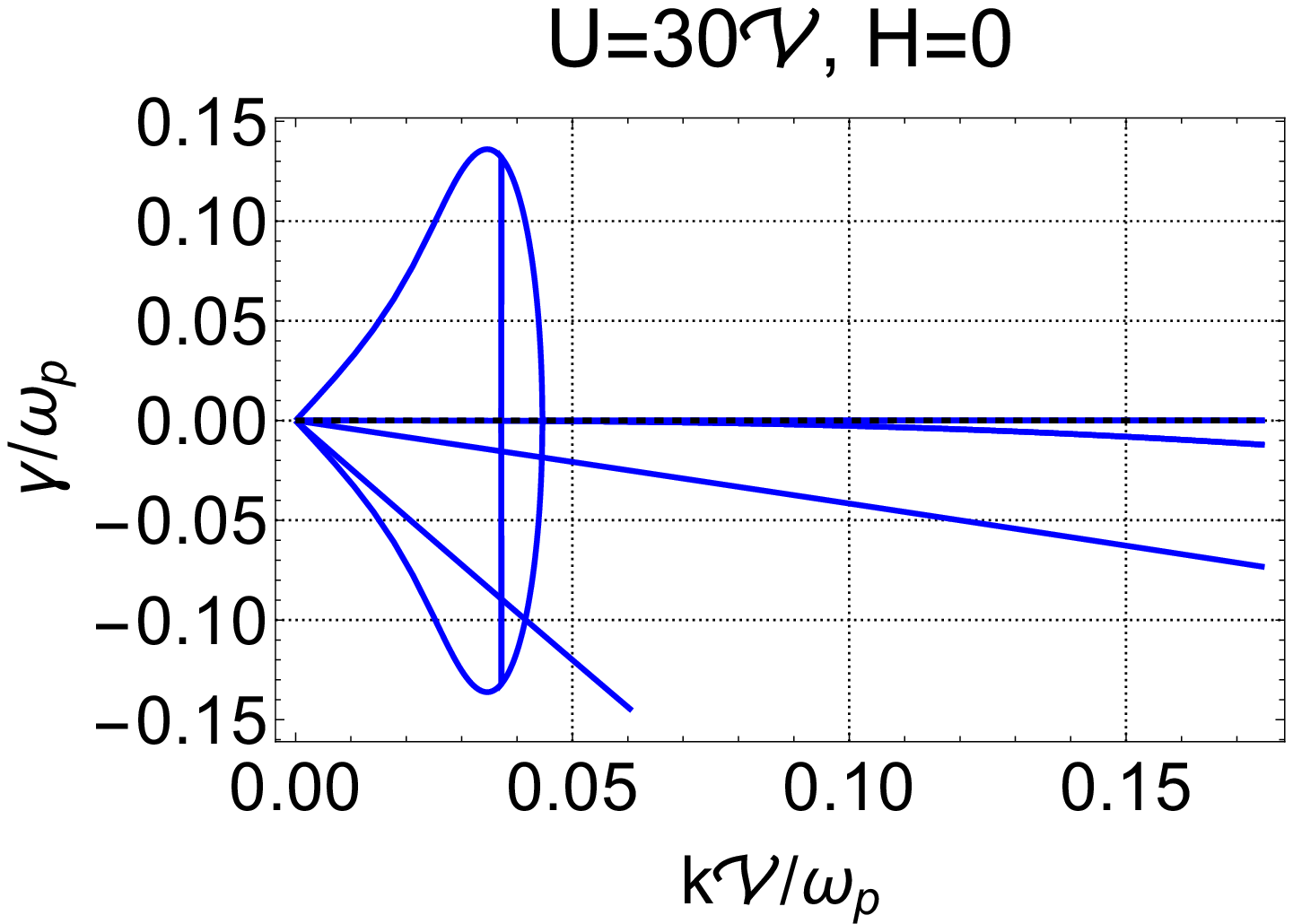}}

\vspace{1mm}
\subfloat[]{\includegraphics[width=0.40\columnwidth]{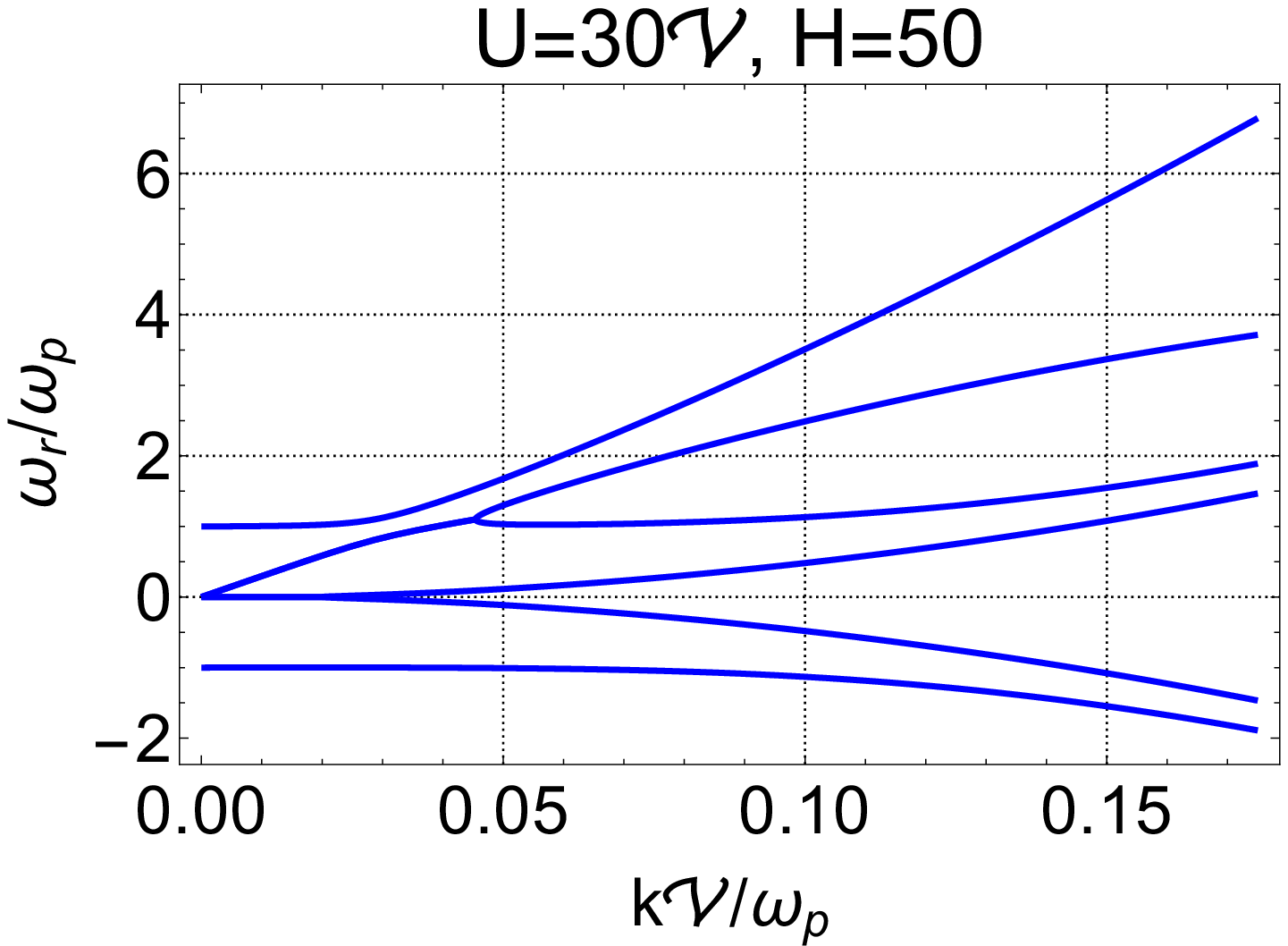}}\hspace{0.02\columnwidth}\subfloat[]{\includegraphics[width=0.40\columnwidth]{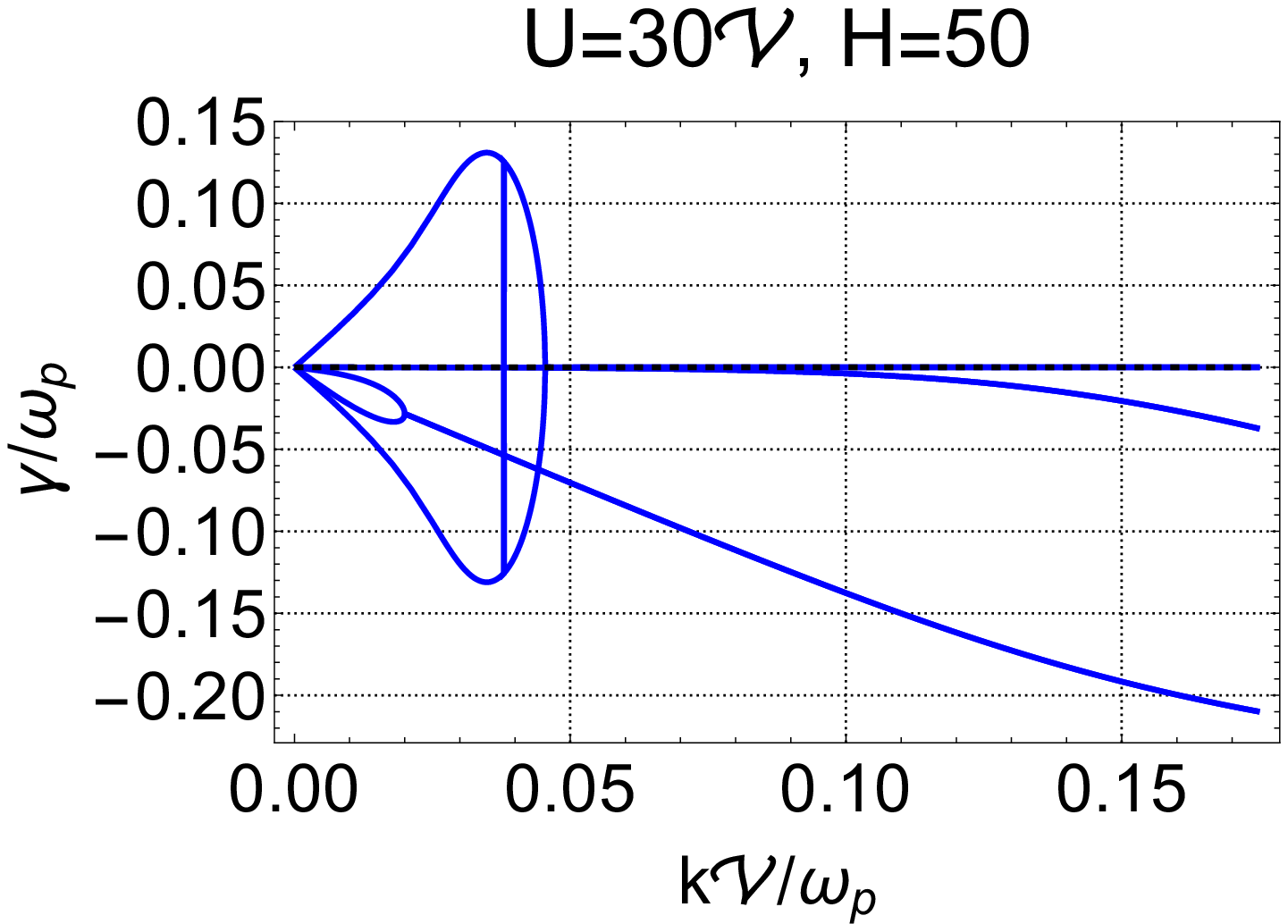}}

\vspace{1mm}
\subfloat[]{\includegraphics[width=0.40\columnwidth]{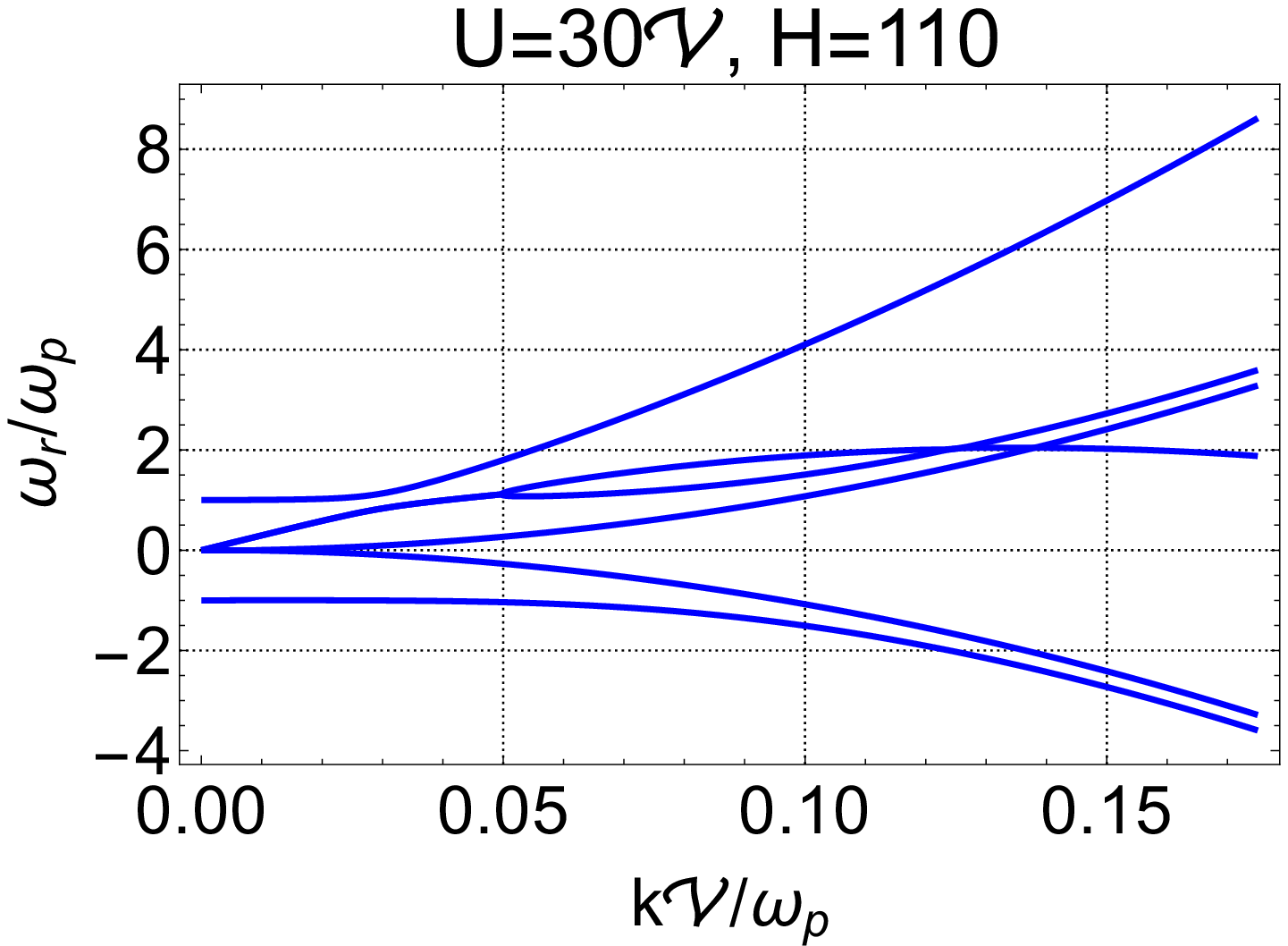}}\hspace{0.02\columnwidth}\subfloat[]{\includegraphics[width=0.40\columnwidth]{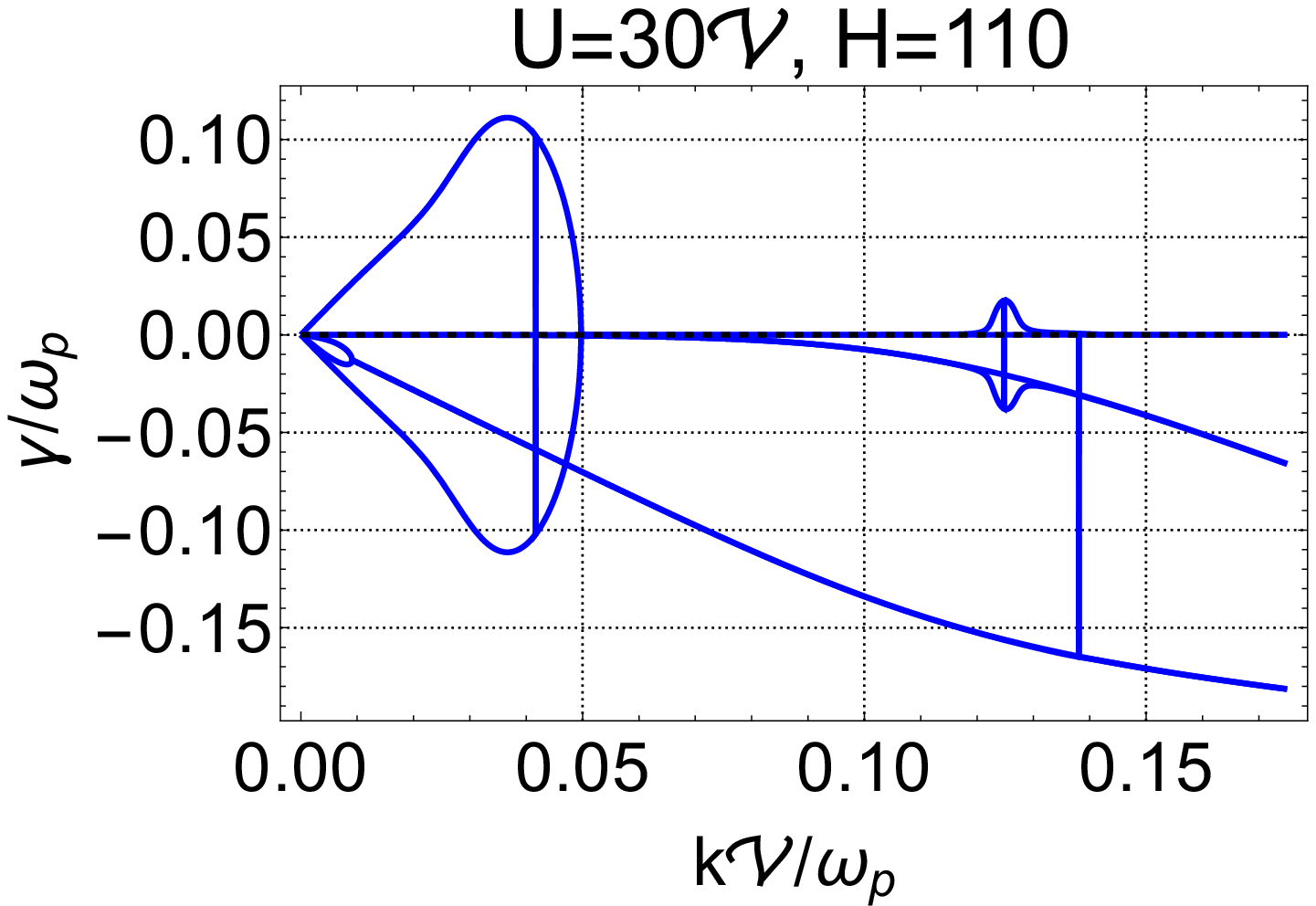}}\end{center}

\caption{Exact dispersion relations for primary inverse-quartic flat-top~$\chi_{4}$
distribution function population with low-density drifting beam with
$n=1/100$ (Case 2). The behaviour is consistent with the general
similarities with the dispersion relation of the squared Cauchy distribution
in Case 1 and the one-population case. \label{fig:cauchyD_instability-1}}
\end{figure}

The full dispersion relation for Case 2 is plotted in figure \ref{fig:cauchyD_instability-1}.
The behaviour here is very similar to that with the squared Cauchy
distribution. Note that the complex behaviour of the normal modes
evidenced in figures \ref{fig:cauchy2_instability} and \ref{fig:cauchyD_instability}
is not apparent for Case 2 in figures \ref{fig:cauchy2_instability-1}
and \ref{fig:cauchyD_instability-1}.

\begin{figure}
\begin{center}%
\begin{tabular}{c}
\includegraphics[width=0.5\columnwidth]{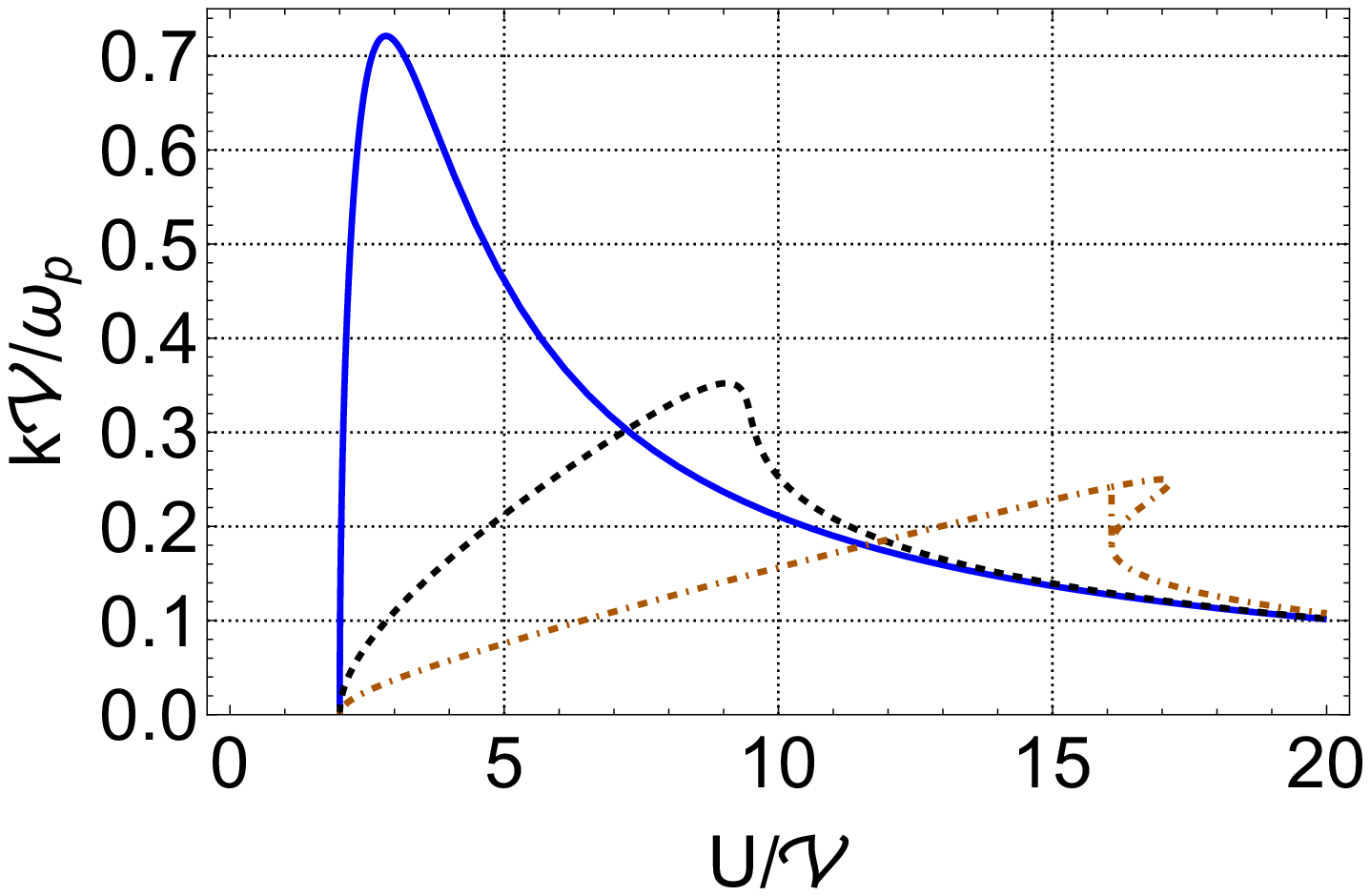}\tabularnewline
\end{tabular}\end{center}

\caption{Region of instability for counter-streaming inverse-quartic distributions
with $H=0$, (blue solid), $H=10$ (black dashed) and $H=30$ (orange
dot-dashed). As in the squared Cauchy system, the second region of
instability is seen to emerge for smaller~$H$ than the Cauchy system.
Aside from differing values of~$U_{{\rm crit}}$, the figure is almost
identical to the squared Cauchy instability region shown in figure
\ref{fig:cauchy2_inst_region}. \label{fig:cauchyD_inst_region}}
\end{figure}

\section{Discussion\label{sec:discussion}}

The phenomena studied in this paper overlap in part with other studies.
Delta function and Cauchy distributions have been used to study quantum
plasma instabilities by \citet{Haas2009a} and \citet{Haas2001},
respectively, while analysis of the two more complicated distribution
functions (squared Cauchy and inverse-quartic) in the context of quantum
plasmas has not until now appeared in the literature, to the best
of our knowledge. The dispersion relation derived by \citet{Haas2009a}
agrees with that of this paper, including the existence of the second
region of instability at larger wavenumbers~$k$. In \citet{Haas2001}
the solutions for the region of instability are in partial agreement
with ours, but that paper does not explicitly solve for the dispersion
relation and does not note the existence of the second region of instability,
which we have shown is still present for the Cauchy distribution and
its generalisations. Our paper is also relevant to the results in
\citet{Bonitz1993} in which streaming instabilities in degenerate
Fermi-Dirac plasmas are considered, but no second region of instability
is reported.

The quantum longitudinal dielectric function used in this paper has
been derived \citep{Klimontovich1960,Lindhard1954} under the assumptions
of immobile ions, ideal non-interacting electrons, non-relativistic
speeds, absence of a background magnetic field, and spinless
electrons. Despite these limitations, the resulting problem is rich
in complexity and has revealed interesting new physics. However, these
assumptions in principle may be relaxed in order to more obtain a
more comprehensive understanding and extend the region of applicability.
Several steps have already been taken to do this in the instance of
single-population plasmas. The more realistic case of Fermi-Dirac
electrons with an immobile ion background has been considered by \citet{Rightley2016}.
The introduction of ion motions through the inclusion of a classical
ion susceptibility with arbitrarily degenerate quantum electrons has
been carried out in the case by \citet{Melrose2010}, under the assumptions
of weak damping and long wavelength. The equation of motion for the
Wigner function including an arbitrary magnetic field has been derived
recently by \citet{Tyshetskiy2011}, and the dielectric tensor for
a uniformly  quantum plasma has been derived by other means by \citet{Canuto1972}.
Additionally, the filamentation instability in quantum plasmas has
been analyzed using a fluid approach by \citet{Bret2007} and using
a kinetic approach by \citet{Bret2011}. The quantum Weibel instability
has been studied by \citet{Haas2008} using a fluid approach, and
a kinetic approach has been taken by \citet{Haas2008a}. A framework
for modelling waves in relativistic quantum plasmas has been derived
by \citet{Melrose2013} and the quantum relativistic longitudinal
dielectric function has been presented by \citet{Melrose2010}, although
only the non-relativistic limit is considered in the bulk of the work.
Spin effects are of interest in sufficiently  quantum plasmas, and
their possibility has been considered in a quantum fluid or MHD framework
\citep{Marklund2007,Brodin2008} but there has also been progress
in spin kinetic theory \citep{Andreev2015,Andreev2016,Andreev2017,Iqbal2016,Marklund2010}.
Certain kinetic and fluid spin models are reviewed by \citet{Brodin2011}.
Nonlinear waves in quantum plasmas have mostly been studied as fluid
phenomena \citep{Shukla2010}, but the nonlinear regime of Landau
damping has been investigated by \citet{Daligault2014,Brodin2015}.

The results obtained in this paper are useful for the continued study
of instabilities in quantum plasmas by building a knowledge base.
Specifically, they provide a baseline for comparison of an in-progress
detailed study of streaming instabilities in plasmas with arbitrarily
degenerate Fermi-Dirac electrons employing numerical solutions of
the dispersion relation. Additionally, further general insight into
these phenomena can be gained by utilising the generalised Cauchy-type
distribution functions mentioned in section \ref{sec:distributions}.
In this manner the influence of complex poles in the distribution
function on the roots of the dielectric function can be systematically
analysed. This is relevant because of the presence of branch cuts
in the arbitrarily degenerate Fermi-Dirac distribution function, as
discussed by \citet{Vladimirov2011}.

Aside from the issue of obtaining a correct theoretical understanding
of Landau damping and streaming instabilities in degenerate plasmas,
the topic of quantum linear waves and instabilities is relevant to
studies of warm dense matter, white dwarf interiors \citep{Akbari-Moghanjoughi2013},
and solid state plasmas, in which the dielectric properties of the
electrons are of importance. Furthermore, the quiescent x-ray emission
of magnetars may be due to the dissipation of magnetospheric currents
penetrating into the upper layers of the neutron star's surface \citep{Beloborodov2007,Beloborodov2009},
where the current-carrying energetic electron-positron pairs deposit
their energy into the layer by exciting Langmuir turbulence \citep{Thompson2005,Uzdensky2014}.

\section{Conclusions\label{sec:conclusions}}

In summary, in this work we have used the established quantum longitudinal
susceptibility to study the complex dispersion relation for electrostatic
waves in plasmas consisting of one and two populations of electrons
with uniform stationary ion background. We have considered four Wigner
distribution functions that lend themselves to convenient analysis:
the delta-function distribution, the Cauchy distribution, the squared
Cauchy distribution, and the inverse-quartic distribution. Other studies
have established dispersion relations for instabilities in two-component
plasmas with both components having either delta-function \citep{Haas2009}
or Cauchy \citep{Haas2001} distributions, but have considered only
the case of symmetrical counter-propagating electron populations with
equal particle densities. This work extends these results by additionally
considering the case of a primary electron population impinged upon
by a delta-function beam of low density.

We have found that the normal mode structure in a given plasma becomes
increasingly complex for more complicated distribution functions,
but that for each distribution function considered, there is a normal
plasmon mode, and a single unstable mode at small wavenumbers~$k$
for two populations separated by sufficient drift velocity. However,
unlike in the classical situation, there can exist a second region
of instability for larger~$k$ that is due entirely to quantum effects.
This has been initially noted for the case of two counter-propagating
cold beams by \citet{Haas2009a}, and we have shown that this effect
carries over to distribution functions with finite temperatures. Additionally,
we have shown that the boundary of the region of instability at small~$k$
for large drift velocity~$U$ is affected by quantum effects at fifth
order in $\mathcal{V}/U$, where~$\mathcal{V}$ is the thermal velocity.
The results for the case with a cold beam penetrating a hot plasma
with an analytically convenient distribution function are generally
similar for the Cauchy, squared Cauchy, and inverse-quartic distribution
functions. For each type of distribution function, the existence of
one region of instability for small or zero quantum parameter~$H$,
and of a second unstable window for sufficiently large~$H$, are
preserved.

While our analysis accounts for the effects of quantum recoil, the
analytically convenient distribution functions used in this paper
do not account for quantum statistics. A more realistic description
of quantum plasma instabilities would include a Fermi-Dirac (FD) background
distribution function. The intent of this paper is to pave the way
for studies of instabilities in FD plasmas, in which analytical results
will be limited by the presence of branch cuts of the FD distribution
function in complex-velocity space. For FD plasmas, it is therefore
necessary to obtain the general dispersion relation using numerical
methods, as has been performed for single-population plasmas in our
previous work \citep{Rightley2016}. This has been carried out in
tandem with the present study, with results to be published in the
near future. Ultimately, a complete understanding of Landau damping
and streaming instabilities in degenerate electron plasmas will have
applications to phenomena which are sensitive to the dielectric properties
of the electrons in environments such as warm dense matter, dense
astrophysical plasmas, and solid state plasmas. Furthermore, an understanding
of nonlinear physics in these systems will be facilitated by a solid
foundation in the linear theory. Finally, quantum effects introduce
a rich complexity to the topic of linear waves and instabilities in
plasmas, and have opened up new avenues of research in this direction.

\bibliographystyle{apsrev4-1}
\bibliography{MasterReferences}

\begin{thebibliography}{51}%
\makeatletter
\providecommand \@ifxundefined [1]{%
 \@ifx{#1\undefined}
}%
\providecommand \@ifnum [1]{%
 \ifnum #1\expandafter \@firstoftwo
 \else \expandafter \@secondoftwo
 \fi
}%
\providecommand \@ifx [1]{%
 \ifx #1\expandafter \@firstoftwo
 \else \expandafter \@secondoftwo
 \fi
}%
\providecommand \natexlab [1]{#1}%
\providecommand \enquote  [1]{``#1''}%
\providecommand \bibnamefont  [1]{#1}%
\providecommand \bibfnamefont [1]{#1}%
\providecommand \citenamefont [1]{#1}%
\providecommand \href@noop [0]{\@secondoftwo}%
\providecommand \href [0]{\begingroup \@sanitize@url \@href}%
\providecommand \@href[1]{\@@startlink{#1}\@@href}%
\providecommand \@@href[1]{\endgroup#1\@@endlink}%
\providecommand \@sanitize@url [0]{\catcode `\\12\catcode `\$12\catcode
  `\&12\catcode `\#12\catcode `\^12\catcode `\_12\catcode `\%12\relax}%
\providecommand \@@startlink[1]{}%
\providecommand \@@endlink[0]{}%
\providecommand \url  [0]{\begingroup\@sanitize@url \@url }%
\providecommand \@url [1]{\endgroup\@href {#1}{\urlprefix }}%
\providecommand \urlprefix  [0]{URL }%
\providecommand \Eprint [0]{\href }%
\providecommand \doibase [0]{http://dx.doi.org/}%
\providecommand \selectlanguage [0]{\@gobble}%
\providecommand \bibinfo  [0]{\@secondoftwo}%
\providecommand \bibfield  [0]{\@secondoftwo}%
\providecommand \translation [1]{[#1]}%
\providecommand \BibitemOpen [0]{}%
\providecommand \bibitemStop [0]{}%
\providecommand \bibitemNoStop [0]{.\EOS\space}%
\providecommand \EOS [0]{\spacefactor3000\relax}%
\providecommand \BibitemShut  [1]{\csname bibitem#1\endcsname}%
\let\auto@bib@innerbib\@empty
\bibitem [{\citenamefont {{Bohm}}\ and\ \citenamefont
  {{Pines}}(1950)}]{Bohm1950}%
  \BibitemOpen
  \bibfield  {author} {\bibinfo {author} {\bibfnamefont {D.}~\bibnamefont
  {{Bohm}}}\ and\ \bibinfo {author} {\bibfnamefont {D.}~\bibnamefont
  {{Pines}}},\ }\href {\doibase 10.1103/PhysRev.80.903.2} {\bibfield  {journal}
  {\bibinfo  {journal} {Phys. Rev.}\ }\textbf {\bibinfo {volume} {80}},\
  \bibinfo {pages} {903} (\bibinfo {year} {1950})}\BibitemShut {NoStop}%
\bibitem [{\citenamefont {{Silin}}(1958)}]{Silin1958}%
  \BibitemOpen
  \bibfield  {author} {\bibinfo {author} {\bibfnamefont {V.~P.}\ \bibnamefont
  {{Silin}}},\ }\href@noop {} {\bibfield  {journal} {\bibinfo  {journal}
  {Soviet Journal of Experimental and Theoretical Physics}\ }\textbf {\bibinfo
  {volume} {6}},\ \bibinfo {pages} {387} (\bibinfo {year} {1958})}\BibitemShut
  {NoStop}%
\bibitem [{\citenamefont {Pines}(1961)}]{Pines1961}%
  \BibitemOpen
  \bibfield  {author} {\bibinfo {author} {\bibfnamefont {D.}~\bibnamefont
  {Pines}},\ }\href {http://stacks.iop.org/0368-3281/2/i=1/a=301} {\bibfield
  {journal} {\bibinfo  {journal} {J. Nucl. Energy. Part C}\ }\textbf {\bibinfo
  {volume} {2}},\ \bibinfo {pages} {5} (\bibinfo {year} {1961})}\BibitemShut
  {NoStop}%
\bibitem [{\citenamefont {Marklund}\ \emph {et~al.}(2008)\citenamefont
  {Marklund}, \citenamefont {Brodin}, \citenamefont {Stenflo},\ and\
  \citenamefont {Liu}}]{Marklund2008}%
  \BibitemOpen
  \bibfield  {author} {\bibinfo {author} {\bibfnamefont {M.}~\bibnamefont
  {Marklund}}, \bibinfo {author} {\bibfnamefont {G.}~\bibnamefont {Brodin}},
  \bibinfo {author} {\bibfnamefont {L.}~\bibnamefont {Stenflo}}, \ and\
  \bibinfo {author} {\bibfnamefont {C.~S.}\ \bibnamefont {Liu}},\ }\href
  {http://stacks.iop.org/0295-5075/84/i=1/a=17006} {\bibfield  {journal}
  {\bibinfo  {journal} {Europhys. Lett.}\ }\textbf {\bibinfo {volume} {84}},\
  \bibinfo {pages} {17006} (\bibinfo {year} {2008})}\BibitemShut {NoStop}%
\bibitem [{\citenamefont {{Shukla}}\ and\ \citenamefont
  {{Eliasson}}(2011)}]{Shukla2011}%
  \BibitemOpen
  \bibfield  {author} {\bibinfo {author} {\bibfnamefont {P.~K.}\ \bibnamefont
  {{Shukla}}}\ and\ \bibinfo {author} {\bibfnamefont {B.}~\bibnamefont
  {{Eliasson}}},\ }\href {\doibase 10.1103/RevModPhys.83.885} {\bibfield
  {journal} {\bibinfo  {journal} {Rev. Mod. Phys.}\ }\textbf {\bibinfo {volume}
  {83}},\ \bibinfo {pages} {885} (\bibinfo {year} {2011})},\ \Eprint
  {http://arxiv.org/abs/1009.5215} {arXiv:1009.5215 [physics.plasm-ph]}
  \BibitemShut {NoStop}%
\bibitem [{\citenamefont {{Tyshetskiy}}\ \emph {et~al.}(2011)\citenamefont
  {{Tyshetskiy}}, \citenamefont {{Vladimirov}},\ and\ \citenamefont
  {{Kompaneets}}}]{Tyshetskiy2011}%
  \BibitemOpen
  \bibfield  {author} {\bibinfo {author} {\bibfnamefont {Y.}~\bibnamefont
  {{Tyshetskiy}}}, \bibinfo {author} {\bibfnamefont {S.~V.}\ \bibnamefont
  {{Vladimirov}}}, \ and\ \bibinfo {author} {\bibfnamefont {R.}~\bibnamefont
  {{Kompaneets}}},\ }\href {\doibase 10.1063/1.3659025} {\bibfield  {journal}
  {\bibinfo  {journal} {Phys. Plasmas}\ }\textbf {\bibinfo {volume} {18}},\
  \bibinfo {eid} {112104} (\bibinfo {year} {2011})},\ \Eprint
  {http://arxiv.org/abs/1108.0988} {arXiv:1108.0988 [physics.plasm-ph]}
  \BibitemShut {NoStop}%
\bibitem [{\citenamefont {Vladimirov}\ and\ \citenamefont
  {Tyshetskiy}(2011)}]{Vladimirov2011}%
  \BibitemOpen
  \bibfield  {author} {\bibinfo {author} {\bibfnamefont {S.~V.}\ \bibnamefont
  {Vladimirov}}\ and\ \bibinfo {author} {\bibfnamefont {Y.~O.}\ \bibnamefont
  {Tyshetskiy}},\ }\href {http://stacks.iop.org/1063-7869/54/i=12/a=A03}
  {\bibfield  {journal} {\bibinfo  {journal} {Physics-Uspekhi}\ }\textbf
  {\bibinfo {volume} {54}},\ \bibinfo {pages} {1243} (\bibinfo {year}
  {2011})}\BibitemShut {NoStop}%
\bibitem [{\citenamefont {{Glenzer}}\ \emph {et~al.}(2007)\citenamefont
  {{Glenzer}}, \citenamefont {{Landen}}, \citenamefont {{Neumayer}},
  \citenamefont {{Lee}}, \citenamefont {{Widmann}}, \citenamefont {{Pollaine}},
  \citenamefont {{Wallace}}, \citenamefont {{Gregori}}, \citenamefont
  {{H{\"o}ll}}, \citenamefont {{Bornath}}, \citenamefont {{Thiele}},
  \citenamefont {{Schwarz}}, \citenamefont {{Kraeft}},\ and\ \citenamefont
  {{Redmer}}}]{Glenzer2007}%
  \BibitemOpen
  \bibfield  {author} {\bibinfo {author} {\bibfnamefont {S.~H.}\ \bibnamefont
  {{Glenzer}}}, \bibinfo {author} {\bibfnamefont {O.~L.}\ \bibnamefont
  {{Landen}}}, \bibinfo {author} {\bibfnamefont {P.}~\bibnamefont
  {{Neumayer}}}, \bibinfo {author} {\bibfnamefont {R.~W.}\ \bibnamefont
  {{Lee}}}, \bibinfo {author} {\bibfnamefont {K.}~\bibnamefont {{Widmann}}},
  \bibinfo {author} {\bibfnamefont {S.~W.}\ \bibnamefont {{Pollaine}}},
  \bibinfo {author} {\bibfnamefont {R.~J.}\ \bibnamefont {{Wallace}}}, \bibinfo
  {author} {\bibfnamefont {G.}~\bibnamefont {{Gregori}}}, \bibinfo {author}
  {\bibfnamefont {A.}~\bibnamefont {{H{\"o}ll}}}, \bibinfo {author}
  {\bibfnamefont {T.}~\bibnamefont {{Bornath}}}, \bibinfo {author}
  {\bibfnamefont {R.}~\bibnamefont {{Thiele}}}, \bibinfo {author}
  {\bibfnamefont {V.}~\bibnamefont {{Schwarz}}}, \bibinfo {author}
  {\bibfnamefont {W.-D.}\ \bibnamefont {{Kraeft}}}, \ and\ \bibinfo {author}
  {\bibfnamefont {R.}~\bibnamefont {{Redmer}}},\ }\href {\doibase
  10.1103/PhysRevLett.98.065002} {\bibfield  {journal} {\bibinfo  {journal}
  {Phys. Rev. Lett.}\ }\textbf {\bibinfo {volume} {98}},\ \bibinfo {eid}
  {065002} (\bibinfo {year} {2007})}\BibitemShut {NoStop}%
\bibitem [{\citenamefont {{Chabrier}}\ \emph {et~al.}(2002)\citenamefont
  {{Chabrier}}, \citenamefont {{Douchin}},\ and\ \citenamefont
  {{Potekhin}}}]{Chabrier2002}%
  \BibitemOpen
  \bibfield  {author} {\bibinfo {author} {\bibfnamefont {G.}~\bibnamefont
  {{Chabrier}}}, \bibinfo {author} {\bibfnamefont {F.}~\bibnamefont
  {{Douchin}}}, \ and\ \bibinfo {author} {\bibfnamefont {A.~Y.}\ \bibnamefont
  {{Potekhin}}},\ }\href {\doibase 10.1088/0953-8984/14/40/307} {\bibfield
  {journal} {\bibinfo  {journal} {Journal of Physics Condensed Matter}\
  }\textbf {\bibinfo {volume} {14}},\ \bibinfo {pages} {9133} (\bibinfo {year}
  {2002})},\ \Eprint {http://arxiv.org/abs/physics/0211089} {physics/0211089}
  \BibitemShut {NoStop}%
\bibitem [{\citenamefont {{Bohm}}\ and\ \citenamefont
  {{Pines}}(1953)}]{Bohm1953}%
  \BibitemOpen
  \bibfield  {author} {\bibinfo {author} {\bibfnamefont {D.}~\bibnamefont
  {{Bohm}}}\ and\ \bibinfo {author} {\bibfnamefont {D.}~\bibnamefont
  {{Pines}}},\ }\href {\doibase 10.1103/PhysRev.92.609} {\bibfield  {journal}
  {\bibinfo  {journal} {Phys. Rev.}\ }\textbf {\bibinfo {volume} {92}},\
  \bibinfo {pages} {609} (\bibinfo {year} {1953})}\BibitemShut {NoStop}%
\bibitem [{\citenamefont {Pines}\ and\ \citenamefont
  {Schrieffer}(1962)}]{Pines1962}%
  \BibitemOpen
  \bibfield  {author} {\bibinfo {author} {\bibfnamefont {D.}~\bibnamefont
  {Pines}}\ and\ \bibinfo {author} {\bibfnamefont {J.~R.}\ \bibnamefont
  {Schrieffer}},\ }\href {\doibase 10.1103/PhysRev.125.804} {\bibfield
  {journal} {\bibinfo  {journal} {Phys. Rev.}\ }\textbf {\bibinfo {volume}
  {125}},\ \bibinfo {pages} {804} (\bibinfo {year} {1962})}\BibitemShut
  {NoStop}%
\bibitem [{\citenamefont {Bonitz}\ \emph
  {et~al.}(1993{\natexlab{a}})\citenamefont {Bonitz}, \citenamefont {Binder},
  \citenamefont {Scott}, \citenamefont {Koch},\ and\ \citenamefont
  {Kremp}}]{Bonitz1993b}%
  \BibitemOpen
  \bibfield  {author} {\bibinfo {author} {\bibfnamefont {M.}~\bibnamefont
  {Bonitz}}, \bibinfo {author} {\bibfnamefont {R.}~\bibnamefont {Binder}},
  \bibinfo {author} {\bibfnamefont {D.~C.}\ \bibnamefont {Scott}}, \bibinfo
  {author} {\bibfnamefont {S.~W.}\ \bibnamefont {Koch}}, \ and\ \bibinfo
  {author} {\bibfnamefont {D.}~\bibnamefont {Kremp}},\ }\href {\doibase
  10.1002/ctpp.2150330528} {\bibfield  {journal} {\bibinfo  {journal}
  {Contributions to Plasma Physics}\ }\textbf {\bibinfo {volume} {33}},\
  \bibinfo {pages} {536} (\bibinfo {year} {1993}{\natexlab{a}})}\BibitemShut
  {NoStop}%
\bibitem [{\citenamefont {{Haas}}\ \emph {et~al.}(2001)\citenamefont {{Haas}},
  \citenamefont {{Manfredi}},\ and\ \citenamefont {{Goedert}}}]{Haas2001}%
  \BibitemOpen
  \bibfield  {author} {\bibinfo {author} {\bibfnamefont {F.}~\bibnamefont
  {{Haas}}}, \bibinfo {author} {\bibfnamefont {G.}~\bibnamefont {{Manfredi}}},
  \ and\ \bibinfo {author} {\bibfnamefont {J.}~\bibnamefont {{Goedert}}},\
  }\href {\doibase 10.1103/PhysRevE.64.026413} {\bibfield  {journal} {\bibinfo
  {journal} {Phys. Rev. E}\ }\textbf {\bibinfo {volume} {64}},\ \bibinfo {eid}
  {026413} (\bibinfo {year} {2001})},\ \Eprint
  {http://arxiv.org/abs/physics/0211076} {physics/0211076} \BibitemShut
  {NoStop}%
\bibitem [{\citenamefont {{Haas}}\ \emph {et~al.}(2009)\citenamefont {{Haas}},
  \citenamefont {{Bret}},\ and\ \citenamefont {{Shukla}}}]{Haas2009a}%
  \BibitemOpen
  \bibfield  {author} {\bibinfo {author} {\bibfnamefont {F.}~\bibnamefont
  {{Haas}}}, \bibinfo {author} {\bibfnamefont {A.}~\bibnamefont {{Bret}}}, \
  and\ \bibinfo {author} {\bibfnamefont {P.~K.}\ \bibnamefont {{Shukla}}},\
  }\href {\doibase 10.1103/PhysRevE.80.066407} {\bibfield  {journal} {\bibinfo
  {journal} {Phys. Rev. E}\ }\textbf {\bibinfo {volume} {80}},\ \bibinfo {eid}
  {066407} (\bibinfo {year} {2009})},\ \Eprint {http://arxiv.org/abs/0907.3061}
  {arXiv:0907.3061 [physics.plasm-ph]} \BibitemShut {NoStop}%
\bibitem [{\citenamefont {Melrose}\ and\ \citenamefont
  {Mushtaq}(2010{\natexlab{a}})}]{Melrose2010}%
  \BibitemOpen
  \bibfield  {author} {\bibinfo {author} {\bibfnamefont {D.~B.}\ \bibnamefont
  {Melrose}}\ and\ \bibinfo {author} {\bibfnamefont {A.}~\bibnamefont
  {Mushtaq}},\ }\href {\doibase 10.1103/PhysRevE.82.056402} {\bibfield
  {journal} {\bibinfo  {journal} {Phys. Rev. E}\ }\textbf {\bibinfo {volume}
  {82}},\ \bibinfo {pages} {056402} (\bibinfo {year}
  {2010}{\natexlab{a}})}\BibitemShut {NoStop}%
\bibitem [{\citenamefont {Bonitz}\ \emph {et~al.}(1994)\citenamefont {Bonitz},
  \citenamefont {Binder}, \citenamefont {Scott}, \citenamefont {Koch},\ and\
  \citenamefont {Kremp}}]{Bonitz1994}%
  \BibitemOpen
  \bibfield  {author} {\bibinfo {author} {\bibfnamefont {M.}~\bibnamefont
  {Bonitz}}, \bibinfo {author} {\bibfnamefont {R.}~\bibnamefont {Binder}},
  \bibinfo {author} {\bibfnamefont {D.~C.}\ \bibnamefont {Scott}}, \bibinfo
  {author} {\bibfnamefont {S.~W.}\ \bibnamefont {Koch}}, \ and\ \bibinfo
  {author} {\bibfnamefont {D.}~\bibnamefont {Kremp}},\ }\href@noop {}
  {\bibfield  {journal} {\bibinfo  {journal} {Phys. Rev. E}\ }\textbf {\bibinfo
  {volume} {49}} (\bibinfo {year} {1994})}\BibitemShut {NoStop}%
\bibitem [{\citenamefont {{Eliasson}}\ and\ \citenamefont
  {{Shukla}}(2010)}]{Eliasson2010}%
  \BibitemOpen
  \bibfield  {author} {\bibinfo {author} {\bibfnamefont {B.}~\bibnamefont
  {{Eliasson}}}\ and\ \bibinfo {author} {\bibfnamefont {P.~K.}\ \bibnamefont
  {{Shukla}}},\ }\href {\doibase 10.1017/S0022377809990316} {\bibfield
  {journal} {\bibinfo  {journal} {J. Plasma Phys.}\ }\textbf {\bibinfo {volume}
  {76}},\ \bibinfo {pages} {7} (\bibinfo {year} {2010})},\ \Eprint
  {http://arxiv.org/abs/0911.4594} {arXiv:0911.4594 [physics.plasm-ph]}
  \BibitemShut {NoStop}%
\bibitem [{\citenamefont {{Krivitskii}}\ and\ \citenamefont
  {{Vladimirov}}(1991)}]{Krivitskii1991}%
  \BibitemOpen
  \bibfield  {author} {\bibinfo {author} {\bibfnamefont {V.~S.}\ \bibnamefont
  {{Krivitskii}}}\ and\ \bibinfo {author} {\bibfnamefont {S.~V.}\ \bibnamefont
  {{Vladimirov}}},\ }\href@noop {} {\bibfield  {journal} {\bibinfo  {journal}
  {Zhurnal Eksperimentalnoi i Teoreticheskoi Fiziki}\ }\textbf {\bibinfo
  {volume} {100}},\ \bibinfo {pages} {1483} (\bibinfo {year}
  {1991})}\BibitemShut {NoStop}%
\bibitem [{\citenamefont {{Rightley}}\ and\ \citenamefont
  {{Uzdensky}}(2016)}]{Rightley2016}%
  \BibitemOpen
  \bibfield  {author} {\bibinfo {author} {\bibfnamefont {S.}~\bibnamefont
  {{Rightley}}}\ and\ \bibinfo {author} {\bibfnamefont {D.}~\bibnamefont
  {{Uzdensky}}},\ }\href {\doibase 10.1063/1.4943870} {\bibfield  {journal}
  {\bibinfo  {journal} {Phys. Plasmas}\ }\textbf {\bibinfo {volume} {23}},\
  \bibinfo {eid} {030702} (\bibinfo {year} {2016})},\ \Eprint
  {http://arxiv.org/abs/1506.05494} {arXiv:1506.05494 [physics.plasm-ph]}
  \BibitemShut {NoStop}%
\bibitem [{\citenamefont {{Stix}}(1992)}]{Stix1992}%
  \BibitemOpen
  \bibfield  {author} {\bibinfo {author} {\bibfnamefont {T.~H.}\ \bibnamefont
  {{Stix}}},\ }\href@noop {} {\emph {\bibinfo {title} {Waves in plasmas , by
  Stix, Thomas Howard.; Stix, Thomas Howard.~ New York : American Institute of
  Physics, c1992.}}}\ (\bibinfo {year} {1992})\BibitemShut {NoStop}%
\bibitem [{\citenamefont {Bonitz}\ \emph
  {et~al.}(1993{\natexlab{b}})\citenamefont {Bonitz}, \citenamefont {Binder},
  \citenamefont {Scott}, \citenamefont {Koch},\ and\ \citenamefont
  {Kremp}}]{Bonitz1993}%
  \BibitemOpen
  \bibfield  {author} {\bibinfo {author} {\bibfnamefont {M.}~\bibnamefont
  {Bonitz}}, \bibinfo {author} {\bibfnamefont {R.}~\bibnamefont {Binder}},
  \bibinfo {author} {\bibfnamefont {D.~C.}\ \bibnamefont {Scott}}, \bibinfo
  {author} {\bibfnamefont {S.~W.}\ \bibnamefont {Koch}}, \ and\ \bibinfo
  {author} {\bibfnamefont {D.}~\bibnamefont {Kremp}},\ }\href {\doibase
  10.1002/ctpp.2150330528} {\bibfield  {journal} {\bibinfo  {journal}
  {Contributions to Plasma Physics}\ }\textbf {\bibinfo {volume} {33}},\
  \bibinfo {pages} {536} (\bibinfo {year} {1993}{\natexlab{b}})}\BibitemShut
  {NoStop}%
\bibitem [{\citenamefont {{Lapuerta}}\ and\ \citenamefont
  {{Ahedo}}(2002)}]{Lapuerta2002}%
  \BibitemOpen
  \bibfield  {author} {\bibinfo {author} {\bibfnamefont {V.}~\bibnamefont
  {{Lapuerta}}}\ and\ \bibinfo {author} {\bibfnamefont {E.}~\bibnamefont
  {{Ahedo}}},\ }\href {\doibase 10.1063/1.1464893} {\bibfield  {journal}
  {\bibinfo  {journal} {Phys. Plasmas}\ }\textbf {\bibinfo {volume} {9}},\
  \bibinfo {pages} {1513} (\bibinfo {year} {2002})}\BibitemShut {NoStop}%
\bibitem [{\citenamefont {Wigner}(1932)}]{Wigner1932}%
  \BibitemOpen
  \bibfield  {author} {\bibinfo {author} {\bibfnamefont {E.}~\bibnamefont
  {Wigner}},\ }\href {\doibase 10.1103/PhysRev.40.749} {\bibfield  {journal}
  {\bibinfo  {journal} {Phys. Rev.}\ }\textbf {\bibinfo {volume} {40}},\
  \bibinfo {pages} {749} (\bibinfo {year} {1932})}\BibitemShut {NoStop}%
\bibitem [{\citenamefont {Moyal}(1949)}]{Moyal1949}%
  \BibitemOpen
  \bibfield  {author} {\bibinfo {author} {\bibfnamefont {J.~E.}\ \bibnamefont
  {Moyal}},\ }\href {\doibase 10.1017/S0305004100000487} {\bibfield  {journal}
  {\bibinfo  {journal} {Math. Proc. Cambridge Phil. Soc.}\ }\textbf {\bibinfo
  {volume} {45}},\ \bibinfo {pages} {99} (\bibinfo {year} {1949})}\BibitemShut
  {NoStop}%
\bibitem [{\citenamefont {{Liboff}}(2003)}]{Liboff2003}%
  \BibitemOpen
  \bibfield  {author} {\bibinfo {author} {\bibfnamefont {R.~L.}\ \bibnamefont
  {{Liboff}}},\ }\href@noop {} {\emph {\bibinfo {title} {Kinetic theory :
  classical, quantum, and relativistic descriptions / Richard L.~Liboff.~ New
  York : Springer, c2003.~(Graduate texts in contemporary physics)}}}\
  (\bibinfo {year} {2003})\BibitemShut {NoStop}%
\bibitem [{\citenamefont {Klimontovich}\ and\ \citenamefont
  {Silin}(1960)}]{Klimontovich1960}%
  \BibitemOpen
  \bibfield  {author} {\bibinfo {author} {\bibfnamefont {Y.~L.}\ \bibnamefont
  {Klimontovich}}\ and\ \bibinfo {author} {\bibfnamefont {V.~P.}\ \bibnamefont
  {Silin}},\ }\href {http://stacks.iop.org/0038-5670/3/i=1/a=R04} {\bibfield
  {journal} {\bibinfo  {journal} {Soviet Physics Uspekhi}\ }\textbf {\bibinfo
  {volume} {3}},\ \bibinfo {pages} {84} (\bibinfo {year} {1960})}\BibitemShut
  {NoStop}%
\bibitem [{\citenamefont {Lindhard}(1954)}]{Lindhard1954}%
  \BibitemOpen
  \bibfield  {author} {\bibinfo {author} {\bibfnamefont {J.}~\bibnamefont
  {Lindhard}},\ }\href@noop {} {\bibfield  {journal} {\bibinfo  {journal} {Kgl.
  Danske Videnskab. Selskab Mat.-fys. Medd.}\ }\textbf {\bibinfo {volume} {Vol:
  28, No. 8}} (\bibinfo {year} {1954})}\BibitemShut {NoStop}%
\bibitem [{\citenamefont {Melrose}\ and\ \citenamefont
  {Mushtaq}(2010{\natexlab{b}})}]{Melrose2010a}%
  \BibitemOpen
  \bibfield  {author} {\bibinfo {author} {\bibfnamefont {D.~B.}\ \bibnamefont
  {Melrose}}\ and\ \bibinfo {author} {\bibfnamefont {A.}~\bibnamefont
  {Mushtaq}},\ }\href {\doibase 10.1063/1.3528272} {\bibfield  {journal}
  {\bibinfo  {journal} {Phys. Plasmas}\ }\textbf {\bibinfo {volume} {17}},\
  \bibinfo {eid} {122103} (\bibinfo {year} {2010}{\natexlab{b}})}\BibitemShut
  {NoStop}%
\bibitem [{\citenamefont {{Canuto}}\ and\ \citenamefont
  {{Ventura}}(1972)}]{Canuto1972}%
  \BibitemOpen
  \bibfield  {author} {\bibinfo {author} {\bibfnamefont {V.}~\bibnamefont
  {{Canuto}}}\ and\ \bibinfo {author} {\bibfnamefont {J.}~\bibnamefont
  {{Ventura}}},\ }\href {\doibase 10.1007/BF00645284} {\bibfield  {journal}
  {\bibinfo  {journal} {Astrophysics and Space Science}\ }\textbf {\bibinfo
  {volume} {18}},\ \bibinfo {pages} {104} (\bibinfo {year} {1972})}\BibitemShut
  {NoStop}%
\bibitem [{\citenamefont {{Bret}}(2007)}]{Bret2007}%
  \BibitemOpen
  \bibfield  {author} {\bibinfo {author} {\bibfnamefont {A.}~\bibnamefont
  {{Bret}}},\ }\href {\doibase 10.1063/1.2759886} {\bibfield  {journal}
  {\bibinfo  {journal} {Phys. Plasmas}\ }\textbf {\bibinfo {volume} {14}},\
  \bibinfo {pages} {084503} (\bibinfo {year} {2007})},\ \Eprint
  {http://arxiv.org/abs/0706.4374} {arXiv:0706.4374 [physics.plasm-ph]}
  \BibitemShut {NoStop}%
\bibitem [{\citenamefont {{Bret}}\ and\ \citenamefont
  {{Haas}}(2011)}]{Bret2011}%
  \BibitemOpen
  \bibfield  {author} {\bibinfo {author} {\bibfnamefont {A.}~\bibnamefont
  {{Bret}}}\ and\ \bibinfo {author} {\bibfnamefont {F.}~\bibnamefont
  {{Haas}}},\ }\href {\doibase 10.1063/1.3605470} {\bibfield  {journal}
  {\bibinfo  {journal} {Physics of Plasmas}\ }\textbf {\bibinfo {volume}
  {18}},\ \bibinfo {eid} {072108} (\bibinfo {year} {2011})},\ \Eprint
  {http://arxiv.org/abs/1103.4880} {arXiv:1103.4880 [physics.plasm-ph]}
  \BibitemShut {NoStop}%
\bibitem [{\citenamefont {{Haas}}\ and\ \citenamefont
  {{Lazar}}(2008)}]{Haas2008}%
  \BibitemOpen
  \bibfield  {author} {\bibinfo {author} {\bibfnamefont {F.}~\bibnamefont
  {{Haas}}}\ and\ \bibinfo {author} {\bibfnamefont {M.}~\bibnamefont
  {{Lazar}}},\ }\href {\doibase 10.1103/PhysRevE.77.046404} {\bibfield
  {journal} {\bibinfo  {journal} {Phys. Rev. E}\ }\textbf {\bibinfo {volume}
  {77}},\ \bibinfo {eid} {046404} (\bibinfo {year} {2008})},\ \Eprint
  {http://arxiv.org/abs/0801.4009} {arXiv:0801.4009 [physics.plasm-ph]}
  \BibitemShut {NoStop}%
\bibitem [{\citenamefont {{Haas}}(2008)}]{Haas2008a}%
  \BibitemOpen
  \bibfield  {author} {\bibinfo {author} {\bibfnamefont {F.}~\bibnamefont
  {{Haas}}},\ }\href {\doibase 10.1063/1.2829071} {\bibfield  {journal}
  {\bibinfo  {journal} {Physics of Plasmas}\ }\textbf {\bibinfo {volume}
  {15}},\ \bibinfo {eid} {022104} (\bibinfo {year} {2008})},\ \Eprint
  {http://arxiv.org/abs/0711.0851} {arXiv:0711.0851 [physics.plasm-ph]}
  \BibitemShut {NoStop}%
\bibitem [{\citenamefont {{Melrose}}(2013)}]{Melrose2013}%
  \BibitemOpen
  \bibinfo {editor} {\bibfnamefont {D.}~\bibnamefont {{Melrose}}},\ ed.,\ \href
  {\doibase 10.1007/978-1-4614-4045-1} {\emph {\bibinfo {title} {Lecture Notes
  in Physics, Berlin Springer Verlag}}},\ \bibinfo {series} {Lecture Notes in
  Physics, Berlin Springer Verlag}, Vol.\ \bibinfo {volume} {854}\ (\bibinfo
  {year} {2013})\BibitemShut {NoStop}%
\bibitem [{\citenamefont {{Marklund}}\ and\ \citenamefont
  {{Brodin}}(2007)}]{Marklund2007}%
  \BibitemOpen
  \bibfield  {author} {\bibinfo {author} {\bibfnamefont {M.}~\bibnamefont
  {{Marklund}}}\ and\ \bibinfo {author} {\bibfnamefont {G.}~\bibnamefont
  {{Brodin}}},\ }\href {\doibase 10.1103/PhysRevLett.98.025001} {\bibfield
  {journal} {\bibinfo  {journal} {Phys. Rev. Lett.}\ }\textbf {\bibinfo
  {volume} {98}},\ \bibinfo {eid} {025001} (\bibinfo {year} {2007})},\ \Eprint
  {http://arxiv.org/abs/physics/0612062} {physics/0612062} \BibitemShut
  {NoStop}%
\bibitem [{\citenamefont {{Brodin}}\ and\ \citenamefont
  {{Marklund}}(2008)}]{Brodin2008}%
  \BibitemOpen
  \bibfield  {author} {\bibinfo {author} {\bibfnamefont {G.}~\bibnamefont
  {{Brodin}}}\ and\ \bibinfo {author} {\bibfnamefont {M.}~\bibnamefont
  {{Marklund}}},\ }\href {\doibase 10.1088/1367-2630/10/11/115031} {\bibfield
  {journal} {\bibinfo  {journal} {New Journal of Physics}\ }\textbf {\bibinfo
  {volume} {10}},\ \bibinfo {eid} {115031} (\bibinfo {year} {2008})},\ \Eprint
  {http://arxiv.org/abs/0806.0912} {arXiv:0806.0912 [physics.plasm-ph]}
  \BibitemShut {NoStop}%
\bibitem [{\citenamefont {{Andreev}}(2015)}]{Andreev2015}%
  \BibitemOpen
  \bibfield  {author} {\bibinfo {author} {\bibfnamefont {P.~A.}\ \bibnamefont
  {{Andreev}}},\ }\href {\doibase 10.1063/1.4922662} {\bibfield  {journal}
  {\bibinfo  {journal} {Phys. Plasmas}\ }\textbf {\bibinfo {volume} {22}},\
  \bibinfo {eid} {062113} (\bibinfo {year} {2015})},\ \Eprint
  {http://arxiv.org/abs/1404.4899} {arXiv:1404.4899 [physics.plasm-ph]}
  \BibitemShut {NoStop}%
\bibitem [{\citenamefont {{Andreev}}(2016)}]{Andreev2016}%
  \BibitemOpen
  \bibfield  {author} {\bibinfo {author} {\bibfnamefont {P.~A.}\ \bibnamefont
  {{Andreev}}},\ }\href {\doibase 10.1142/S0217984916501803} {\bibfield
  {journal} {\bibinfo  {journal} {Mod. Phys. Lett. B}\ }\textbf {\bibinfo
  {volume} {30}},\ \bibinfo {eid} {1650180-162} (\bibinfo {year}
  {2016})}\BibitemShut {NoStop}%
\bibitem [{\citenamefont {{Andreev}}(2017)}]{Andreev2017}%
  \BibitemOpen
  \bibfield  {author} {\bibinfo {author} {\bibfnamefont {P.~A.}\ \bibnamefont
  {{Andreev}}},\ }\href {\doibase 10.1063/1.4975015} {\bibfield  {journal}
  {\bibinfo  {journal} {Physics of Plasmas}\ }\textbf {\bibinfo {volume}
  {24}},\ \bibinfo {eid} {022115} (\bibinfo {year} {2017})},\ \Eprint
  {http://arxiv.org/abs/1611.00046} {arXiv:1611.00046 [physics.plasm-ph]}
  \BibitemShut {NoStop}%
\bibitem [{\citenamefont {{Iqbal}}\ and\ \citenamefont
  {{Andreev}}(2016)}]{Iqbal2016}%
  \BibitemOpen
  \bibfield  {author} {\bibinfo {author} {\bibfnamefont {Z.}~\bibnamefont
  {{Iqbal}}}\ and\ \bibinfo {author} {\bibfnamefont {P.~A.}\ \bibnamefont
  {{Andreev}}},\ }\href {\doibase 10.1063/1.4954908} {\bibfield  {journal}
  {\bibinfo  {journal} {Physics of Plasmas}\ }\textbf {\bibinfo {volume}
  {23}},\ \bibinfo {eid} {062320} (\bibinfo {year} {2016})},\ \Eprint
  {http://arxiv.org/abs/1602.08640} {arXiv:1602.08640 [physics.plasm-ph]}
  \BibitemShut {NoStop}%
\bibitem [{\citenamefont {{Marklund}}\ \emph {et~al.}(2010)\citenamefont
  {{Marklund}}, \citenamefont {{Zamanian}},\ and\ \citenamefont
  {{Brodin}}}]{Marklund2010}%
  \BibitemOpen
  \bibfield  {author} {\bibinfo {author} {\bibfnamefont {M.}~\bibnamefont
  {{Marklund}}}, \bibinfo {author} {\bibfnamefont {J.}~\bibnamefont
  {{Zamanian}}}, \ and\ \bibinfo {author} {\bibfnamefont {G.}~\bibnamefont
  {{Brodin}}},\ }\href {\doibase 10.1080/00411450.2011.566502} {\bibfield
  {journal} {\bibinfo  {journal} {Transport Theory and Statistical Physics}\
  }\textbf {\bibinfo {volume} {39}},\ \bibinfo {pages} {502} (\bibinfo {year}
  {2010})},\ \Eprint {http://arxiv.org/abs/1002.0426} {arXiv:1002.0426
  [quant-ph]} \BibitemShut {NoStop}%
\bibitem [{\citenamefont {{Brodin}}\ \emph {et~al.}(2011)\citenamefont
  {{Brodin}}, \citenamefont {{Marklund}}, \citenamefont {{Zamanian}},\ and\
  \citenamefont {{Stefan}}}]{Brodin2011}%
  \BibitemOpen
  \bibfield  {author} {\bibinfo {author} {\bibfnamefont {G.}~\bibnamefont
  {{Brodin}}}, \bibinfo {author} {\bibfnamefont {M.}~\bibnamefont
  {{Marklund}}}, \bibinfo {author} {\bibfnamefont {J.}~\bibnamefont
  {{Zamanian}}}, \ and\ \bibinfo {author} {\bibfnamefont {M.}~\bibnamefont
  {{Stefan}}},\ }\href {\doibase 10.1088/0741-3335/53/7/074013} {\bibfield
  {journal} {\bibinfo  {journal} {Plasma Physics and Controlled Fusion}\
  }\textbf {\bibinfo {volume} {53}},\ \bibinfo {eid} {074013} (\bibinfo {year}
  {2011})},\ \Eprint {http://arxiv.org/abs/1010.0572} {arXiv:1010.0572}
  \BibitemShut {NoStop}%
\bibitem [{\citenamefont {Shukla}\ and\ \citenamefont
  {Eliasson}(2010)}]{Shukla2010}%
  \BibitemOpen
  \bibfield  {author} {\bibinfo {author} {\bibfnamefont {P.~K.}\ \bibnamefont
  {Shukla}}\ and\ \bibinfo {author} {\bibfnamefont {B.}~\bibnamefont
  {Eliasson}},\ }\href {http://stacks.iop.org/1063-7869/53/i=1/a=R02}
  {\bibfield  {journal} {\bibinfo  {journal} {Physics-Uspekhi}\ }\textbf
  {\bibinfo {volume} {53}},\ \bibinfo {pages} {51} (\bibinfo {year}
  {2010})}\BibitemShut {NoStop}%
\bibitem [{\citenamefont {Daligault}(2014)}]{Daligault2014}%
  \BibitemOpen
  \bibfield  {author} {\bibinfo {author} {\bibfnamefont {J.}~\bibnamefont
  {Daligault}},\ }\href {\doibase 10.1063/1.4873378} {\bibfield  {journal}
  {\bibinfo  {journal} {Phys. Plasmas}\ }\textbf {\bibinfo {volume} {21}},\
  \bibinfo {eid} {040701} (\bibinfo {year} {2014})}\BibitemShut {NoStop}%
\bibitem [{\citenamefont {Brodin}\ \emph {et~al.}(2015)\citenamefont {Brodin},
  \citenamefont {Zamanian},\ and\ \citenamefont {Mendonca}}]{Brodin2015}%
  \BibitemOpen
  \bibfield  {author} {\bibinfo {author} {\bibfnamefont {G.}~\bibnamefont
  {Brodin}}, \bibinfo {author} {\bibfnamefont {J.}~\bibnamefont {Zamanian}}, \
  and\ \bibinfo {author} {\bibfnamefont {J.~T.}\ \bibnamefont {Mendonca}},\
  }\href {http://stacks.iop.org/1402-4896/90/i=6/a=068020} {\bibfield
  {journal} {\bibinfo  {journal} {Phys. Scr.}\ }\textbf {\bibinfo {volume}
  {90}},\ \bibinfo {pages} {068020} (\bibinfo {year} {2015})}\BibitemShut
  {NoStop}%
\bibitem [{\citenamefont
  {{Akbari-Moghanjoughi}}(2013)}]{Akbari-Moghanjoughi2013}%
  \BibitemOpen
  \bibfield  {author} {\bibinfo {author} {\bibfnamefont {M.}~\bibnamefont
  {{Akbari-Moghanjoughi}}},\ }\href {\doibase 10.1063/1.4820802} {\bibfield
  {journal} {\bibinfo  {journal} {Physics of Plasmas}\ }\textbf {\bibinfo
  {volume} {20}},\ \bibinfo {eid} {092902} (\bibinfo {year}
  {2013})}\BibitemShut {NoStop}%
\bibitem [{\citenamefont {{Beloborodov}}\ and\ \citenamefont
  {{Thompson}}(2007)}]{Beloborodov2007}%
  \BibitemOpen
  \bibfield  {author} {\bibinfo {author} {\bibfnamefont {A.~M.}\ \bibnamefont
  {{Beloborodov}}}\ and\ \bibinfo {author} {\bibfnamefont {C.}~\bibnamefont
  {{Thompson}}},\ }\href {\doibase 10.1086/508917} {\bibfield  {journal}
  {\bibinfo  {journal} {The Astrophysical Journal}\ }\textbf {\bibinfo {volume}
  {657}},\ \bibinfo {pages} {967} (\bibinfo {year} {2007})},\ \Eprint
  {http://arxiv.org/abs/astro-ph/0602417} {astro-ph/0602417} \BibitemShut
  {NoStop}%
\bibitem [{\citenamefont {{Beloborodov}}(2009)}]{Beloborodov2009}%
  \BibitemOpen
  \bibfield  {author} {\bibinfo {author} {\bibfnamefont {A.~M.}\ \bibnamefont
  {{Beloborodov}}},\ }\href {\doibase 10.1088/0004-637X/703/1/1044} {\bibfield
  {journal} {\bibinfo  {journal} {The Astrophysical Journal}\ }\textbf
  {\bibinfo {volume} {703}},\ \bibinfo {pages} {1044} (\bibinfo {year}
  {2009})},\ \Eprint {http://arxiv.org/abs/0812.4873} {arXiv:0812.4873}
  \BibitemShut {NoStop}%
\bibitem [{\citenamefont {{Thompson}}\ and\ \citenamefont
  {{Beloborodov}}(2005)}]{Thompson2005}%
  \BibitemOpen
  \bibfield  {author} {\bibinfo {author} {\bibfnamefont {C.}~\bibnamefont
  {{Thompson}}}\ and\ \bibinfo {author} {\bibfnamefont {A.~M.}\ \bibnamefont
  {{Beloborodov}}},\ }\href {\doibase 10.1086/432245} {\bibfield  {journal}
  {\bibinfo  {journal} {The Astrophysical Journal}\ }\textbf {\bibinfo {volume}
  {634}},\ \bibinfo {pages} {565} (\bibinfo {year} {2005})},\ \Eprint
  {http://arxiv.org/abs/astro-ph/0408538} {astro-ph/0408538} \BibitemShut
  {NoStop}%
\bibitem [{\citenamefont {Uzdensky}\ and\ \citenamefont
  {Rightley}(2014)}]{Uzdensky2014}%
  \BibitemOpen
  \bibfield  {author} {\bibinfo {author} {\bibfnamefont {D.~A.}\ \bibnamefont
  {Uzdensky}}\ and\ \bibinfo {author} {\bibfnamefont {S.}~\bibnamefont
  {Rightley}},\ }\href {http://stacks.iop.org/0034-4885/77/i=3/a=036902}
  {\bibfield  {journal} {\bibinfo  {journal} {Reports on Progress in Physics}\
  }\textbf {\bibinfo {volume} {77}},\ \bibinfo {pages} {036902} (\bibinfo
  {year} {2014})}\BibitemShut {NoStop}%
\bibitem [{\citenamefont {{Haas}}\ and\ \citenamefont
  {{Shukla}}(2009)}]{Haas2009}%
  \BibitemOpen
  \bibfield  {author} {\bibinfo {author} {\bibfnamefont {F.}~\bibnamefont
  {{Haas}}}\ and\ \bibinfo {author} {\bibfnamefont {P.~K.}\ \bibnamefont
  {{Shukla}}},\ }\href {\doibase 10.1103/PhysRevE.79.066402} {\bibfield
  {journal} {\bibinfo  {journal} {Phys. Rev. E}\ }\textbf {\bibinfo {volume}
  {79}},\ \bibinfo {eid} {066402} (\bibinfo {year} {2009})},\ \Eprint
  {http://arxiv.org/abs/0902.3584} {arXiv:0902.3584 [physics.plasm-ph]}
  \BibitemShut {NoStop}%
\end{thebibliography}%

\end{document}